\documentclass[12 pt]{JHEP3}
\usepackage[centertags]{amsmath}
\usepackage{amssymb}
\newcommand{\half}{{1 \over 2}}
\newcommand{\del}{\partial}
\def\bft{\begin{footnote}}
\def\eft{\end{footnote}}
\def\gap#1{\vspace{#1 ex}}
\def\be{\begin{equation}}
\def\ee{\end{equation}}
\def\ba{\begin{array}{l}}
\def\ea{\end{array}}
\def\bea{\begin{eqnarray}}
\def\eea{\end{eqnarray}}
\def\beas{\begin{eqnarray*}}
\def\eeas{\end{eqnarray*}}
\def\eq#1{(\ref{#1})}

\def\nn{\nonumber\\}

\def\c#1{{\hat{#1}}}

\def\eps{{\epsilon}}
\def\ads{$AdS_5 \times S^5$}
\newcommand{\asq}{{\cal A}^2}
\def\mysec#1{\gap1 {\bf #1}\gap1} 
\def\one{{\hbox{1\kern-.88mm l}}} 
\date{}
\title {Supersymmetric Giant Graviton Solutions in $AdS_3$}
\author{Gautam Mandal$^{a}$,
Suvrat Raju$^{a,b}$ and
Mikael Smedb\"ack$^{a,c}$ \\
\small{\emph{$^{a}$Department of Theoretical Physics,
                   Tata Institute of Fundamental Research,}}\\
\small{\emph{Homi Bhabha Road, Mumbai 400005, India}}  \\
\small{\emph{$^{b}$Department of Physics, Harvard University, Cambridge MA 02138, USA}}\\
\small{\emph{$^{c}$Department of Particle Physics,
                   Weizmann Institute of Science, Rehovot 76100, Israel}} }
\preprint{
  TIFR/TH/07-16 \\
    HUTP-07/A0004 \\
    WIS/13/07-AUG-DPP\\
    \texttt{arXiv:0709.1168 [hep-th]}
}

\abstract{
We parameterize all classical probe brane configurations that preserve
4 supersymmetries in (a) the extremal D1-D5 geometry, (b) the extremal
D1-D5-P geometry, (c) the smooth D1-D5 solutions proposed by Lunin and
Mathur and (d) global $AdS_3 \times S_3 \times T^4/K3$. These
configurations consist of D1 branes, D5 branes and bound states of D5
and D1 branes with the property that a particular Killing vector is
tangent to the brane worldvolume at each point. We show that the
supersymmetric sector of the D5 brane worldvolume theory may be
analyzed in an effective 1+1 dimensional framework that places it on the
same footing as D1 branes. In global AdS and the corresponding 
Lunin-Mathur solution, the solutions we describe are `bound' to the
 center of AdS for generic parameters and cannot
escape to infinity. We show that these probes only exist on the
submanifold of moduli space where the background $B_{NS}$ field and
theta angle vanish. We quantize these probes in the near horizon region of
the extremal D1-D5 geometry and obtain the theory of long strings discussed by
Seiberg and Witten.}

\begin{document}
%\maketitle
%\tableofcontents
\section{Introduction}

Despite many advances, quantizing string theory in non-trivial
spacetime backgrounds remains a difficult task. In the past few years,
some progress has been made by approaching this problem using
canonical methods
\cite{Grant:2005qc,Mandal:2005wv,Maoz:2005nk,Rychkov:2005ji,Biswas:2006tj,Mandal:2006tk,Martelli:2006vh,Basu:2006id}. The principle behind these studies is that if one can
understand a subsector of the {\em classical} theory well
enough it may be possible to quantize it autonomously and obtain a
sector of the Hilbert space of the full quantum theory. This procedure
can only work if the canonical structure of the classical phase space
`decouples' this sector from the rest of the theory. The
studies above suggest that supersymmetric sectors, such as the one we
will study here, often satisfy this criterion.

Since the space of all classical solutions of a theory is isomorphic
to its classical phase-space, it is of interest if one can obtain a
complete parameterization of even a special subsector of classical
solutions. This subsector can then be quantized using the methods of
\cite{dedecker1953cvf}(See \cite{Lee:1990nz} for a review).  In this
paper, we pursue this programme by parameterizing all classical
supersymmetric brane probes moving in (a) the extremal D1-D5 background, 
(b) the extremal D1-D5-P background, (c) the smooth geometries 
proposed in \cite{Lunin:2001fv,Lunin:2002bj,Lunin:2002iz} with the 
same charges as the D1-D5 system and (d) global $AdS_3
\times S^3 \times T^4/K3$.

The physical significance of these backgrounds is as follows. The AdS/CFT conjecture\cite{Maldacena:1997re,Aharony:1999ti} relates type IIB
string theory on global $AdS_3$ to the NS sector of a 1+1 dimensional 
CFT on its boundary. The solutions in global AdS we find below correspond to 
the $1/4$ BPS
sector of the CFT of the Higgs branch.
On the boundary, the NS and R sectors are related by an operation called `spectral flow'. Performing this operation on the supergravity solution for global $AdS$ yields the near horizon region of one of the 
solutions of Lunin and Mathur \cite{Lunin:2002bj}. This corresponds to the specific Ramond ground state obtained by spectrally flowing the NS vacuum. 
Other Ramond vacua are described by other solutions in \cite{Lunin:2002bj}.
The zero mass BTZ black hole which is the near-horizon of the extremal D1-D5 geometry, on the other hand, has been argued to be an `average' over all 
Ramond ground states.

The giant graviton brane probes we find comprise D1 branes, D5 branes
and bound states of D1 and D5 branes. As we make more precise in
section \ref{probe} we find that these 
supersymmetric
probes have the property that a certain Killing vector is tangent to
the brane worldvolume at each point. Hence, given the shape of the
brane at any one point of time, one can translate it in time along the
integral curves of this Killing vector to obtain the entire brane 
worldvolume. The set of all
solutions is parameterized by the set of all initial shapes. This
simple prescription is sufficient to describe supersymmetric probes in
all the backgrounds we mentioned above.

Surprisingly, we find that the symplectic structure on these classical
solutions is such that we can describe all the solutions above,
including supersymmetric solutions to the DBI action on the 6
dimensional D5 brane worldvolume, in a unified 1+1 dimensional
framework. It is well known that the infra-red limit of the world
volume theory of a bound state of D1 branes and D5 branes, in flat
space, is given by a 1+1 dimensional sigma model. However, our result
which we emphasize is classical, is valid in curved backgrounds and
does not rely on taking the infra-red limit.
% above does not rely on taking the infra-red limit and is also valid
%in curved backgrounds.  {\underline{Classically}},  it suffices
%to focus on supersymmetric configurations. 

Our probes exist on the submanifold of moduli space where the background
 NS-NS fluxes and theta angle are set to zero. On this submanifold, the 
boundary theory is known to be singular because the stack of D1 and D5 branes
that make up the background can separate at no cost in energy \cite{Seiberg:1999xz}.  
One may wonder then, whether the probes we find are artifacts 
of this singularity, {\em i.e}, whether they merely represent breakaway
D1-D5 subsystems which can escape to infinity. In global $AdS$, and 
in the Ramond sector solution dual to global $AdS$, this is not the case.
 In these geometries, for generic
parameters, the 1/4 BPS giant gravitons that we describe, are `bound'
to the center of AdS and cannot escape to infinity. This indicates
that they correspond to discrete states and not to states in a 
continuum.  In the boundary theory this
means that they correspond to BPS states that are {\em not}
localized about the singularities of the Higgs branch. 
Averaging over the Ramond vacua to produce the zero mass BTZ black hole, however,
washes out the structure of these discrete bound states and the only solutions 
we are left with are at the bottom of a continuum of non-supersymmetric states.

We prove that no BPS probes survive if we turn on a
small NS-NS field. This is not a contradiction for it merely means
that the ${1 \over 4}$ BPS partition function jumps as we move off
this submanifold of moduli space. Further investigation of this issue
in the quantum theory and of protected quantities, like the elliptic
genus and the spectrum of chiral-chiral primaries is left to
\cite{Raju}.

Giant gravitons in $AdS_3$ have been considered previously
\cite{McGreevy:2000cw,Grisaru:2000zn,Lunin:2002bj,Huang:2006te} and it was noted that
regular 1/2 BPS brane configurations exist only for specific values
of the charges. These are precisely the values at which the giant
gravitons we describe can escape to `infinity' in global AdS. The
moduli space of 1/4 BPS giant gravitons, however, is far richer and
this is what we will concern ourselves with in this paper.

A brief outline of this paper is as follows. In Section
\ref{Killingspinor}, we perform a Killing spinor and kappa symmetry
analysis to determine the conditions that 
D brane probes, in the four backgrounds above, must obey
in order to be supersymmetric. Using this insight, in
section \ref{susydstringsdbi} we explicitly construct supersymmetric
D1 brane solutions in these backgrounds and verify that they satisfy
the BPS bound. Then, in section \ref{boundstates} we show how bound
states of D1 and D5 branes(represented by D5 branes with gauge fields
turned on in their worldvolume) can also be described in the framework
of section \ref{susydstringsdbi}. In section \ref{movingoffspecial} we
discuss the effect of turning on background NS-NS fluxes.  In
section \ref{quantizationsection} we discuss the quantization of
probes moving in the near horizon region of the D1-D5 background. In section
\ref{discussionsection}, we conclude with a summary of our results and
their implications. Appendices \ref{miscdetails}--\ref{sec:gauge} discuss 
some technical details while in Appendices \ref{appD1D5} and \ref{appglobal} we discuss Killing spinor equations for various D1-D5 geometries and global AdS.

\section{Killing spinor and kappa symmetry analysis\label{Killingspinor}}

We consider type IIB superstring theory compactified on $S^1 \times
{\cal K}$ where ${\cal K}$ is $T^4$ or $K3$. We will concentrate on
the case of $T^4$, unless otherwise stated. Let us parameterize
$S^1$ by the coordinate $x_5$, $T^4$ by $x^6,x^7,x^8,x^9$ and the
noncompact spatial directions by $x^1, x^2, x^3, x^4$. We will use
coordinate indices $x^M, M=0,1,\ldots, 9$;
$x^m, m=1,2,3,4$; $x^{a} ~{\rm or}~ x^{i}, a,i = 6,7,8,9$.  We will parameterize the 32 supersymmetries
of IIB theory by two real constant chiral spinors $\epsilon_1$ and
$\epsilon_2$, or equivalently by a single complex chiral spinor
$\eps=\eps_1+i\eps_2$.

In Section \ref{background} we will review the preserved supersymmetries, or
the Killing spinors, of the backgrounds (a) D1-D5, (b) D1-D5-P, (c)
Lunin-Mathur geometries and (d) Global AdS$_3 \times S^3$.  In Section
\ref{probe} we will describe the construction of supersymmetric probe
branes, using a kappa-symmetry analysis, which preserve a certain subset
of the supersymmetries of the background geometry.

\subsection{Review of supersymmetry of the backgrounds\label{background}}

\subsubsection{\label{flat-d1-d5}
SUSY of D1-D5 and D1-D5-P in the Flat space approximation}

We first consider the D1-D5 system, which 
consists of $Q_1$ D1 branes wrapped on the $S^1$ and
$Q_5$ D5 branes wrapped on $S^1\times T^4$. 
Let us first compute the supersymmetries of the background ignoring
back-reaction.  In this approximation we
regard the $Q_1$ D1 branes and the $Q_5$ D5 branes as placed in
flat space.  The residual supersymmetries of the
system can be figured out in the following way. A D1 brane wrapped on
the $S^1$ preserves the supersymmetry
\footnote{\label{ft:gamma}
We will denote by $\Gamma_{\c M}$ the flat space
Gamma-matrices satisfying $[\Gamma_{\c M}, \Gamma_{\c N}]= 2 \eta_{\c
M, \c N}$, By contrast, Gamma matrices in a curved space, $\Gamma_M$
will defined by $\Gamma_M= \Gamma_{\c M}e^{\c M}_M$ where $e^{\c M}$
are the vielbeins. In the flat space approximation, $\Gamma_M
= \Gamma_{\c M}.$}
\begin{equation}\label{dosusy}
\Gamma_{\c 0}\Gamma_{\c 5} \eps =- i \eps^*.
\end{equation} 
Similarly, a D5 brane wrapped on $S^1 \times T^4$ 
preserves the supersymmetry 
\begin{equation}
\label{dfsusy}
\Gamma_{\c0}\Gamma_{\c5} \Gamma_{\c6}\Gamma_{\c7}
\Gamma_{\c8}\Gamma_{\c9} \eps = - i \eps^*.
\end{equation} 
The above equations can be derived by 
considering the BPS relations
arising from IIB SUSY algebra or by considering the $\kappa$-symmetry
condition on the DBI description of a D1 or D5 brane. A combined 
system of D1 and D5 branes will therefore preserve eight supersymmetries
given by $\eps$'s which satisfy both \eq{dosusy} and \eq{dfsusy}.

For later reference, we set up some notation.  The eight residual
supersymmetries of the D1-D5 system can be described as 
satisfying  either
\bea
\Gamma_{\c6}\Gamma_{\c7}
\Gamma_{\c8}\Gamma_{\c9} \eps = \eps,
\Gamma_{\c 0}\Gamma_{\c 5} \eps =- \eps, \, \eps=i\eps^* 
\label{left-susy}
\eea 
or
\bea
\Gamma_{\c6}\Gamma_{\c7}
\Gamma_{\c8}\Gamma_{\c9} \eps = \eps,
\Gamma_{\c 0}\Gamma_{\c 5} \eps = \eps, \, \eps=-i\eps^* .
\label{right-susy}
\eea 
The two conditions above are called left- and right-moving
supersymmetries, respectively. Thus the D1-D5 system has
(4,4) (left,right) supersymmetries.

\mysec{D1-D5-P}

If we add to the D1-D5 system $P$ units of left-moving momentum along
the $S^1$, the resulting D1-D5-P system has (0,4) supersymmetry
(defined by \eq{right-susy}), in the notation of the previous
paragraph.\begin{footnote}{We adopt the slightly unusual terminology that
a wave rotating counterclockwise on the $S^1$ is
left-moving.}\end{footnote} In the flat space limit and for non-compact
$x_5$, a left-moving momentum can be seen as arising from applying an
infinite boost to the D1-D5 system in the $t$-$x_5$ plane. It is easy
to see that the right-moving supersymmetries are invariant under such
a boost while the left-moving supersymmetries are not. Since the
supersymmetry conditions are local, the argument can be extended to
the case where $x_5$ is compact.

\subsubsection{SUSY of the full D1-D5 and D1-D5-P geometry}

It has been assumed above that the $Q_1$ D1 branes and $Q_5$ D5 branes
are in flat space. For $Q_1, Q_5$ large, the metric, dilaton and the
RR fields get deformed. The modified background geometry, applying
standard constructions, is given by the `D1-D5' geometry, described in
Table \eqref{donefive}.  This geometry should be thought of as describing 
an `ensemble' rather than any particular microstate of the D1-D5 system. 
In case of the D1-D5-P the backreacted metric
is given in \eq{donefivep} (the dilaton and RR fields are given by
Table \eqref{donefive}).

To analyze unbroken supersymmetries of these backgrounds and the others 
to follow, we need to solve the Killing spinor equations in these
backgrounds. These Killing spinors were considered, in fact for a much
larger class of metrics, in
\cite{Bergshoeff:1992cw,Bergshoeff:1994dg}. We quote the results of
this analysis here, with a very brief introduction, and explain
details, for each of the cases, in Appendix \ref{appD1D5}.

In case of the D1-D5 geometry and the other geometries we consider
below, the metric may always be written in terms of vielbeins, as:
\begin{equation}
d s^2 = -(e^{\hat t})^2 + (e^{\hat 5})^2 + e^{\hat{m}} 
e^{\hat{m}} + e^{\hat{a}} e^{\hat{a}}.
\end{equation}
The coordinate indices are as explained in the beginning of Section
\ref{Killingspinor}. The $\c{()}$ represents a flat space index
(vielbein label).  Spinors are defined with respect to a specific
choice of vielbeins and they transform in the spinorial representation
under a $SO(1,9)$ rotation of the vielbeins. The precise form of the
vielbein, in the geometries we consider, may be found in Appendix
\ref{allvielbeins}.

Finding the residual supersymmetries of a
particular background amounts to solving the Killing spinor equations
which are obtained by setting to zero the dilatino variation
\eq{dilatinoequation} and the gravitino variation
\eq{gravitinoequation}. The analysis in Appendix
\ref{appD1D5} tells us that \eq{dosusy}, \eq{dfsusy} continue
to describe the supersymmetries of the D1-D5 geometry, while
\eq{right-susy} continues to describe the
supersymmetries of the D1-D5-P geometry.

\subsubsection{\label{sec:lunin-mathur-bgd}SUSY of Lunin-Mathur geometries}

It was explained in a sequence of papers \cite{Maldacena:2000dr,Balasubramanian:2000rt,Lunin:2001fv,Lunin:2002bj,Lunin:2002iz} that the geometry of Table \ref{donefive} should be treated as an `average' over several allowed D1-D5 microstates. The gravity solution dual to any particular Ramond groundstate was described by Lunin and Mathur \cite{Lunin:2001fv,Lunin:2002bj}. The analysis of \cite{Bergshoeff:1992cw,Bergshoeff:1994dg} and Appendix \ref{appD1D5} shows that even these solutions preserve the supersymmetries given by  \eqref{dosusy} and \eqref{dfsusy}.

\subsubsection{\label{background-global}SUSY of Global AdS$_3\times S^3
\times T^4$}

Type IIB string theory on global AdS$_3$ is dual to the NS sector of the 
CFT on the boundary. If we take the geometry to be $AdS_3 \times S^3 \times T^4$, the boundary CFT has $(4,4)$ superconformal symmetry. 
We will describe these supersymmetries below.

Global $AdS_3 \times S^3$ is described by the metric
\begin{equation}
\label{globmet} 
ds^2=-\cosh^2 \rho dt^2 +\sinh^2 \rho d
\theta^2 + d\rho^2+ \cos^2 \zeta d\phi_1^2 + \sin^2 \zeta d \phi_2^2
+ d \zeta^2.
\end{equation} 
We will find the bulk Killing spinors of this background
in two ways. In Appendix \ref{appglobal}, we will find them
by explicitly solving the IIB Killing spinor equations in a manner similar
to \cite{Lu:1998nu}.
Below we will find them in an alternative method, due to Mikhailov
\cite{Mikhailov:2000ya}, which is quite illuminating.

The metric \eq{globmet} arises by embedding (a) $AdS_3$ in flat
$R^{2,2}$ by the equations $X^{-1}=\cosh \rho \cos t$, $X^0=\cosh \rho
\sin t$, $X^1= \sinh \rho \cos \theta$, $X^2=\sinh \rho \sin \theta$
and (b) $S^3$ in flat $R^4$ by the equations $Y^1= \cos \zeta \cos
\phi^1$, $Y^2= \cos \zeta \sin \phi_1$, $Y^3= \sin \zeta \cos \phi_2$,
$Y^4= \sin \zeta \sin \phi_2$. We can therefore regard $AdS_3 \times
S^3 \times T^4$ as embedded in $R^{2,10}$ as a codimension two
submanifold.

Now consider $R^{2,10}$ spinors that are simultaneously real and
chiral. Regard $R^{2,10}$ as a product of $R^{2,2} (\supset AdS_3)$,
$R^4 (\supset S^3)$, and $R^4$ (which we compactify to get the
$T^4$). The spinors now should be regarded as transforming under
$SO(2,2) \times SO(4) \times SO(4)$.  It is possible to consistently
restrict attention to a subclass of these spinors, namely those that
are chiral under the last $SO(4)$ (this is consistent because complex
conjugation does not change $SO(4)$ spinor chirality). We now have a
set of 16 real or 8 complex spinors. These spinors are chiral in
$R^{2, 6}$ as well as in $R^4$. We will denote these spinors by
$\chi$.

Let us denote by $\tilde \Gamma_A, A=-1,0,1,..,10$ the $R^{2,10}$
gamma-matrices. We define by $N_{AdS}$ the vector in $R^{2, 2}$ which
is normal to the AdS$_3$ submanifold and by $N_S$ the vector in
$R^4$ which is the normal to $S^3$.  The prescription of
\cite{Mikhailov:2000ya}  is that the Killing
spinors are given by
\begin{equation} 
\eps= \left(1+ \left( \tilde{\Gamma} \cdot N_{AdS}\right) 
\left(\tilde{\Gamma} \cdot N_{S}\right) \right) \chi.
\label{global-susy}
\end{equation} 
where $\chi$ are the $R^{2,10}$ spinors constrained as in the previous
paragraph. The two normal gamma matrices are explicitly given by
$\tilde\Gamma \cdot N_{AdS}= (X^{-1} \tilde \Gamma_{-1} + X^0 \tilde
\Gamma_{0} + X^1 \tilde \Gamma_1 + X^2 \tilde \Gamma_2)$ and
$\tilde\Gamma \cdot N_{S}= X^{3} \tilde \Gamma_{3} + X^4 \tilde \Gamma_{4} +
X^5 \tilde \Gamma_5 + X^6 \tilde \Gamma_6$.
 
In Appendix \ref{appglobal} we show that the 16 real spinors defined by
\eq{global-susy} are the same as the ones obtained from
directly solving the IIB Killing spinor equations.  

\subsection{Construction of supersymmetric probes\label{probe}}

\subsubsection{D1 probe in D1-D5/D1-D5-P background: flat space
approximation}

We first construct supersymmetric D1 brane probes in the D1-D5
background, in the approximation
described in Sec \ref{flat-d1-d5}.  
Consider a probe D-string executing some motion in this
background.

In this subsection we demonstrate that this probe preserves all the
right-moving supercharges of the background  (corresponding
to supersymmetry transformations \eq{right-susy}), provided 
its motion is such that:
\begin{description}
\item[1.] The vector 
\be
{\bf n}={\partial \over \partial t}+{\partial \over \partial x_5}
\label{n-d1d5}
\ee 
is tangent to the brane worldvolume at every point.
\item[2.] 
The brane always maintains a positive orientation with respect to the 
branes that make up the background. 
\end{description} 

We will first prove these statements, and then return, at the end of this 
subsection, to an elaboration of their meaning. 

According to assumption 1 above, ${\bf n}$ 
is tangent to the worldvolume at every 
point. A second, linearly
independent, tangent vector may be chosen at each point  so that the 
coefficient of ${\partial \over \partial t}$ is zero; making this 
choice this normalized vector may be written as ${\bf v_2}= 
\sin \alpha {\partial \over \partial x_5} + \cos \alpha 
\ {\bf u}$ where ${\bf u}$ represents 
a spacelike unit vector orthogonal to $x_5$. By assumption
2, we have $\sin\alpha > 0$\footnote{When 
$\sin \alpha$ is less than zero the ${\bf v_1}$ and ${\bf v_2}$ are not
appropriately oriented. Also $\alpha \neq 0$, because in that case, the determinant
of the induced worldsheet metric would vanish.}. In general the direction of  ${\bf u}$ 
 and the value of $\alpha$ will vary as a function of world volume 
coordinates. Although ${\bf n},{\bf v_2}$ are linearly independent, they
are not an orthonormal set since ${\bf n}$ is a
null vector. We can construct an orthonormal basis of vectors
${\bf v_1}, {\bf v_2}$ at each point of the world volume by the Gram-Schmidt  
method, yielding 
\begin{equation}
\label{v1v2def}
{\bf v_1}= {\bf n}/ \sin
\alpha - {\bf v_2}= 1/\sin \alpha 
\left({\partial \over \partial t} + 
\cos^2 \alpha {\partial \over \partial x_5} -\cos\alpha \sin
\alpha\ {\bf u} \right). 
\end{equation}
For the probe to preserve some
supersymmetry $\eps$ we must have, at each point of the
world-volume, 
\begin{equation}\label{probedo}
\Gamma_{{\bf v_1}} \Gamma_{{\bf v_2}} \eps =- i \eps^*.
\end{equation}
The above equation is equivalent to
\be
\left[ \Gamma_{\hat{0}}\Gamma_{\hat{5}} - \frac{\Gamma_{\bf u}}{\sin\alpha} \left(
\cos\alpha \Gamma_{\hat{0}} + (\sin^2\alpha\cos \alpha 
+ \cos^3\alpha)\Gamma_{\hat{5}} \right)
\right] \eps =- i \eps^* .
\ee
This is clearly satisfied by spinors that satisfy \eq{right-susy} since
\eq{right-susy} implies that $\Gamma_0\Gamma_5 \eps=\eps$ 
which ensures 
$\Gamma_0 \eps= -\Gamma_5 \eps$ and a consequent
vanishing of the coefficient of $\Gamma_{\bf u}$ above.  Note that in flat
space the  $\Gamma_{\c M}=\Gamma_M$.\footnote{This
 derivation does not work for left-moving supercharges where \eqref{left-susy} implies $\Gamma_0 \epsilon = + \Gamma_5 \epsilon$. Left moving supercharges
are symmetries for D1-branes that move at the speed of light to the
left (branes whose tangent space includes $(1,-1,0, \ldots 0)$).}

The conditions 1 and 2, listed at the beginning of this subsection are 
easily solved by choosing a world-sheet parameterization in terms of 
coordinates $\sigma, \tau$, such that
\bea
&
x^M = {\bf n}^M \tau + x^M(\sigma),
\nn
&
x^0= \tau,
x_5=x_5(\sigma)+ \tau, x^q = x^q(\sigma), q=1,2,3,4,6,7,8,9
\label{parameterization}
\eea
where
$x_5(\sigma), x^q(\sigma)$ are arbitrary functions, 
except that  $\del_\sigma x_5> 0$. To connect with the earlier discussion,
we identify ${\bf v_2}$ as the unit vector along 
${\bf s}^M \equiv \del_\sigma x^M $. Note that by condition (2) above 
we need 
$
\del_\sigma x_5=({\bf n}, {\bf s})> 0 
$
which is equivalent to our earlier condition $\sin\alpha > 0$. 
This constraint 
together with the periodicity of configurations in $\sigma$, implies that 
$\int d \sigma x_5(\sigma) = 2 \pi R w$, where $R$ is the radius of the $x_5$ 
circle, and $w$ is a positive integer that we will refer to as the winding 
number. The configurations described in this paragraph are easy to visualize. 
They consist of D-strings with arbitrary transverse profiles, winding the $x_5$
direction $w$ times, and moving bodily at the speed of light in the positive 
$x_5$ direction. 

Eqn. \eq{probedo} is equivalent to the $\kappa$-symmetry
projection, which can alternatively be written as 
\bea 
& \Gamma \eps=
i \eps^*,~~ \Gamma:= \half \Gamma_{MN}\del_\alpha x^M \del_\beta x^N
\eps^{\alpha\beta}/\sqrt{-h} 
\nn
&
=\half[\Gamma_{\bf n}, \Gamma_{{\bf s}}]/\sqrt{-h} = 
\Gamma_{{\bf v_1}} \Gamma_{{\bf v_2}} ,
\label{kappa}
\eea
where $h$ is the determinant of the induced metric on the
world volume in the $\sigma, \tau$ coordinates above. 
In the second line we have used the parameterization
\eq{parameterization}. This is equivalent to \eq{probedo}
by using $\sqrt{-h}= \sin\alpha |{\bf s}| $.

Since all we needed in the above discussion is the (0,4) supersymmetry
\eq{right-susy} of the background, the above discussion goes through
unchanged for D1 probes in the D1-D5-P background in the flat space
approximation.

\subsubsection{D1 probe in D1-D5/D1-D5-P background}

We now  consider the curved D1-D5-P background, described in 
\eq{donefivep}. The specialization to the D1-D5 background
is straightforward (we just need to put $r_p=0$).  
We will show that \eq{parameterization}, or
equivalently, the condition that ${\bf n}
= \del_t + \del_5$ is tangent to the world volume,
again ensures the appropriate supersymmetry of the probe.  For this,
we need to show that \eq{kappa} is valid in this background. 
We find that (see, \eq{inducedeterminant})
\bea
&
\sqrt{-h} = \dot X \cdot X' \equiv {\bf n} \cdot {\bf s}
= x_5' (g_{05} + g_{55}) ,
\nn
&
\Gamma \eps =1/(2\sqrt{-h})
[\Gamma_{\bf n}, \Gamma_{{\bf s}}]\eps
= \frac{1}{(g_{05} + g_{55}) x_5'} \left( \Gamma_{0 5} x_5' +
(\Gamma_0 + \Gamma_5)\Gamma_q x_q' \right)\eps .
\label{sqrt-h-d1-in15p}
\eea 
To show that $\Gamma \eps= \eps$ we need 
\bea
&
\Gamma_0 \eps =- \Gamma_5 \eps, 
\nn
&
(g_{05} + g_{55})^{-1}\left( \Gamma_0 \Gamma_5 \right)\eps = \eps .
\label{d1-5-probe-conditions}
\eea
The first line is equivalent to
\bea
&
e^{\c0}_0 \Gamma_{\c 0} \eps = - \left(
e^{\c0}_5 \Gamma_{\c 0} + e^{\c5}_5\Gamma_{\c5} \right) \eps .
\eea
After explicitly inserting the vielbeins using equations 
\eqref{allvielbeinsd1d5} and \eqref{allvielbeinsd1d5p} we are
left with
\bea
&
\Gamma_{\c0} \eps =- \Gamma_{\c5} \eps ,
\eea
which is equivalent to $\Gamma_{\c0} \Gamma_{\c5}\eps = \eps$.
The second line of \eq{d1-5-probe-conditions}  gives rise to
the same condition
\bea  
\Gamma_{\c0} \Gamma_{\c5}\eps = \eps ,
\eea
by using $e^{\c0}_0 e^{\c5}_5= g_{05} + g_{55}$.

Thus, we have shown that a D1 brane probe moving such that ${\bf n} =
 \del_t + \del_5$ is always  tangent to the world-volume,
equivalently satisfying Eqn. \eq{parameterization}, 
preserves the supersymmetry \eq{right-susy}. 

\subsubsection{D1 probe in Lunin-Mathur background}

We now show that the same condition as in the previous subsection,
namely that ${\bf n}$ should be everywhere tangent to the
world-volume of the D1 brane (alternatively, that the D1 brane  
embedding can be expressed as in \eq{parameterization})
is valid for supersymmetry of D1 probes in the background 
\eq{luninmetric}, discussed in Section \ref{sec:lunin-mathur-bgd} above.
This analysis is fairly similar to the one above. In this
case, Eqn. \eq{sqrt-h-d1-in15p}
changes to
\bea
&
\sqrt{-h} = \dot X \cdot X'
\equiv {\bf n} \cdot {\bf s} = x_5' g_{55} + x_m' (g_{0m} + g_{5m}).
\eea
Hence
\bea
&\Gamma \eps&=1/(2\sqrt{-h})
[\Gamma_{\bf n}, \Gamma_{{\bf s}}]\eps
=\left(x_5' g_{55} + x_m' (g_{0m} + g_{5m})
\right)^{-1}\left(\Gamma_{0 5}x_5' +
\half x_q'[(\Gamma_0 + \Gamma_5),\Gamma_q] \right)\eps
\nn
&&
=\left(x_5' g_{55} + x_m' ((g_{0m} + g_{5m})\right)^{-1}
\left(\Gamma_{\c0 \c5}x_5'g_{55} +  x_q' ((g_{0q} + g_{5q})
- \Gamma_q (\Gamma_0 + \Gamma_5)  \right)\eps)
\label{lunin-gamma-eps}
\eea 
Thus, if $\Gamma_{\c0\c5}\eps
=\eps$, as in \eq{right-susy}, (which also implies
$(\Gamma_0 + \Gamma_5) \eps=0$, using
$e^{\c0}_0 = e^{\c5}_5$), the expression
\eq{lunin-gamma-eps}, evaluates to $\Gamma\eps=\eps$.
For spinors satisfying \eq{right-susy} this also
implies  $\Gamma\eps= i\eps^*$ which is the
kappa-symmetry projection condition. In the
last step of  \eq{lunin-gamma-eps} we have used
\[
\Gamma_{05} = g_{55}\Gamma_{\c0 \c5},\;
\half [\Gamma_0 + \Gamma_5,\Gamma_m]=
\half\{\Gamma_0 + \Gamma_5,\Gamma_m\} - \Gamma_m (\Gamma_0 + \Gamma_5)
= 
(g_{0m} + g_{5m}) - \Gamma_m (\Gamma_0 + \Gamma_5)
\]

\subsubsection{\label{d1-in-global}D1 probe in  Global $AdS_3 \times S^3$}

We will use the description of supersymmetries of the
background as in Section \ref{background-global}.
We will show in this section that D1 strings with 
world volumes, to which  
\bea
{\bf n}=\partial_t +  \partial_\theta +
\partial_{\phi_1} +\partial_{\phi_2}
\label{global-tangent}
\eea
is everywhere tangent, preserve 4 supercharges.

We will first mention the geometric significance of ${\bf n}$.  Let us
group the $R^{2,6}$ (see Section \ref{background-global}) coordinates into
complex numbers as $X^{-1}+i X^0$, $X^1+iX^2$, $Y^1+iY^2$,
$Y^3+iY^4$. This defines a complex structure $I$ on $R^{2,6}$. In
Section \ref{background-global}, we have defined $N_{AdS}$ as the normal to
$AdS_3$ in $R^{2,2}$ and $N_S$ as the normal to $S^3$ in $R^4$.  It is
easy to check that the complex partner of $N_{AdS}$ is
$I(N_{AdS})=-\partial_t - \partial_\theta$, which generates (twice) the
right-moving conformal spin $2h_r$.  Similarly, the complex 
partner of $N_S$ is $I(N_{S})= \partial_{\phi^1} +\partial_{\phi_2}$,
which generates (twice) the $z$ component of angular momentum in the
right moving $SU(2)$ (out of $SO(4)= SU(2) \times SU(2)$). The vector
${\bf n}$ therefore generates, $-2(h_r - J_r)$.
\footnote{It is not difficult to check that 
$2h_L - 2j_L$ is generated by the vector
field ${\bf n}'= -\partial_t + \partial_\theta - \partial_{\phi_1} +
\partial_{\phi_2}$.}.

Note, first,  that ${\bf n}$ is a null vector (its two components are,
respectively, unit timelike and unit spacelike vectors). Let ${\bf n}_s=
K( \partial_\theta +
\partial_{\phi_1} +\partial_{\phi_2})$ (the purely spatial component of ${\bf n}$) with the normalization $K$ chosen to give ${\bf n}_s$ unit norm.
Consider a positively oriented purely spatial vector ${\bf v}_2$ at a particular point $p$ on the string
at constant time. We may
decompose ${\bf v}_2$ as 
\begin{equation}\label{ldecomp} {\bf v}_2 = \sin \alpha {\bf n}_s + \cos \alpha {\bf u},
\end{equation} 
where
${\bf u}$ is some purely spatial unit vector orthogonal to ${\bf n}_s$. Let us assume that the string evolves in time so that the 
vector ${\bf n}$ is always tangent to
its world volume. It follows that, at the point $P$, the world
volume of the string is spanned by ${\bf n}$ and ${\bf v_2}$. These two vectors
are not orthogonal, but it is easy to check that with
\begin{equation}
\label{v1global}
{\bf v_1} = {{\bf n} \over \sin \alpha} - {\bf v_2},
\end{equation}
 $\{ {\bf v_1}, {\bf v_2} \}$ form an orthonormal set, with the first vector
timelike. The D-string preserves those supersymmetries of \eqref{global-susy}, that satisfy: 
\begin{equation}
\label{probeinglobal}
{\tilde{\Gamma}}_{{\bf v_1}} {\tilde{\Gamma}}_{{\bf v_2}} \epsilon = \epsilon.
\end{equation}

Before proceeding further, let us introduce some terminology. Consider
a complex vector $u$, say $X_1 + i X_2$. A spinor that is
annihilated by ${\tilde{\Gamma}}_u$ is said to have spin $-$ under rotation in
the $X_{1}$-$X_2$ plane, while a spinor annihilated by ${\tilde{\Gamma}}_{{\bar
u}}$ has positive spin (consequently, the spin operator is $i {\tilde{\Gamma}}_1
{\tilde{\Gamma}}_2$), with similar definitions for the other directions.
Let us now consider constant spinors $\chi$ whose spins(eigenvalues under this `spin' operator) in $R^{2,2}$ and $R^4$, respectively, are $(++) (--)$ or $(--) (++)$. The spins in $T^4$
could be either $(++)$ or $(--)$ -- this gives a total of 4 spinors --
or two sets of complex conjugate pairs of spinors.  We will now
demonstrate that any giant graviton whose world volume tangent space
contains the vector \eq{global-tangent} preserves all 4 of these supersymmetries.

To avoid cluttering the notation below, we define:
\begin{equation}
{\tilde{\Gamma}}_{AdS} = \tilde{\Gamma}\cdot N_{AdS}, ~~~ {\tilde{\Gamma}}_{S} = \tilde{\Gamma}\cdot N_{S}, ~~~ 
{\tilde{\Gamma}}_{I(N_{AdS})} = \tilde{\Gamma}\cdot I(N_{AdS}), ~~~{\tilde{\Gamma}}_{I(N_S)} = \tilde{\Gamma}\cdot I(N_S).
\end{equation}
Now consider \begin{equation}\label{wvsatis}
\begin{split}
 A&= ({\tilde{\Gamma}}_{{\bf v_1}} {\tilde{\Gamma}}_{{\bf v_2}} - 1)(1 + {\tilde{\Gamma}}_{AdS} {\tilde{\Gamma}}_{S}) \chi\\
& = \left( {1 \over \sin \alpha}
{\tilde{\Gamma}}_{\bf n} - {\tilde{\Gamma}}_{\bf v_2} \right)
 {\tilde{\Gamma}}_{\bf v_2} \left(1+{\tilde{\Gamma}}_{AdS}{\tilde{\Gamma}}_{S} \right) \chi -
\left(1+{\tilde{\Gamma}}_{AdS} {\tilde{\Gamma}}_{S} \right) \chi \\
 &= -{1 \over \sin \alpha} {\tilde{\Gamma}}_{\bf v_2} {\tilde{\Gamma}}_{\bf n} \left(1+{\tilde{\Gamma}}_{AdS} {\tilde{\Gamma}}_{S} \right) \chi \\
&=  -{1 \over \sin \alpha} {\tilde{\Gamma}}_{\bf v_2} \tilde{\Gamma}_{I(N_S)} \left[\left(1 + \tilde{\Gamma}_{I(N_S)} \tilde{\Gamma}_{I(N_{AdS})}\right)\left(1 + \tilde{\Gamma}_{AdS} \tilde{\Gamma}_{S}\right) \right] \chi \\
&=-{1 \over \sin \alpha} {\tilde{\Gamma}}_{\bf v_2} \tilde{\Gamma}_{I(N_S)}\left( 1 +  \tilde{\Gamma}_{I(N_S)} \tilde{\Gamma}_{I(N_{AdS})} \right)\left[1 + \tilde{\Gamma}_{I(N_S)} \tilde{\Gamma}_{I(N_{AdS})} \tilde{\Gamma}_{AdS} \tilde{\Gamma}_{S}\right] .
\end{split}
\end{equation}
where we have used $\tilde{\Gamma}_{I(N_S)}^2= 1=-\tilde{\Gamma}_{I(N_{AdS})}^2$.

It is now relatively simple to check that \eqref{wvsatis} vanishes
when $\chi$ is any of the four spinors $(++)(--)(++)$,
$(++)(--)(--)$, $(--)(++)(++)$, $(--)(++)(--)$. \footnote{ The first
and second of these spinors are $Qs$ while the third and fourth of
these are complex conjugate $Ss$.} Recall that a positive spin is
annihilated by ${\tilde{\Gamma}}_{S} -i {\tilde{\Gamma}}_{I(N_S)}$ and by the equivalent
$AdS$ expression. Using ${\tilde{\Gamma}}_{S}^2=-{\tilde{\Gamma}}_{AdS}^2 =1$ we find
\begin{equation}\label{gammaproof} \begin{split}
  {\tilde{\Gamma}}_{AdS} {\tilde{\Gamma}}_{I(N_{AdS})} \chi_{(++)(..) } = +i \chi_{(++)(..)},  \\
  {\tilde{\Gamma}}_{S} {\tilde{\Gamma}}_{I(N_{S})} \chi_{(..)(++)} = -i \chi_{(..)(++)},  \\
  {\tilde{\Gamma}}_{AdS} {\tilde{\Gamma}}_{I(N_{AdS})} \chi_{(--)(..)} = -i \chi_{(--)(..)},  \\
  {\tilde{\Gamma}}_{S} {\tilde{\Gamma}}_{I(N_{S})} \chi_{(..)(--)} = +i \chi_{(..)(--)}.
\end{split}
\end{equation} 
from which \eqref{probeinglobal} follows for all the spinors listed
above.

We conclude that any D1 brane world volume, to which the vector ${\bf n}$
is always tangent, preserves the 4 supersymmetries listed above. The same
is true of a D5-brane world volume that wraps the 4-torus.

\subsubsection{D1-D5 bound state probe}
\label{kappadonefive}
Now, we consider D5 branes that wrap the 4-torus, and move so as
to keep the vector ${\bf n}$ tangent to their worldvolume at all points, 
but also have gauge fields on their
worldvolume. These gauge fields, in a configuration with non-zero
instanton number, can represent bound states of D1 and D5 branes. Our
analysis here is valid for all four backgrounds considered above. 

Consider a D5 brane with a non-zero 2-form BI field strength $F$, that
wraps the $S^1 \times T^4$.  We denote the world-volume coordinates by
$\sigma^\alpha= \sigma^{1,2,6,7,8,9} \equiv \{ \tau, \sigma,z^1, z^2,
z^3, z^4\}$. The embedding of the world volume, as before, will be denoted
by $x^M(\sigma^\alpha)$ and the induced metric, by
$h_{\alpha \beta} = G_{M N} \partial_{\alpha} X^{M} \partial_{\beta} X^{N}$.
For a non-degenerate world-volume (det $h \ne 0$) the 
tangent vectors $\del_\alpha x^M$ are linearly independent and
provide a basis for the tangent space at each point of the world-volume. 
It is clearly possible to introduce an orthonormal (in the
spacetime metric $G_{MN}$) basis of six vectors 
${\bf v}_{\hat{\alpha}}$, related to the $\del_\alpha x^M$
by $\del_\alpha x^M = e^{\c\alpha}_\alpha {\bf v}_{\hat{\alpha}}$
such that 
\[
G_{MN} {\bf v}_{\hat{\alpha}}^M {\bf v}_{\hat{\beta}}^N 
= \tilde\eta_{\hat{\alpha}\hat{\beta}}.
\]
The invertible matrix $e^{\c\alpha}_\alpha$ defines 6-beins
of the induced metric:
\bea
h_{\alpha \beta} 
&&= G_{M N}  \partial_{\alpha} X^{M} \partial_{\beta}
X^N 
\nn
&&=
G_{M N} e^{\c\alpha}_\alpha  e^{\c\beta}_\beta
{\bf v}_{\hat{\alpha}}^M {\bf v}_{\hat{\beta}}^N 
= \tilde\eta_{\hat{\alpha}\hat{\beta}}
e^{\c\alpha}_\alpha  e^{\c\beta}_\beta .
\eea
Here  $\tilde{\eta}$ is 6 dimensional and $\alpha, \beta$ run
over the worldvolume coordinates. We will define below
\[
\gamma_{\c\alpha} = {\bf v}_{\hat{\alpha}}^M \Gamma_M .
\]
We take ${\bf v}_1, {\bf v}_2$ to be the same as in the previous
subsections.  The other four vectors point along the internal
manifold, ${\bf v}_{i} \propto {\partial \over \partial x^i},
i=6,7,8,9$. 

The condition for branes with worldvolume gauge fields 
to be supersymmetric was considered in
\cite{Bergshoeff:1996tu,Bergshoeff:1997kr}.
Using the two component notation for spinors
\begin{equation}\label{spinor2comptext1}
  \epsilon=\left(
    \begin{array}{cc}
      \epsilon_1 \\
      \epsilon_2
    \end{array}
  \right),
\end{equation} 
the BPS condition is (see Eqn. (13) of \cite{Bergshoeff:1997kr})
\bea
\label{addSUSYreq}
&& 
{\bf R} {\gamma}_{\c1 \c2 \c6 \c7 \c8 \c9} \epsilon = 
\epsilon,
\nn
&&
{\bf R}  = {1 \over \sqrt{-\det\{\tilde{\eta}_{\hat{\alpha} \hat{\beta}} + F_{\hat{\alpha} \hat{\beta}}\}}} \sum_{n=0}^\infty \frac{(-1)^n}{2^n n!}\gamma^{\c\alpha_1 \c\beta_1
...\c\alpha_n \c\beta_n} F_{\c\alpha_1 \c\beta_1}...F_{\c\alpha_n \c\beta_n} \sigma_3^{n + 1} i \sigma_2,
\eea
where we have expressed the world-volume gauge fields in the
local orthonormal frame: 
$  F_{\alpha\beta}= F_{\c\alpha \c\beta} e^{\c\alpha}_\alpha
e^{\c\beta}_\beta$. 
Note, that the product in \eqref{addSUSYreq}, terminates at $n=3$ because
the indices are anti-symmetrized. From the $n=0$ term we find, using 
the analysis of the previous subsections that the condition \eqref{addSUSYreq}
can be met only
for spinors that obey \eqref{right-susy}. The spinors \eqref{right-susy}
are eigenspinors of $\sigma_1$. Since $i \sigma_2$ appears in the $n=1$
term, this term must vanish. Hence,
 the gauge fields 
must be of the form 
\begin{equation}
\label{prelimFform}
F_{{\c1} {\c2}} = 0, ~~~ F_{{\c1} {\c i}} = -F_{{\c2} {\c i}}, 
~~~F_{{\c i} {\c j}} = \epsilon_{{\c i} {\c j}}^{{\c k} {\c l}} 
F_{{\c k} {\c l}}.
\end{equation}
For a gauge field of this kind, the determinant above is calculated in \eqref{detD} and
\[
  \sqrt{-\det\{\tilde{\eta}_{\hat{\alpha} \hat{\beta}} + F_{\hat{\alpha} \hat{\beta}}\}} = 1 + {F_{\hat{i} \hat{j}} F^{\hat{i} \hat{j}} \over 4}.
\] The $n=2$ term gives us the right factor in
the numerator to cancel this and the $n=3$ term vanishes as a virtue of \eqref{prelimFform}.

In the world-volume curved basis, our result implies (see \eqref{v1v2def}, \eqref{v1global}) that
\begin{equation}\label{Ftotal}
 F = F_{\sigma i} d \sigma \wedge dx^i + {1 \over 2} F_{i j} dx^i \wedge dx^j,
\end{equation}
and is self-dual on the torus, i.e
\begin{equation}\label{Fselfdual}
 F_{ij} \epsilon^{i j}_{~k l}=F_{kl}.
\end{equation} 
For `wavy instantons' where the gauge fields depend on $\sigma$ and the 
field strength is of the form \eqref{Ftotal}, the Gauss law and equation \eqref{Fselfdual} are enough to gaurantee that $F$ solves the equations of motion \cite{Eto:2005sw}. 
%that
%Hence, using that $\Gamma_6 \Gamma_7 \Gamma_8 \Gamma_9 \psi = \psi$,
%we end up with $Y_{ij} \epsilon^{i j}_{~k l}=Y_{kl}$, which, 
%after using the above mentioned relation between $Y$ and $F$, yields
% \begin{equation}\label{Fselfdual}
%  F_{ij} \epsilon^{i j}_{~k l}=F_{kl},
% \end{equation} 
% i.e.
% $F$ has to be self-dual in the torus directions. 

% Now, if we turn on 
% a $F_{\sigma i}$ component, and allow the brane to move while keeping
% the vector ${\bf n}$ tangent to its worldvolume at all points,
% it is easy to see, by repeating the analysis above, that 
% this brane only preserves supersymmetries of the form \eqref{right-susy}.\begin{footnote}{For `wavy' instantons of this type, we need to check the Gauss law. See \cite{Eto:2005sw}}\end{footnote}
% To summarize, supersymmetry requires $F$ to be of the form 
% \begin{equation}\label{Ftotal}
%  F = F_{\sigma i} d \sigma \wedge dx^i + F_{i j} dx^i \wedge dx^j,
% \end{equation}
% where $i$ and $j$ run over torus directions, and $F_{i j}$ obeys \eqref{Fselfdual}. 
The form of $F$ in \eqref{Ftotal} is adequate to 
guarantee supersymmetry in all the four backgrounds considered previously. 
For the sake of completeness, we mention that
the explicit embedding of the D5 brane in spacetime is 
described by the functions $X^{M}(\tau, \sigma, z^{1...4})$
satisfying
\begin{equation}
{\partial X^{M}(\tau, \sigma, z^{1...4}) \over \partial \tau} = {\bf n}^{M} .
\end{equation}
In the coordinate systems that we will discuss, ${\bf n}^{M}$ is a
constant and in such a coordinate system we again have:
\begin{equation}
X^{M}(\tau, \sigma, z^{1...4}) = X^{M}(\sigma) + {\bf n}^{M} \tau .
\label{bound-embed}
\end{equation}
Using the value of ${\bf n}$ \eq{n-d1d5} in 
the D1-D5, D1-D5-P and Lunin-Mathur geometries, the above
equation translates to:
\begin{equation}
\label{embeddingd1d5}
t = \tau,~~ x_5 = x_5(\sigma) + \tau,~~x^m = x^m(\sigma), 
~~x^6 = z^1,~ \ldots~,~~ x^9 = z^4,
\end{equation}
while, in global $AdS_3 \times S^3 \times T^4$, using 
\eq{global-tangent}, the brane motion is:
\begin{equation}
\label{embeddinglobal}
\begin{split}
t &= \tau, ~~ \theta = \theta(\sigma) + \tau, ~~ \rho = \rho(\sigma), ~~ \zeta = \zeta(\sigma), ~~ \phi_1 = \phi_1(\sigma) + \tau, ~~ \phi_2 = \phi_2(\sigma)+\tau,\\
x^6&=z^1,~\ldots~,~~ x^9 = z^4.
\end{split}
\end{equation}
We are assuming, in the embedding above, that the brane wraps the internal manifold only once. 
The case of multiple
wrapping is identical to the case of multiple brane probes, each wrapping the internal manifold
once and is discussed in more detail in Section \ref{nonabelianextension}.

The field strength above gives rise to an induced D1 charge, $p$, on the D5 brane worldvolume, which is proportional to the second Chern class and is given by
\begin{equation}
\label{instantonumber}
p = {1 \over (2 \pi \sqrt{\alpha'})^4} \int_{T^4}{{\rm Tr}\left(F \wedge F\right) \over 2},
\end{equation}
and also to an induced D3 brane charge on the 2 cycles of the $T^4$ (which we denote by $C_2$ below),  proportional to the first Chern class, given by
\begin{equation}
\label{firstchernclass}
p^3_{C_2} = {1 \over (2 \pi \sqrt{\alpha'})^2} \int_{C_2} {\rm Tr}(F).
\end{equation}
This D5 brane configuration with worldvolume gauge fields then represents a D1-D3-D5 bound state. This bound state has the property that whenever we wrap a  D3 brane on a two-cycle, we need to put an equal amount of D3 brane charge on the dual two-cycle. It may be surprising that a probe of this kind, with induced D3 brane charge, is mutually supersymmetric with the D1-D5 background.

However, this fact may be familiar to the reader from another perspective. Consider a configuration of $Q_1$ D1 branes, $Q_5$ D5 branes, $Q_3$ D3 branes and $Q_3'$ D3' branes,  wrapping the $5, 56789, 567, 589$ directions respectively. Following the standard BPS analysis, of say Chapter 13 in \cite{Polchinski:1998rr}, the BPS bound for this configuration is:
\begin{equation}
M \geq \sqrt{(Q_1 + Q_5)^2 + (Q_3 - Q_3')^2}.
\end{equation}
When $Q_3 = Q_3'$, this bound becomes $M \geq Q_1 + Q_5$ and it may further be shown that this configuration preserves the same supersymmetries as the D1-D5 system. 

Nevertheless, we will not be interested in probes with non-vanishing first Chern class in this paper. The AdS/CFT conjecture requires us to sum over all geometries with fixed boundary conditions for the fields at $\infty$. When we consider a D1 or D5 probe, we can reduce the D1 or D5 charge in the background so that the total D1 and D5 charge remains constant at $\infty$. 
A probe with non-vanishing $p^3_{C_2}$ will lead to some finite D3 charge at $\infty$ and turning on an anti-D3 charge in the background will render the probe non-supersymmetric. So, such probes must be excluded from a consideration of the supersymmetric excitations of the pure D1-D5 system. Henceforth, we will set $p^3_{C_2}$ to zero on all 2-cycles $C_2$ of the $T^4$.

\section{Charge Analysis: D strings}
\label{susydstringsdbi}

From the Killing spinor analysis above, we conclude that in all the
four different backgrounds we will consider, D-strings that move so as
to keep a particular null Killing vector field tangent to their
worldvolume at each point preserve 4 supersymmetries.  This means, as
we mentioned, that given the initial shape of the D-string we can
translate it along the integral curves of this vector field to
generate the entire worldvolume.  In this section, we will use this
fact to explicitly parameterize all supersymmetric D-string probes in
terms of their initial profile functions.  We will then use the DBI
action to calculate the spacetime momenta of these configurations and
verify the saturation of the BPS bound.

In the first subsection below, we present a general formalism that is
applicable to all the examples we consider. We then proceed to apply
this formalism to the extremal D1-D5 background, the D1-D5-P background,
the smooth geometries of \cite{Lunin:2001fv} and finally global AdS.

\subsection{Supersymmetric D1 Probe Solutions}
\label{notationdone}

We introduce coordinates, $\tau$ and $\sigma$, on the D1 brane
worldvolume.  We use $X^{M}(\sigma, \tau)$ to describe the embedding
of the worldsheet in spacetime, with $t \equiv X^{0}$ denoting time.
We will use $\dot{X}^{M} \equiv {\partial X^{M} \over \partial
\tau}$ and $(X^{M})' \equiv {\partial X^{M} \over \partial \sigma}$.
The special null vector, discussed above, is denoted by ${\bf n}^{M}$
(see also Section \ref{probe}). We
will always work with the string frame metric $G_{MN}$.  This is
the metric we use while calculating dot products. For example, $X'
\cdot X' = G_{MN} X'^{M} X'^{N}$.  The Ramond-Ramond 3 form
field strength is denoted by $G^{(3)}_{M N P}$ and the 2 form
potential is denoted by $C^{(2)}_{MN}$.  The dilaton is $\phi$.
The induced worldsheet metric is $h_{\alpha \beta} = G_{M
N}\partial_{\alpha} X^{M} \partial_{\beta} X^{N}$.  In all the
cases that we consider in this section, the $NS$-$NS$ two form is set to
zero.

With this notation, the bosonic part of the D1 brane action is:
\begin{equation}
\label{probebrane} 
S=\int {\cal L}_{\rm brane} d \sigma d \tau = -{1 \over 2 \pi \alpha'} \int
e^{-\phi} \sqrt{-h} d \sigma d \tau + {1 \over 2 \pi \alpha'} 
\int C^{(2)}_{M N}
\partial_\alpha X^M \partial_\beta X^N {\epsilon^{\alpha \beta}
\over 2} d \sigma d \tau,
\end{equation}
where 
\begin{equation}
h = {\rm Det}[h_{\alpha \beta}] = (X' \cdot X')(\dot{X} \cdot \dot {X}) - 
(X' \cdot \dot{X})^2.
\label{induced-metric}
\end{equation}
We take $\epsilon^{\tau \sigma} = -\epsilon^{\sigma \tau}=+1$.  In
line with the analysis presented above, we take our solutions to have
the property:
\begin{equation}
\label{solutionpropert}
{\partial X^{M}(\sigma, \tau) \over \partial \tau} = {\bf n}^{M}.
\end{equation}

In the examples in this section, we will be using a coordinate system
where ${\bf n}^{M}$ is constant. When this happens, we may solve
\eqref{solutionpropert} via (see \eqref{parameterization}, \eqref{embeddinglobal})
\begin{equation}
\label{solutions}
X^{M}(\sigma,\tau) = X^{M}(\sigma) + {\bf n}^{M} \tau.
\end{equation}
As we explained above, the set of supersymmetric worldvolumes is
parameterized by the set of initial shapes $X^{M}(\sigma)$.

On these solutions, we find 
\begin{equation}
\label{inducedeterminant}
\sqrt{-h} = \left| X' \cdot \dot{X} \right|.
\end{equation}
From the action \eqref{probebrane}, we can then derive the momenta
\begin{equation}
\label{momenta}
\begin{split}
P_{M} &= {\partial {\cal L}_{\rm brane} \over \partial {\dot X^{M}}} \cr
&= {-e^{-\phi} \over 2 \pi \alpha'} \left[(G_{M N} - e^{\phi} 
C^{(2)}_{M N} ) X'^{N} - {\bf n}_{M} {(X' \cdot X') \over X' \cdot 
\dot{X}}\right].
\end{split}
\end{equation}

Since these momenta are independent of $\tau$ the equations of
motion reduce to
\begin{equation}
\label{eqmotion}
-{\partial {\cal L}_{\rm brane} \over \partial X^{P}} =  \left({\partial(e^{-\phi} G_{MN}) \over \partial X^{P}} + {\partial C^{(2)}_{MN} \over \partial X^{P}}\right) (X'^{M} \dot{X}^{N} - \dot{X}^{M} \dot{X}^{N} {X' \cdot X' \over X' \cdot \dot{X}}) = 0.
\end{equation}

Before we apply this general formalism to specific cases, we would
like to make two comments.
\begin{enumerate}
\item
First, as noted above, we find that $\sqrt{-h}=+|X'\cdot\dot{X}|$. If
we do not put the absolute value sign, a worldsheet that folds on
itself could have zero area. If we now work out the equations of
motion carefully, taking into account that no such absolute value sign
occurs in the coupling to the RR 2-form, then we find that unless $X'
\cdot \dot{X}$ maintains a constant sign, our configurations are not
solutions to the equations of motion. Here, we have taken $|X' \cdot
\dot{X}| = +X' \cdot \dot{X}$.  The other choice of sign, would have
led to anti-branes which would not be supersymmetric in the
backgrounds we consider.

\item
The worldsheet may be parameterized by two coordinates, $\sigma$ and
$\tau$.  In many of the examples that we will consider, the vector
${\bf n}$ is a constant in our preferred coordinate system(see, tables
\ref{donefive} and \ref{global}). In such cases, we may take $t =
\tau$. Now, given the profile of the string at any fixed $\tau$, we
can translate each point on that profile by the integral curves of
${\bf n}$, to obtain the entire worldsheet. We may then use $\sigma$
to label these various integral curves of ${\bf n}$.

\end{enumerate}

\subsection{Supersymmetric Solutions in the D1-D5 background}
Consider $Q_1$ D1 branes and $Q_5$ D5 branes wrapping an internal
$T^4$ with sides of length $2 \pi (\alpha')^{1 \over 2} v^{1 \over 4}$
 and an $S^1$ of length $2 \pi$ that we take to
be along $x_5$. Table 1 describes the geometry of this background.
Notice that the 3-form fluxes are normalized so that
\begin{equation}\label{fluxes}
{1 \over 2 \pi} \int _{S^3} {G^{(3)} \over \alpha'}= 2 \pi Q_5 ~~~~ {1
\over 2 \pi} \int_{S_3 \times M_{int}} {\star_{10} G^{(3)} \over \alpha'}
= 2 \pi Q_1.
\end{equation}
If we take the near-horizon limit of the solution above, we find the
geometry of $AdS_3$ in the Poincare patch, with $x_5$ identified on a
circle. This is nothing but the zero mass BTZ black-hole. Although the
probe solutions we present below are valid in the entire D1-D5 geometry, it will turn out that quantization of these solutions in
Section \eqref{quantizationsection} is only tractable when the
probe-branes are in the near-horizon region.

The equations of motion, \eqref{eqmotion} reduce, on the solutions of
\eqref{solutions} to
\begin{equation}
\label{eqcheck}
{\partial (e^{-\phi} G_{55} + C^{(2)}_{5 t})\over \partial X^{P}} = 0.
\end{equation}
and these are manifestly satisfied since $e^{-\phi} G_{5 5} + C^{(2)}_{5
t} = 0$.

Table \ref{donefive} explicitly lists the solutions \eqref{solutions} and the
conserved charges.  The $RR$ 2-form potential in Table \ref{donefive}
has a gauge ambiguity(the coefficient $b$).  The canonical momenta
$P_{\phi_{1,2}}$, to begin with, depend on $b$; However, the momenta
$\tilde P_{\phi_{1,2}}$ appearing in the both Table \ref{donefive} and
Table \ref{global}(that deals with probe D-strings in global AdS) are
the gauge-invariant momenta which figure in the BPS relations and do
not have a gauge-ambiguity. This issue is discussed in detail in
Appendix \ref{sec:gauge}.  Note that the gauge-ambiguity is only in
the magnetic part and not in case of the electric part. The reason is
that it is possible to have a globally defined electric part of the
potential while it is impossible to do so for the magnetic part (for
reasons similar to the case of the Dirac monopole).

We now apply the general analysis presented above to obtain Table
\ref{donefive}.

%\begin{table}[!h]
\TABLE[h!]{
\caption{D1-D5 system}
\label{donefive}
\begin{tabular}{|l|}
\hline
{\bf Geometry:} \\
$ds^2 = f_1^{{-\half}}f_5^{-\half} \left( -dt^2 +(d x_5)^2 \right) + f_1^\half f_5^\half \left( dr^2+ r^2 (d \zeta^2 + \cos^2 \zeta d\phi_1^2 + \sin^2 \zeta d \phi_2^2)\right) + {e^{\phi} \over g} ds^2_{{\rm int}} $\\ 
$e^{-2 \phi}= {1\over g^2}{f_5 \over  f_1}, ~~ f_1 =1+{g \alpha' Q_1 \over v r^2}, ~~ f_5 =1+{g \alpha' Q_5 \over r^2}, ~~ v ={V \over (2 \pi)^4 \alpha'^2}$\\ 
${G^{(3)} \over \alpha'} =Q_5 \sin 2 \zeta d \zeta \wedge d \phi_1 \wedge d \phi_2 - {2 Q_1 \over v f_1^2 r^3} d r \wedge d t \wedge d x_5$ \\
${{C^{(2)}} \over \alpha'}=-{Q_5 \over 2} (\cos 2 \zeta + b) \zeta d \phi_1 \wedge d
\phi_2 + {1 \over g f_1 \alpha'} dt \wedge dx_5$ \\
{\bf BPS Condition}\\
$E - L = -\int P_t d \sigma - \int P_{5} d \sigma = 0$\\
{\bf Null Vector tangent to worldvolume:} \\
${\bf n}^{M} = {\partial \over \partial t} + {\partial \over \partial x_5}$\\
{\bf Solution}\\
$t=\tau~~x_5=x_5(\sigma)+\tau~~r=r(\sigma)$ \\
$\zeta=\zeta(\sigma)~~\phi_1=\phi_1(\sigma)~~\phi_2=\phi_2(\sigma)$ \\
$z^a_{\rm int} = z^{a}_{\rm int}(\sigma)$ \\
{\bf Momenta:}\\
$P_t  =  {1 \over 2 \pi \alpha' g} \left[{x_5' \over  f_1}  - \sqrt{f_5 \over f_1} {X' \cdot X' \over x_5'}\right] $\\
$P_5 =  -{1 \over 2 \pi \alpha' g} \left[{x_5' \over  f_1} - \sqrt{f_5 \over f_1} {X' \cdot X' \over x_5'}\right]$\\
$P_{r}=-{1 \over 2 \pi \alpha' } \left[{f_5 \over g} r'\right] $ \\
$P_{\zeta}= -{1 \over 2 \pi \alpha'} \left[{f_5 r^2 \zeta' \over g}\right]$ \\
$\tilde P_{\phi_1}=-{1 \over 2 \pi \alpha'}\left[{f_5 r^2 \cos^2 \zeta \phi_1'\over g} + {Q_5 \alpha' \over 2} [\cos(2 \zeta)-1] \phi_2'\right] $\\
$\tilde P_{\phi_2}=-{1 \over 2 \pi \alpha'}\left[{f_5 r^2 \sin^2 \zeta \phi_2' \over g} - {Q_5 \alpha' \over 2} [\cos(2 \zeta)+1] \phi_1'\right]$ \\
$P_{z^{a}} = -{1 \over 2 \pi \alpha' g}\left[g^{int}_{a b} z^{b'}\right] ~~ {\rm (internal~manifold)}$\\
\hline
\end{tabular} }
%\end{table}

\subsection{Supersymmetric Solutions in the D1-D5-P background}
The D1-D5 system above may be generalized by adding a third charge
using purely left-moving excitations which gives the `D1-D5-P'
system. The field strengths and dilaton are exactly as in Table
\ref{donefive} but the metric is altered as follows:
\bea
\label{donefivep}
&& ds^2 =  f_1^{{-\half}}f_5^{-\half} \left( -dt^2 +d x_5^2  + {r_p^2 \over r^2} (dt - dx_5)^2\right) +
\nn
&&~~~~~~~~~~~~~~~~~~
f_1^\half f_5^\half \left( dr^2+
r^2 (d \zeta^2 + \cos^2 \zeta d\phi_1^2 + \sin^2 \zeta d \phi_2^2)\right) 
+ {e^{\phi} \over g} ds^2_{{\rm int}} 
\eea
Here $r_p^2 = c_p g^2 P$, where $P$ is the quantized momentum along 
$x_5$ and $c_p$ is a numerical constant which is not 
important for our purpose here.

It is easy to repeat the supersymmetry analysis above, for this background. In particular, we find that:
\begin{equation}
\begin{split}
& P_t  =  {1 \over 2 \pi \alpha' g} \left[(1 + {r_p^2 \over r^2}){x_5' \over  f_1}  - \sqrt{f_5 \over f_1} {X' \cdot X' \over x_5'}\right], \\
& P_5 =  -{1 \over 2 \pi \alpha' g} \left[(1 + {r_p^2 \over r^2}){x_5' \over  f_1} - \sqrt{f_5 \over f_1} {X' \cdot X' \over x_5'}\right], \\
& P_t + P_{5} = 0.
\end{split}
\end{equation}
The rest of Table \ref{donefive} remains valid. 

%\enlargethispage{\baselineskip}
\subsection{Supersymmetric Solutions in the Lunin-Mathur Geometries}
\label{luninmathursection}
In this subsection, we describe supersymmetric D-string probes in the smooth 2 charge geometries of Lunin and Mathur\cite{Lunin:2001jy,Lunin:2001fv}.
The geometry is as follows
\begin{equation}
\label{luninmetric}
\begin{split}
ds^2 &= \sqrt{H \over 1+K} [-(dt - A_m dx^{m})^2 + (d x_5 + B_{m} dx^{m})^2] + \sqrt{1 + K \over H} d \vec{x} \cdot d \vec{x}  \\
&+ \sqrt{H(1 + K)} d \vec{z} \cdot d \vec{z}, \\
e^{2 \phi} &= H(1 + K), ~~C^{(2)}_{t {m}} = {-B_{m} \over 1 + K}, ~~ C^{(2)}_{t 5} = {1 \over 1 + K}, ~~ C^{(2)}_{{m} 5} = {A_{m} \over 1 + K}, \\
C^{(2)}_{{m} n} &= C_{{m} n} + {A_m B_n - A_n B_m \over 1 + K},~~ d B = - * d A, ~~ d C = - * d H^{-1}, \\
\end{split}
\end{equation}
where $H=H(\vec{x})$, $A=A(\vec{x})$ and $K=K(\vec{x})$ are three harmonic functions that are determined by 4 `string-profile' functions $F_m(v)$ as follows:
\begin{equation}
\begin{split}
H^{-1} &= 1 + {1 \over 2 \pi} \int_0^{2 \pi Q_5} {d v \over |x - F(v)|^2}, ~~ K = {1 \over 2 \pi} \int_0^{2 \pi Q_5} {|\dot{F}|^2 \, d v \over |x - F(v)|^2}\\
A_m & =-{1 \over 2 \pi} \int_0^{2 \pi Q_5} {\dot{F}_m \, d v \over |x - F(v)|^2 }.
\end{split}
\end{equation}
We have added 1 to $C^{(2)}_{t 5}$ to be consistent with our conventions where the energy of a probe D-string infinitely far away from the parent stack of D1-D5 branes is zero. Comparing conventions with Table \ref{donefive}, we see that the parameter $g$ has been absorbed into an additive shift of the dilaton and is set to $1$.

The vector ${\bf n} = {\partial \over \partial t} + {\partial \over \partial x_5}$ is null and we choose our solutions so that this vector is always tangent to the D-string worldvolume. We may apply the formalism of section \ref{notationdone} here to obtain
\begin{equation}
\begin{split}
\label{susymathur}
P_t &= -{1 \over 2 \pi \alpha'}(e^{-\phi} G_{t M} - C^{(2)}_{t M}) (X^{M})' - {\bf n}_{t} \gamma , \\
P_5 &= -{1 \over 2 \pi \alpha'} (e^{-\phi} G_{5 M} - C^{(2)}_{5 M}) (X^{M})' - {\bf n}_{5} \gamma , \\
\end{split}
\end{equation}
where we have defined $\gamma = {(X')^2 \over X' \cdot \dot{X}}$. We now only need to notice that ${\bf n}_{t} + {\bf n}_{5} = 0,~e^{-\phi} G_{5 5} - C^{(2)}_{t 5} = 0,~e^{-\phi} (G_{t m} + G_{5 m}) + (C^{(2)}_{t m} + C^{(2)}_{5 m}) = 0$ to see that the BPS condition $P_t + P_5 = 0$ is satisfied.

We comment on the relation of these geometries to global AdS in Section \ref{globaltolunin}.

\subsection{Supersymmetric Solutions in Global AdS}
We now consider a probe D1 string propagating in global $AdS_3 \times S^3 \times M_{\rm int}$. This geometry is described in Table \ref{global}. In particular, the metric is:
\begin{equation}
\label{globaladsmetric}
\begin{split}
&ds^2= G_{M N} d x^{M} d x^{N} \\
&= g \sqrt{{Q_1 Q_5 \over v}} \alpha' \left[-\cosh^2{\rho} dt^2 + \sinh^2{\rho} d \theta^2  + d \rho^2 + d \zeta^2 + \cos^2 \zeta d \phi_1^2 + \sin^2 \zeta d \phi_2^2\right] \\
&+ \sqrt{Q_1 \over Q_5 v} \alpha' d s^2_{\rm int} .\\
\end{split}
\end{equation}
$ds_{\rm int}^2$ is the metric on the internal manifold. $g, v, Q_1, Q_5$ are parameters that determine the string coupling constant, volume of the internal manifold and the electric and magnetic parts of the 3-form RR field strength according to the formulae summarized in Table \ref{global} below. We are following the notation of \cite{Maldacena:1998bw}. 
We parameterize the internal manifold using the coordinate $z^{1 \ldots 4}$. 

In terms of this coordinate system, the Killing spinor analysis of section \ref{d1-in-global} tell us that probe branes that preserve the Killing vector
$$
{\bf n}={\partial \over \partial t} + {\partial \over \partial \theta} + {\partial \over \partial \phi_1} + {\partial \over \partial \phi_2}
$$ (i.e. branes that have ${\bf n}$ everywhere tangent to their
world-volume) will preserve $4$ of the background $16$
supersymmetries.

We can now proceed as above to obtain Table \ref{global}
%\begin{table}[!h]
\TABLE[!h]{
\caption{D branes in Global AdS}
\label{global}
\begin{tabular}{|l|}
\hline
{\bf Geometry} \\
${ds^2 \over \alpha'} = l^2 \left[-\cosh^2{\rho} dt^2 + \sinh^2{\rho} d \theta^2  + d \rho^2 + d \zeta^2 + \cos^2 \zeta d \phi_1^2 + \sin^2 \zeta d \phi_2^2\right] + \sqrt{Q_1 \over Q_5 v} {d s^2_{\rm int} \over \alpha'}$ \\
$e^{-2 \phi} = {Q_5 v \over g^2 Q_1}, l^2 = {g \over \sqrt{v}} \sqrt{Q_1 Q_5}$ \\
${G^{(3)} \over \alpha'} = {*G^{(7)} \over \alpha'} = {d C^{(2)} \over \alpha'} =  Q_5 \sin{2 \zeta} d \zeta \wedge d \phi_1 \wedge d \phi_2 +  Q_5 \sinh(2 \rho) d \rho \wedge d t \wedge d \theta$ \\
${C^{(2)} \over \alpha'}=-{Q_5 \over 2 } \left[(\cos 2 \zeta + b) d \phi_1 \wedge d
\phi_2 - (\cosh(2 \rho) - 1) d t \wedge d \theta \right]$ \\
{\bf BPS Condition}\\
$E - L - J_1 - J_2 = -\int(P_t + P_{\theta} +
\tilde P_{\phi_1} +\tilde  P_{\phi_2})\, d \sigma = 0$\\
{\bf Null Vector tangent to worldvolume:} \\
${\bf n}^{M} = {\partial \over \partial t} + {\partial \over \partial \theta} + {\partial \over \partial \phi_1} + {\partial \over \partial \phi_2}$\\
{\bf Solution}\\
$t=\tau~~\theta=\theta(\sigma)+\tau~~\rho=\rho(\sigma)$ \\
$\zeta=\zeta(\sigma)~~\phi_1=\phi_1(\sigma)+\tau~~\phi_2=\phi_2(\sigma)+\tau$ \\
$z^a_{\rm int} = z^{a}_{\rm int}(\sigma)$ \\
{\bf Momenta:}\\
$\gamma = {\sinh^2 \rho \theta^{'2} + \cos^2 \zeta \phi_1^{'2} +\sin^2 \zeta \phi_2^{'2} + \zeta'^2 + \rho^{'2} + {1 \over g \alpha' Q_5} g^{\rm int}_{ab}z^{a'} z^{b'} \over  \cos^2 \zeta \phi_1'+ \sin^2 \zeta \phi_2'+ \sinh^2 \rho \theta'}$ \\
$P_{t} ={Q_5 \over 2 \pi} \left[ -\gamma \cosh^2
\rho + \sinh^2{\rho} \theta' \right]$  \\
$P_\theta = {-Q_5 \over 2 \pi} \left[ \left(-\gamma + \theta' \right) \sinh^2 \rho \right]$ \\  
$\tilde P_{\phi_1}={-Q_5 \over 2 \pi} \left[  \left(-\gamma  + \phi_1' \right) \cos^2 \zeta + 
\half \left( \cos 2 \zeta -1 \right) \phi_2' \right] $\\ 
$\tilde P_{\phi_2} ={-Q_5 \over 2 \pi} \left[ \left( -\gamma + \phi_2' \right) \sin^2 \zeta - \half \left( \cos 2 \zeta + 1 \right) \phi_1' \right]$ \\
$P_{\rho} = {-Q_5 \over 2 \pi} \rho'$\\
$P_{\zeta}= {-Q_5 \over 2 \pi} \zeta'$ \\
$P_{z^{a}} = {-1 \over 2 \pi \alpha' g}\left[g^{int}_{a b} z^{b'}\right] ~~ {\rm (internal~manifold)}$\\
\hline
\end{tabular}}
%\end{table}

\subsubsection{Spectral Flow}
\label{globaltolunin}
The Global AdS geometry above corresponds to the $NS$ vacuum of the boundary CFT. The geometries considered in section \ref{luninmathursection} correspond, on the other hand to the different Ramond ground states of this CFT. Now, the NS-sector and Ramond sector in CFT with at least $(2,2)$ supersymmetry are related by an operation called spectral flow, where the Virasoro generators $L_n$ and R-symmetry current modes $J_n$ change as follows (see, e.g., \cite{David:2002wn}
for a review):
\begin{equation}
\label{spectralflow} 
L_{n}^{NS} = L_n^{R} + J_n^{R} + {c \over 24} \delta_{n,0}, ~~~ 
J_{n}^{NS} = J_{n}^{R} + {c \over 12} \delta_{n,0},
\end{equation}
and the moding of the fermions changes from integral to half-integral. $c$ is the central charge of the theory which, for the boundary CFT, is $6 Q_1 Q_5$.

Under spectral flow, the NS vacuum maps to the Ramond vacuum with the smallest possible $U(1)$ charge of $J_0^{R} = - {Q_1 Q_5 \over 2}$. 
It was shown in \cite{Lunin:2002bj}, that in the set of solutions \eqref{luninmetric}, this corresponds to the profile function $F_1(v) = a \sin(w v),~~F_2(v) = -a \cos(w v),~~ F_3(v) = F_4(v) = 0$. In our conventions, $a = \sqrt{Q_1 Q_5}, w = {1 \over Q_5}$. 
After choosing this profile function, we make the coordinate redefinitions 
\begin{equation}
\begin{split}
x_1 &= a \cosh{\rho} \sin{\zeta} \cos{\phi_1},~~~ x_2=a \cosh{\rho} \sin{\zeta} \sin{\phi_1}, \\ x_3 &= a \sinh{\rho} \cos{\zeta} \cos{\phi_2},~~~ x_4 = a \sinh{\rho} \cos{\zeta} \sin{\phi_2},
\end{split}
\end{equation}
and take the near-horizon limit(i.e drop the $1$ in the harmonic functions) to obtain the metric and 3-form field strength:
\begin{equation}
\label{spectralflowedglobal}
\begin{split}
ds^2 &= \sqrt{Q_1 Q_5} \left[-\cosh^2{\rho} d t^2 + \sinh^2{\rho} d x_5^2 + d \rho^2 + d \zeta^2 + \cos^2 \zeta (d \phi_1 + d x_5)^2 + \sin^2 \zeta (d \phi_2 + t)^2 \right] \\ &+ \sqrt{Q_1 \over Q_5} d z^i d z^i , \\
G^{3} &= Q_5 \sinh(2 \rho) d t \wedge d \theta \wedge d \rho + Q_5 \sin(2 \zeta) d \zeta \wedge (d \phi_1 + d x_5) \wedge (d \phi_2 + d t) .
\end{split}
\end{equation}
The dual of the `spectral flow' \eqref{spectralflow} on the boundary in supergravity is the coordinate redefinition \cite{Lunin:2002bj}
\begin{equation}
t_{NS} = t_{R}, ~~ \theta_{NS} = (x_5)_{R},~~(\phi_1)_{NS} = (\phi_1)_R + (x_5)_R,~~ (\phi_2)_{NS} = (\phi_2)_{R} + t_R . 
\end{equation}
Under this mapping the solution above turns into global AdS! Moreover, going around the $\theta$ circle, once in the $NS$ sector, causes us to also go around the $(\phi_1)_{NS}$ circle to stay at constant $(\phi_1)_R$. Hence, fermions
which are anti-periodic in the $NS$ sector, become periodic in the $R$ sector. One may also check that the coordinate transformation above takes:
\begin{equation}
{\partial \over \partial t_R} + {\partial \over \partial (x_5)_R} = {\partial \over \partial t_{NS}} + {\partial \over \partial \theta_{NS}} + {\partial \over \partial (\phi_1)_{NS}} + {\partial \over \partial (\phi_2)_{NS}} .
\end{equation}
Thus this mapping maps the null Killing vector ${\bf n}$ of the Ramond sector to the special null Killing vector ${\bf n}$ of the NS sector. It also takes us from solutions that satisfy $E - L = 0$ to solutions that satisfy $E - L - (J_1 + J_2) = 0$. 

This one to one mapping between global $AdS$ and the corresponding Lunin Mathur solution implies that
everything that we say below regarding probes in global $AdS$ is also true (with appropriate redefinitions) for probes in this Lunin-Mathur geometry.

\subsubsection{Bound States\label{sec:btz-comparison}}
The probe solutions, in global $AdS$ above have a salient feature that we wish to point out.
Consider, a  D-string near the boundary of $AdS$. Such a string can have finite energy only if the flux through the string almost cancels its tension. Hence, it must wrap the $\theta$ direction and we can use our freedom to redefine $\sigma$  to set $\theta'=w$. For such a string, if we take the strict $\rho \rightarrow \infty$ limit, we obtain
\begin{equation}
\begin{split}
E - L &=  {Q_5 \over 2 \pi}\int \gamma d \sigma \\
 &= {Q_5 \over 2 \pi} \int  \left[{\sinh^2 \rho \theta^{'2} + \cos^2 \zeta \phi_1^{'2} +\sin^2 \zeta \phi_2^{'2} + \rho^{'2} + G_{ab}X^{a'} X^{b'} \over  \cos^2 \zeta \phi_1'+ \sin^2 \zeta \phi_2'+ \sinh^2 \rho \theta'}\right] d \sigma = {Q_5 w} .
\end{split}
\end{equation}
Thus, we notice that for strings stretched close to the boundary, the
quantity $E-L$ must be quantized in units of $Q_5$. If we wish to have
intermediate values of $E - L$, our strings are `bound' to the center
of AdS. In other words the moduli space of solutions with a value of
$E-L$ other than $Q_5 w$ does not include these long strings. This
leads us to believe that quantum mechanically, the quantization of
these solutions would lead to discrete states and not states in a
continuum. This expectation is validated by the analysis
of \cite{Raju}.

The `spectral flow' operation discussed above tells us that a similar 
statement holds in the geometry described by \eqref{spectralflowedglobal}. 
There, what must be quantized in units of $Q_5$ is the quantity $J_1 + J_2$. 
On the other hand, if we consider the near-horizon of the D1-D5 geometry
(see \eqref{dodfb}), which is the zero mass BTZ black hole, we find that the 
various momenta become independent of the radial direction! This means
 that in that background, all probes can escape to infinity. This implies that
`averaging' over different Ramond vacua to obtain the zero mass BTZ black hole, 
washes out the interesting structure of `bound-states' that we see above. 

Returning now to probes in global $AdS$, those probes that do not wrap the $\theta$ direction cannot go to $\rho
\rightarrow \infty$, yet their energy shows an interesting $\rho$
dependence. Consider the following solution (parameterized by
$w,~\rho_0,~\zeta_0,~\phi_{1_0},~\theta_0$)
\begin{equation}
t = \tau,\theta(\sigma)=\theta_0,~\rho(\sigma)=\rho_0,~\zeta(\sigma)=
\zeta_0,~\phi_1(\sigma)=\phi_{1_0},~\phi_2(\sigma)=w \sigma .
\label{black-hole-config}
\end{equation}
For this solution (using $w>0$ which is necessary for supersymmetry)
\begin{equation}
E = Q_5 w \cosh^2(\rho_0), L = Q_5 w \sinh^2(\rho_0), 
P_{\phi_1}=Q_5 w, P_{\phi_2} = 0 .
\end{equation}
In this subsector, a given set of charges fixes $\rho_0$:
\begin{equation}
\sinh^2{\rho_0} = {L \over w Q_5} .
\label{size-bound-q1}
\end{equation}
The fact that the size of the bound state is larger
for smaller $w$ is intuitively obvious; e.g. the
size of an electron orbit is inversely proportional
to its mass.  

The equation \eq{size-bound-q1} leads to an interesting result.
The extremal BTZ black hole \cite{Banados:1998gg} has
a horizon radius:
\be
\sinh^2\rho_h = 4 MG = 4 J G/l .
\label{horizon}
\ee
Using the values of various constants appearing in the
above equation (cf. \cite{Strominger:1997eq}, p 8)
\bea
l &&= 2\pi\alpha' \sqrt g (Q_1 Q_5)^{1/4} V^{-1/4},
\nn
G^{-1}  &&= 2 (Q_1 Q_5)^{3/4} V^{1/4}/(\pi\alpha' \sqrt g),
\eea
we get for the radius of the horizon
\be
\sinh^2\rho_h= {J \over Q_1 Q_5}.
\label{horizon-value}
\ee
We now make the following identifications:

\hspace{20ex}
\begin{tabular}{|l | l|}
\hline
Probe configuration & BTZ 
 \\
\hline
$L$  & $J$ \\
$w$  &  $Q_1$ \\
$E$ &$ lM + 1$\\
\hline 
\end{tabular}\label{table-btz}

We find that the horizon radius \eq{horizon-value} exactly coincides
with the size of the bound state, \eq{size-bound-q1}, under the above
identifications (the third identification, of energies, follows from
the second one; the extra `1' on the BTZ side owes to the mass
convention used by \cite{Banados:1998gg} in which  AdS$_3$ space
has mass $-1/l$). 

The above agreement would appear to suggest an interpretation of the
BTZ black hole as an ensemble of  bound states of $Q_1$ D-string probes rotating
around the center of the global AdS$_3$ background at a coordinate
distance $\rho_h$, given by \eq{horizon-value}. Since the AdS$_3$
background itself is ``made of'' of $Q_1$ D-strings and $Q_5$ D5
branes, the above configuration is well beyond the
domain of validity of the probe approximation
\footnote{This is similar to the situation with $N$ dual giant 
gravitons in \ads
background, at a fixed value of the global radius $\rho$.} and the
above interpretation should be regarded as tentative. Note that
probe configurations with $w < Q_1$ have a size {\em larger than}
the black hole radius
\be
w < Q_1 \Rightarrow \rho_0 > \rho_h ,
\ee
which, therefore, do not form a black hole.\footnote{This is similar to the situation with a star, e.g. the Sun,
whose size is larger than its Schwarzschild radius and hence does not
form a black hole.} The back-reacted geometry corresponding to such
probe configurations is likely to be some smooth non-singular
configurations. The maximum allowed value of $w (=Q_1)$
corresponds precisely to a threshold for black hole formation ($\rho_0
= \rho_h$).

\subsubsection{Classical lower bound of energy}
It can be shown (see Appendix \eqref{sec:bound-proof}) that, 
in global AdS, the set of solutions that we have described above has 
an `energy gap'.
\begin{equation}\label{bound}
E= -\int P_t \, d \sigma  \ge Q_5 .
\end{equation}

\section{Charge Analysis: D1-D5 bound state probes}
\label{boundstates}

We now consider D5 branes with gauge fields on their worldvolume.
Supersymmetric probes of this kind were discussed in Section
\ref{kappadonefive}. The embedding for such
branes is given by \eq{bound-embed} and 
the gauge fields $A_{{i}}(\sigma)$ are of the
form that gives rise to \eq{Ftotal} 
\be
\label{fdfivebrane}
F = F_{\sigma i} d \sigma \wedge d z^{i} + {1 \over 2} F_{i j} d z^{i} \wedge d z^{j} .
\end{equation}
with the self-duality requirement \eq{Fselfdual}
\begin{equation}
\label{adhm}
F_{i j} = \epsilon_{~~i j}^{k l} F_{k l} .
\end{equation}
In this section we will obtain two results. First, we will verify the
analysis of Section \ref{kappadonefive} by a charge analysis and
confirm that the above configurations are indeed supersymmetric.
Next, we will show that the canonical structure on the space of
supersymmetric solutions of the 5+1 dimensional worldvolume theory of
coincident D5 branes is identical to the canonical structure on the
set of supersymmetric solutions to a 1+1 dimensional theory. For a
probe comprising $p$ D1 branes and $q$ D5 branes, this effective 1+1
dimensional theory is the theory of a D-string propagating in the
geometries discussed above but with the internal manifold $T^4$ or
$K3$ replaced by the instanton moduli space of $p$ instantons in a
$U(q)$ theory on $T^4$(or $K3$).  This is similar to the result
\cite{Vafa:1995zh,Dijkgraaf:1998gf,Seiberg:1999xz} (see, e.g.
\cite{David:2002wn} for a review) that the worldvolume theory of
supersymmetric D5 branes in flat space flows, in the IR, to the sigma
model on the instanton moduli space. However, our result here is for
D5 branes in {\em curved backgrounds} (discussed in Section
\ref{kappadonefive}) and, furthermore, the result holds (as we
will see below) as long as the DBI description is valid and {\em we do not
need to go to the IR fixed point}.

\subsection{ Classical Supersymmetric Bound State Solutions}

We consider, first, a single D5 brane.\bft{We will be eventually
interested in the instanton moduli space only for $q>1$ D5 branes
since the $q=1$ case is rather subtle \cite{Seiberg:1999xz}.  However,
we include the calculations for $q=1$ here for simplicity. The generalization
to $q>1$, which is straightforward, is left to Section 
\ref{nonabelianextension} }\eft Our background has both a
three form flux $G^{(3)} = d {C^{(2)}}$ and a seven form flux
$G^{(7)}=*G^{(3)} = d C^{(6)}$. In all the examples we will consider,
it is possible to define a new two-form $C'^{(2)}$ such that
\begin{equation}
\label{cprimedef}
{C^{(6)}} = C'^{(2)} \wedge d z^1 \wedge \ldots \wedge d z^4 .
\end{equation}
Using this notation, the DBI action becomes
\begin{equation}\label{probebraneF} 
\begin{split} &S= \int {\cal L} d \sigma d \tau \prod_i d z^i \\
&=-{1 \over  (2 \pi)^5 {\alpha'}^3 } 
\int e^{-\phi} \sqrt{-{\rm Det}[D_{\alpha \beta}]} + {1 \over  (2 \pi)^5 {\alpha'}^3}\left[\int {C^{(2)}} \wedge {1
\over 2!} F \wedge F + \int C'^{(2)} \wedge d z^1 \wedge \ldots \wedge d z^4 \right], \\
&D_{\alpha \beta} = h_{\alpha \beta} + F_{\alpha \beta}, \end{split}
\end{equation} 
where as usual $h_{\alpha \beta}$ is the pull-back of the
string-frame metric to the worldvolume, $F_{\alpha \beta} =
\partial_{[\alpha}A_{\beta]}$ is the two-form field strength and
$A_{\alpha}$ is the gauge potential. It is important to note, that we
have normalized $F$ unconventionally which accounts for the absence of
the usual $2 \pi \alpha'$ factor. We have written the action in terms of forms to lighten the notation, but in indices: $C^{(2)} = {1 \over 2} C^{(2)}_{M N} d X^{M} \wedge d X^{N}$.

We will now formally assume that $F$ is of the form \eqref{fdfivebrane} and write:
\begin{equation}\label{Ddefined}
D_{\alpha \beta} = \left( \begin{matrix} 0 & h_{\tau \sigma}
& 0 & 0 & 0 & 0 \\ h_{\tau \sigma} & h_{\sigma \sigma}  & F_{\sigma 1} &
F_{\sigma 2} & F_{\sigma 3} & F_{\sigma 4} \\ 0 & -F_{\sigma 1} &
e^{\phi} / g & F_{12} & F_{13} & F_{14} \\ 0 & -F_{\sigma 2} &
-F_{12} & e^{\phi} / g & F_{14} & -F_{13} \\ 0 & -F_{\sigma 3} &
-F_{13} & -F_{14} & e^{\phi} / g & F_{12} \\ 0 & -F_{\sigma 4} &
-F_{14} & +F_{13} & -F_{12} & e^{\phi} / g \\ \end{matrix} \right),
\end{equation}
where we have assumed an internal $T^{4}$ with a metric $ds^2_{T^4} = {e^{\phi} \over g} \sum_i d z^i d z^i$ and the embedding \eqref{embeddingd1d5} or \eqref{embeddinglobal}.

The Determinant of this matrix is \begin{equation}\label{detD}
\begin{split}
\sqrt{-|D|} &= h_{t \sigma} (\beta^2 + {F_{i j} F^{i j} \over 4}) \equiv h_{t \sigma} (\beta^2 + {|F|^2 \over 2}) ,  
\\ \beta &= {e^{\phi} \over g}.
\end{split}
\end{equation}

Note that:
\begin{equation}
|F|^2  d z^1 \wedge \ldots \wedge d z^4 = F \wedge F .
\end{equation}
The field strength $F$ is derived from the gauge fields $A_i$ via
$F_{\alpha \beta} = \partial_{[\alpha} A_{\beta]}$. Note that the $A_i$ have components
only along the internal manifold. Let us suppose that 
there are solutions to \eq{adhm} characterized by
`moduli' $\zeta^{a}$ (the solutions we are interested
in exist, actually, for $q>1$, so the calculations in this
section and the next are to be understood in a formal sense till
we apply these to $q>1$ in Section \ref{nonabelianextension}).
We can assign $\sigma$ dependence to these moduli consistent
with Gauss's law \cite{Eto:2005sw} and supersymmetry, thus
\begin{equation}
\label{Ainmoduli}
A_i(\sigma) = A_i(\zeta^a(\sigma)) .
\end{equation}
Although the moduli can vary as functions of $\sigma$,
supersymmetry implies that they cannot depend on $\tau$.

To calculate the momenta, we will need the inverse of D. We have listed the relevant components of the inverse in the appendix. Using these, we find:
\begin{equation}\label{momentumwithF} \begin{split} 
P_{M} = {\delta {\cal L} \over \delta \dot{X^{M}}}
 &= {-e^{- \phi} \over ( 2 \pi)^5 \alpha'^3} \left(
  \sqrt{-D} {D^{\tau \beta} + D^{\beta \tau} \over 2}
  { G_{M N} \partial_{\beta}X^{N} - e^{\phi} \partial_\sigma X^{N}
     \left( C^{(2)}_{M N} {|F|^2 \over 2}   +  C'^{(2)}_{M N} \right) }\right) \\
=  {-e^{- \phi} \over ( 2 \pi)^5 \alpha'^3} &\left[ \left((\beta^2 + {|F|^2\over 2}) G_{M N} - {e^{\phi} C^{(2)}_{M N} |F|^2 \over 2} - {e^{\phi} C'^{(2)}_{M N} }\right) \partial_{\sigma} X^{N} 
\right. \\  &\left. - {\beta F_{\sigma i} F_{\sigma}^{i} + h_{\sigma \sigma} (\beta^2 + {|F|^2 \over 2})\over h_{\tau \sigma} } {\bf n}_{M} \right] , \\
P_{Ai} = {\delta {\cal L} \over \delta \partial_{\tau} A_i} 
&=  -{e^{- \phi} \over  (2 \pi)^5 \alpha'^3} \sqrt{-D} {D^{\tau i} - D^{i \tau} \over 2} =  {e^{-\phi} \beta F_{\sigma i}\over (2 \pi)^5 \alpha'^3} = {1 \over  ( 2 \pi)^5 \alpha'^3 g} {\partial A_i \over \partial \zeta^{\alpha}} {\partial \zeta^{\alpha} \over \partial \sigma} .
\end{split}
\end{equation}
In the equation above, $M,N$ run over $0 \ldots 5$. To obtain the conserved charges of the action \eqref{probebraneF}, we need to integrate the momenta above over all $6$ worldvolume coordinates.
We now proceed to show that a D5 brane that keeps the vector ${\bf n}^{M}$ of Section \ref{Killingspinor} tangent to its worldvolume at all points and has a worldvolume field strength of the form \eqref{Ftotal} is supersymmetric in the 4 backgrounds that we have discussed.

\subsubsection{D1-D5 background} 
We will discuss the D1-D5 background in some detail. The calculations required to verify supersymmetry in other backgrounds are almost identical, so we will be brief in later subsections. In the notation above, 

In the D1-D5 background of Table \ref{donefive}
\begin{equation} 
\begin{split}
{G^{(3)} \over \alpha'} &= Q_5 \sin 2 \zeta d \zeta \wedge d \phi_1 \wedge d \phi_2 - {2 Q_1 \over v f_1^2 r^3} d r \wedge d t \wedge d x_5 ,\\
{{C^{(2)}} \over \alpha'}&=-{Q_5 \over 2} \cos 2 \zeta d \phi_1 \wedge d
\phi_2 + {1 \over g f_1 \alpha'} dt \wedge dx_5 ,\\
{G^{(7)} \over \alpha'} &= \left({Q_1 \over v } \sin 2 \zeta d \zeta \wedge d \phi_1 \wedge d \phi_2 - {2 Q_5 \over f_5^2 r^3} d r \wedge d t \wedge d x_5\right)\wedge d z^{1} \wedge d z^{2} \wedge d z^{3} \wedge d z^{4} ,\\
{C^{(6)} \over \alpha'} &=  \left({-Q_1 \over 2 v } \cos 2 \zeta d \phi_1 \wedge d \phi_2 + {1 \over g f_5 \alpha'} dt \wedge d x_5 \right) \wedge d z^{1} \wedge d z^{2} \wedge d z^{3} \wedge d z^{4} .
\end{split}
\end{equation}
With the definition of $C'^{(2)}$ above, we have:
\begin{equation}
{C'^{(2)} \over \alpha'} =  \left({-Q_1 \over 2 v } \cos 2 \zeta d \phi_1 \wedge d \phi_2 + {1 \over g f_5 \alpha'} dt \wedge d x_5 \right) .
\end{equation}
Notice, that in the near horizon limit, we find $C'^{(2)} = {e^{2 \phi} \over g^2} {C^{(2)}}$.

To check the supersymmetry condition, we explicitly calculate $P_{t}$ and $P_{5}$ using \eqref{momentumwithF}.
\begin{equation}\label{momentatxfive} 
\begin{split} (2 \pi)^5 \alpha'^3 P_{t} &= -{F_{\sigma i} F_{\sigma}^{i} \over g x_5'} - {e^{- \phi} h_{\sigma \sigma} (\beta^2 + {|F|^2 \over 2}) \over x_5'}- C^{(2)}_{5 t} (\beta^2 + {|F|^2 \over 2}) x_5' ,\\
(2 \pi)^5 \alpha'^3 P_5 &= {F_{\sigma i} F_{\sigma}^{i} \over g x_5'} +{e^{-\phi}  h_{\sigma \sigma} (\beta^2 + {|F|^2 \over 2})\over x_5'} - (\beta^2 + {|F|^2 \over 2}) e^{-\phi} G_{5 5} x_5' ,
\end{split}
\end{equation}
where we have used that
\begin{equation}
C'^{(2)}_{5 t} = \beta^2 C^{(2)}_{5 t}.
\end{equation}
Using $G_{00} = -G_{55}$ and $e^{-\phi} G_{5 5} + C^{(2)}_{5 t} = 0$(See Table \ref{donefive}), we see that
\begin{equation}
E - L = \int \left( P_{t} + P_{5} \right) \, d \tau d \sigma d z^1 \ldots d z^4 = 0,
\end{equation}
and hence, the BPS relation is satisfied.

If we integrate \eqref{momentumwithF} to obtain the conserved charges
we see that in the near-horizon limit, where $C'^{(2)}={e^{2 \phi}
\over g^2} {C^{(2)}}$, the formulae for the energy, angular momentum
and other charges are almost identical in structure to Table
\ref{donefive} except that \be {1 \over 2 \pi \alpha'} \rightarrow {1
\over 2 \pi \alpha'} \left(\beta^2 v + {1 \over 32 \pi^4
\alpha'^2}\int |F|^2 d^4 z^{i}\right).
\label{tension-renorm}
\ee Hence, turning on the gauge fields simply renormalizes the 
tension according to the `instanton number' \eq{instantonumber}.\footnote{\label{real-inst} This will become the {\em real} instanton
number for $q>1$ in Section \ref{nonabelianextension}} This
equation is the precursor to the more general 
\eq{tension-renorm-general}.

\subsubsection{D1-D5-P Geometry}
The discussion for the D1-D5-P geometry specified by equation \eqref{donefivep} is almost identical to the one above. The only modification is that we find:
\begin{equation}\label{momentatxfivep} 
\begin{split} (2 \pi)^5 \alpha'^3 P_{t} &= -{F_{\sigma i} F_{\sigma}^{i} \over g x_5'} - {e^{- \phi} h_{\sigma \sigma} (\beta^2 + {|F|^2 \over 2}) \over x_5'}- \left(C^{(2)}_{5 t} + e^{-\phi} G_{5 t}\right) (\beta^2 + {|F|^2 \over 2}) x_5' ,\\
(2 \pi)^5 \alpha'^3 P_5 &= {F_{\sigma i} F_{\sigma}^{i} \over g x_5'} +{e^{-\phi}  h_{\sigma \sigma} (\beta^2 + {|F|^2 \over 2})\over x_5'} - (\beta^2 + {|F|^2 \over 2}) e^{-\phi} G_{5 5} x_5' ,
\end{split}
\end{equation}
In the new background \eqref{donefivep}, we have $e^{-\phi}(G_{5 5} + G_{5 t}) + C^{(2)}_{5 t} = 0$. Hence, the BPS relation follows.

\subsubsection{Lunin-Mathur Geometries}

To check the BPS condition for bound state probes in the Lunin-Mathur geometries, we need to derive an expression for $C'^{(2)}$ which is defined by \eqref{cprimedef}. At first sight, this may seem a formidable task, but the result is quite intuitive. In Appendix \ref{appLM} we show that $C'^{(2)}$ is obtained by taking $C^{(2)}$ in \eqref{luninmetric} and performing the substitution $H \leftrightarrow {1 \over 1+K}$. So
\begin{equation}
\label{cprimeluninmathur}
\begin{split}
C'^{(2)}_{t {m}} &= {-B_{m} H } , ~~ C'^{(2)}_{t 5} = {H}, ~~ C'^{(2)}_{{m} 5} = {H A_{m} }, ~~ C'^{(2)}_{{m} n} = C'_{{m} n} + H \left( A_m B_n - A_n B_m\right), \\
d B &= - * d A , ~~ d C' = - * d (1 + K) . \\
\end{split}
\end{equation}
Now, we only need to notice that $C'^{(2)}_{t M} = \beta^2 C^{(2)}_{t M}, C'{(2)}_{5 M} = \beta^2 C^{(2)}_{5 M}, ~ \forall M$\begin{footnote}{
As we mentioned earlier, the conventions of \cite{Lunin:2002bj} differ slightly from \cite{Maldacena:1998bw} and $g$ has been absorbed into a shift of $\phi$. So, here $\beta = e^{\phi}$} \end{footnote} and repeat the argument for the D1-D5 system above to see that $P_{t} + P_{5} = 0$.

\subsubsection{Global AdS}

The analysis, with gauge fields turned on in the D5 brane
worldvolume is almost identical to the analysis in the full D1-D5 background.
Here, we find
\begin{equation}
{C'^{(2)}_{\rm global} \over \alpha'} = {e^{2 \phi} \over g^2} {C^{(2)}_{\rm global} \over \alpha'} = -{Q_1 \over 2 v} \left[\cos 2 \zeta d \phi_1 \wedge d \phi_2 - (\cosh(2 \rho) - 1) d t \wedge d \theta \right] .
\end{equation}

To check the BPS condition, let us use formula \eqref{momentumwithF}
to write down the momenta in the $t, \theta, \phi_1, \phi_2$
directions. In analogy to the analysis for the D-string, we define 
\begin{equation}\label{gammaone}
\gamma_1 = {{1 \over g } F_{\sigma i} F_{\sigma}^{i}+  Q_5 \alpha' \left( \beta^2 + {|F|^2 \over 2} \right) \left(\sinh^2 \rho \theta^{'2} + \cos^2 \zeta \phi_1^{'2} +\sin^2 \zeta \phi_2^{'2} + \zeta'^2 + \rho^{'2} \right) \over  \cos^2 \zeta \phi_1'+ \sin^2 \zeta \phi_2'+ \sinh^2 \rho \theta'} .
 \end{equation} with this definition, we find the momenta
\begin{equation}\label{globalmomentawithf} \begin{split} 
&{(2 \pi)^5 \alpha'^3 } P_{t} = -\gamma_1
\cosh^2(\rho) + Q_5 \alpha'  \theta' \sinh^2(\rho) (\beta^2 + {1
\over 2} |F|^2) ,\\ 
&{(2 \pi)^5 \alpha'^3 } P_{\theta} = \gamma_1
\sinh^2(\rho) - Q_5 \alpha' \theta' \sinh^2(\rho) (\beta^2 + {1 \over 2} |F|^2)
 , \\ 
&{(2 \pi)^5 \alpha'^3} \tilde P_{\phi_1} = \gamma_1 \cos^2 \zeta -
Q_5 \alpha' \left(\beta^2 + {1 \over 2} |F|^2\right) \left(\cos^2 \zeta \phi_1' - \sin^2 \zeta \phi_2'\right)
\phi_2' , \\ 
&{(2 \pi)^5 \alpha'^3 }\tilde P_{\phi_2} = \gamma_1 \sin^2 \zeta +
Q_5 \alpha' \left(\beta^2 + {1 \over 2} |F|^2\right) \left( \cos^2 \zeta \phi_1' - \sin^2 \zeta \phi_2'\right)
\phi_1', \\ 
&P_{t} +P_{\theta} + \tilde{P}_{\phi_1} + \tilde{P}_{\phi_2} = 0,
\end{split}
\end{equation}
which verifies the BPS relation.

\subsection{Obtaining an Effective Two-Dimensional Action}

The space of supersymmetric solutions above, gives us a description of
the supersymmetric sector of the classical phase space of the
worldvolume theory defined by the action \eqref{probebraneF}. Each
solution corresponds to a point in this phase-space. Now, the action
\eqref{probebraneF} gives rise to a canonical symplectic structure on
this phase space. This structure may be encapsulated in terms of a
{\em symplectic form}. See, for example \cite{crnkovic1987cdc} for
details of this construction. We will return to this formalism again
in Section \ref{quantizationsection}. We will now show that, the
classical symplectic structure on the space of supersymmetric
solutions above is identical to the symplectic structure on the space
of supersymmetric solutions of a 1+1 dimensional theory! This $1+1$
dimensional theory will be like the theory of the D-string studied in
\eqref{susydstringsdbi} but propagating on a different space, where
the internal manifold has been replaced by the instanton moduli
space. Furthermore, we will find that the tension of this string is
renormalized by a factor determined by the instanton number.

First consider the gauge fields. Recall, that in \eqref{momentumwithF}, we found that
\begin{equation}
p_{Ai} =  {1 \over  ( 2 \pi)^5 \alpha'^3 g} {\partial A_i \over \partial \zeta^{\alpha}} {\partial \zeta^{\alpha} \over \partial \sigma} .
\end{equation}

The {\em symplectic structure} on the manifold of solutions may be written in 
terms of the symplectic form:
\begin{equation}
\label{omegaonai}
\Omega = \int \delta p_{Ai} \wedge \delta A_i \, d \sigma d^4 z^{i} ,
\end{equation}
where $\delta$ is an exterior derivative on the space of all solutions. $\delta A_i$ is then a 1-form in the cotangent space at the point in phase space specified by the function $A_i$ and the wedge product is taken in this cotangent space.

The $A_i$ are given as a function of the moduli $\zeta^{a}$ by \eqref{Ainmoduli}. We can then rewrite \eqref{omegaonai} as:
\begin{equation}
\label{sympinstan}
\Omega= {1 \over  ( 2 \pi)^5 \alpha'^3 g}\int \delta \left(\int d^4 z^{i} {\partial A_i \over \partial \zeta^a} {\partial A_i \over \partial \zeta^b} \zeta'^{a}\right) \wedge \delta \zeta^{b} .
\end{equation}
If we define a metric on instanton moduli space,
\begin{equation}
\label{instantonmetric}
g^{\rm inst}_{a b} ={1 \over (2 \pi \sqrt{\alpha'})^4 } \int d^4 {z^{i}} {\partial A_i \over \partial \zeta^a} {\partial A_i \over \partial \zeta^b} ,
\end{equation}
then, this is exactly the symplectic structure of the left-moving sector($(\zeta^{a})'(\sigma,\tau) = \dot{\zeta}^{a}(\sigma,\tau)$) of the non-linear sigma model on the instanton moduli space defined by
\begin{equation}
\label{nonlinearinstanton}
S_{\rm inst} = {1 \over 4 \pi \alpha' g } \int g^{\rm inst}_{a b} \left(\dot{\zeta}^{a} \dot{\zeta}^{b} - (\zeta^{a})' (\zeta^{b})' \right) d \sigma d \tau .
\end{equation}

What about the contribution of the gauge fields to the spacetime Hamiltonian? From formula \eqref{momentumwithF} and the expressions in \eqref{Dinv}, we see that the gauge field momenta enter the expression for the spacetime energy only through 
$${1 \over (2 \pi)^5 \alpha'^3 }\int d^4 z^{i} d \sigma \, {F_{\sigma i}F_{\sigma}^{i} \over g} = {1 \over 2 \pi \alpha' g}  \int d \sigma g^{\rm inst}_{a b} \zeta'^{a} \zeta'^{b}
$$
This is exactly the Hamiltonian of the `left-moving' sector of the non-linear 
sigma model \eqref{nonlinearinstanton}.

Finally, we would like to write down an effective action that generates the symplectic structure above both in the D1-D5 system and in global AdS. To do this, first we formally extend our spacetime, by excising the coordinates on the internal manifold and including coordinates on the instanton moduli space. We now define a metric and and B field on this extended space as follows:
\begin{equation}
\label{effectiveonedfivebrane}
\begin{split}
{\cal \chi}^{m}&=\begin{pmatrix} X^{M} \\ \zeta^a \end{pmatrix} ,\\
{\cal G}^1_{m n} &= \begin{pmatrix} e^{-\phi} \left(\beta^2 v + \int d^4 z^{i}~ {|F|^2 \over 8 \pi^2 (2 \pi \alpha')^2} \right) G_{M N} &0 \\
0 & {g^{\rm inst}_{a b} \over g}\end{pmatrix} ,\\
{\cal B}^1 &= \left( C'^{(2)}_{M N} v + C^{(2)}_{M N}  \int d^4 z^{i}~ { |F|^2 \over 8 \pi^2 (2 \pi \alpha')^2} \right) d X^{M} \wedge d X^{N}, \\
{\cal H}^1_{\alpha \beta} &= {\cal G}^1_{m n} \partial_{\alpha} {\cal \chi}^{m} \partial_{\beta} {\cal \chi}^{n} . \\
\end{split}
\end{equation}
In the equation above, $M,N$ runs over $0 \ldots 5$, $a,b$ run over
the coordinates of the instanton moduli space, $m,n$ run over both
these ranges and $\alpha, \beta$ range over $\sigma, \tau$.  Now,
consider a sector with a fixed value of the `instanton number' $\int
d^4 z^{i} {|F|^2 \over 8 \pi^2 (2 \pi \alpha')^2}$
(see \eq{instantonumber}, also footnote \ref{real-inst}). In this
sector, consider the action:
\begin{equation}
\label{effectiveactionsingle}
S^1_{\rm eff} = {1 \over 2 \pi \alpha'} \int \left(-{\rm Det}[{\cal H}^1]\right)^{1 \over 2} d \sigma d \tau + {1 \over 2 \pi \alpha'} \int {\cal B}^1
\end{equation}
If we look for supersymmetric solutions to the action above, we will find that they too have the property that:
\begin{equation}
\label{extendedKilling}
{\partial \chi^m \over \partial \tau} = {\bf n}^{m}
\end{equation}
where we have extended the Killing vector field ${\bf n}^{M}$ of the previous section to this extended space in the natural way by setting its components along ${\partial \over \partial \zeta^a}$ to zero. On {\em these} solutions, the spacetime 
momenta derived from the action above reproduce the momenta \eqref{momentumwithF}. Together with \eqref{sympinstan} this tells us the symplectic structure on
supersymmetric solutions to the action \eqref{probebraneF} is the same as the 
symplectic structure on supersymmetric solutions to the action \eqref{effectiveactionsingle}. 
 The superscript $1$ above indicates that this analysis is valid for a single D5 brane. The formula above is very suggestive and has a natural non-Abelian extension that we now proceed to discuss.

\subsection{Non-Abelian Extensions}
\label{nonabelianextension}
The analysis in the
last two subsections was valid for a single D5 brane. 
It is easy to generalize the salient results to
$q$ D5 branes for $q>1$.  Again, we consider a sector with fixed
\begin{equation}
\label{instantonumberepeat}
p = {1 \over (2 \pi \sqrt{\alpha'})^4} \int_{T^4}{{\rm Tr}\left(F \wedge F\right) \over 2}.
\end{equation} 
$p$ is now a bona-fide instanton number. 
In this sector consider the following natural extension to the 
effective quantities above given by \eqref{effectiveonedfivebrane}:
\begin{equation}
\label{effectivePdfivebrane}
\begin{split}
{\cal \chi}^{m}&=\begin{pmatrix} X^{M} \\ \zeta^a \end{pmatrix} , \\
{\cal G}^{p,q}_{m n} &= \begin{pmatrix} e^{-\phi} \left(q \beta^2 v + p \right) G_{M N} &0 \\
0 & {g^{\rm inst}_{a b} \over g} \end{pmatrix} ,\\
{\cal B}^{p,q} &= \left( q C'^{(2)}_{M N} v  + C^{(2)}_{M N} p \right) d X^{M} \wedge d X^{N} ,\\
{\cal H}^{p,q}_{\alpha \beta} &= {\cal G}^{p,q}_{m n} \partial_{\alpha} {\cal \chi}^{m} \partial_{\beta} {\cal \chi}^{n} . \\
\end{split}
\end{equation}
$\zeta^a$ 
span the moduli space of $p$ instantons in a $U(q)$ theory. We can define an effective two dimensional action for each such value of $p,q$ as:
\begin{equation}
\label{effectiveactionPdfivebrane}
S^{p,q}_{\rm eff} = {1 \over 2 \pi \alpha'} \int \left(-{\rm Det}[{\cal H}^{p,q}]\right)^{1\over2} + {1 \over 2 \pi \alpha'} \int {\cal B}^{p,q} .
\end{equation}
Remarkably, we have found, that we can now apply the entire machinery of section \ref{susydstringsdbi}(which we developed for D1 branes) to bound-states of D1 and D5 branes. 

This result takes an especially pretty form in the near-horizon of the D1-D5 and D1-D5-P system and global AdS. Recall, that for these scenarios:
\begin{equation}
C'^{(2)}_{M N} = \beta^2 C^{(2)}_{M N} = {Q_1 \over Q_5 v} C^{(2)}_{M N}.
\end{equation}
The formula \eqref{effectivePdfivebrane} then tells us that in the near-horizon of the D1-D5 system and in global AdS(and in the corresponding Ramond sector, LM geometry), the formulae for the canonical momenta in
 Tables \ref{donefive} and \ref{global} are {\em quantitatively} correct with the following substitutions:
\begin{description}
\item[1.]
The internal manifold is replaced by the instanton moduli space of $p$ 
instantons in a $U(q)$ theory.
\item[2.]
The tension of the `string' is renormalized by $Q_5 \rightarrow p Q_5' + 
q Q_1'$. Here $Q_5'$ is the D5-charge of the background in Table 
\ref{donefive} and \ref{global} which must be taken to be $Q_5-q$
in case the D5 charge of the probe is $q$ (so that the total charge
at the boundary is kept fixed at $Q_5$). Similarly $Q_1'= Q_1-p$. Thus
\be
 Q_5 \rightarrow p (Q_5-q) + 
q (Q_1-p)
\label{tension-renorm-general}
\ee

\end{description}

\section{Moving off the Special Point in Moduli Space}
\label{movingoffspecial}
We can generalize the simplest D1-D5 system that we have been discussing by turning on a bulk anti self-dual $B_{NS}$ field in the background geometry.\begin{footnote}{Our conventions regarding `self-dual' and `anti-self-dual' are the opposite of \cite{Seiberg:1999xz,Dijkgraaf:1998gf,Dhar:1999ax}.}\end{footnote}
 This is like turning on some dissolved D3 brane charge in the background that we have taken, till now, to have only D1 and D5 charges. We should expect that the BPS solutions we have been discussing above no longer remain BPS, since a D1 or a D5 probe is not, in general, mutually supersymmetric with a D1-D3-D5 bound state (the exception is the system considered in Section \ref{kappadonefive}). In this section, we will verify the expectation above by first performing a Killing spinor analysis and then by verifying our results using the DBI action. 

\subsection{Killing Spinor Analysis}
The explicit extremal D1-D5 supergravity background with a non-zero
$B_{NS}$ fields turned on was calculated in
\cite{Dhar:1999ax,Maldacena:2000vw}. We will follow
\cite{Dhar:1999ax} here. In addition to this $B_{NS}$
field and the usual 3-form RR field strength $G$, this background also
has a 5-form field strength $G^{(5)}$. This solution depends on a
single parameter $\varphi$ that determines the strength of the
anti-self dual $B_{NS}$ field. The metric, dilaton and field strengths
(adapted to our conventions regarding `self-duality', and with
$\alpha'=1$ for simplicity) may be written as follows:
\begin{equation}\label{bnsturnedon}
\begin{split}
  ds^2 & = (f_1 f_5)^{-1/2} \left[ -dt^2 +dx_5^2 \right]
    + (f_1 f_5)^{+1/2} \left( d r^2 + r^2 ( d \zeta^2 + \cos^2 \zeta d \phi_1^2 + \sin^2 \zeta d \phi_2^2 ) \right) \\
    & + (f_1 f_5)^{+1/2} Z^{-1} \left[ (dx_6^2 + dx_8^2) + (dx_7^2 + dx_9^2) \right], \\
    e^{2 \phi} & = \frac{f_1 f_5}{Z^2}, \\
    H&=dB_{\text{NS}}, \\
    B_{\text{NS}}^{(2)} & = \left(Z^{-1} \sin(\varphi)\cos(\varphi)(f_1 - f_5) + {(\mu_5 - \mu_1) \sin{\varphi} \cos{\varphi} \over \mu_5 \cos^2 \varphi - \mu_1 \sin^2 \varphi}\right) \left(dx^6 \wedge dx^8 + d x^7 \wedge d x^9 \right) ,\\
    G^{(3)} & = \cos^2(\varphi) \tilde{K}^{(3)} - \sin^2(\varphi){K}^{(3)}, \\
G^{(5)} &= Z^{-1} \cos \varphi \sin \varphi \left(+f_5  K^{(3)}+ f_1 \tilde{K}^3 \right) \wedge \left(d x^6 \wedge d x^8 + d x^7 \wedge d x^9 \right) , \\
\end{split}
\end{equation}
where we defined
\begin{equation}
\begin{split}
    f_1 &= 1 + {\mu_1 \over r^2}~~~~f_5 = 1 + {\mu_5 \over r^2} ,\\
    \tilde{K}^{(3)} & = - \frac{f_1'}{f_1^2} dr \wedge dx^0 \wedge d x_5 + \mu_5 \sin(2 \zeta) d\zeta \wedge d \phi_1 \wedge d \phi_2, \\
    K^{(3)} & = - \frac{f_5'}{f_5^2} dr \wedge dx^0 \wedge d x_5 + \mu_1 \sin(2 \zeta) d\zeta \wedge d \phi_1 \wedge d \phi_2, \\
    Z & =1+\frac{\mu_1 \sin^2(\varphi)+ \mu_5 \cos^2{\varphi}}{r^2} . \\
\end{split}
\end{equation}
$\mu_1, \mu_5$ are parameters that determine the charges of the system according to the formulae in \cite{Dhar:1999ax}. We alert the reader that our normalizations for $\mu_1, \mu_5$ differ from that paper by a factor of 2.

We start by calculating the bulk Killing spinors that this geometry preserves.
As explained earlier the supersymmetries of the type IIB theory may be written in terms of a two-component spinor
\begin{equation}\label{spinor2compmtext}
  \epsilon=\left(
    \begin{array}{cc}
      \epsilon_1 \\
      \epsilon_2
    \end{array}
  \right),
\end{equation}
which satisfies $\Gamma^{11} \epsilon = - \epsilon$.
The dilatino Killing spinor equation is (see \cite{Papadopoulos:2003jk} and references therein)
\begin{equation}\label{nsdilatinoequation}
  \left[
    \partial_M \phi \Gamma^M + \frac{1}{12} H_{MAB}\Gamma^{MAB} \otimes \sigma_3
    +\frac{1}{4} e^\phi
    \sum_{n=1}^{5} \frac{(-1)^{n-1}(n-3)}{(2n-1)!}
    G_{A_1 \ldots A_{2n-1}} \Gamma^{A_1 \ldots A_{2n-1}} \otimes \lambda_n
  \right] \epsilon = 0,
\end{equation}
where $\lambda_n = \sigma_1$ for $n$ even, and
$\lambda_n = i\sigma_2$ for $n$ odd. The $\{\sigma_i\}$, $i=1,2,3$ are
the Pauli matrices.
$H$ and $G$ are the NS-NS and R-R field strengths, and $\phi$
denotes the dilaton. Our conventions are slightly different from \cite{Papadopoulos:2003jk} because the solution of \eqref{bnsturnedon} has $G_7=*G_3$ and $G_5 = -*G_5$.

The spinors above are defined with respect to a particular local Lorentz frame. In our case, a convenient basis is defined by the following one-forms.
\begin{equation}
\label{nsvielbein}
\begin{split}
e^{\c t} &= (f_1 f_5)^{-{1 \over 4}} d t , \\
e^{\hat{5}} &= (f_1 f_5)^{-{1 \over 4}} d x_5  ,\\ 
e^{\hat{r}} &= (f_1 f_5)^{1 \over 4} d r ,\\
e^{ \hat{\zeta}} &= (f_1 f_5)^{1 \over 4} r d \zeta , \\
e^{ \hat{\phi_1}} &=  (f_1 f_5)^{1 \over 4} r \cos \zeta d \phi_1 , \\ 
e^{ \hat{\phi_2}} &=  (f_1 f_5)^{1 \over 4} r \cos \zeta d \phi_2 , \\
e^{{\hat a}} &=  (f_1 f_5)^{1 \over 4} Z^{-{1 \over 2}} d x^{a} . \\
\end{split}
\end{equation}
Defining spinors with respect to this local Lorentz frame, we find that the Dilatino equation becomes
\begin{equation}\label{dilatinoBNS}
\begin{split}
  & \left[
  f_1^{-5/4} f_5^{-1/4} f_1' \Gamma^{\hat{r}} \left(
    \left( 1-2 \frac{f_1}{f_5} \frac{\sin^2(\varphi)}{\alpha} \right) \one - \Gamma^{\hat{0}\hat{5}} \otimes \sigma_1 - B \left( \Gamma^{\hat{6}\hat{8}}+\Gamma^{\hat{7}\hat{9}}\right)\otimes \sigma_3
  \right) \right] \epsilon \\
  + & \left[ f_5^{-5/4}f_1^{-1/4}f_5' \Gamma^{\hat{r}} \left(
    - \left( 1-2 \frac{f_1}{f_5} \frac{\sin^2(\varphi)}{\alpha} \right) \one - \Gamma^{\hat{r}\hat{\zeta}\hat{\phi_1}\hat{\phi_2}} \otimes \sigma_1 + B \left( \Gamma^{\hat{6}\hat{8}}+\Gamma^{\hat{7}\hat{9}}\right)\otimes \sigma_3
  \right)
  \right] \epsilon = 0 ,
\end{split}
\end{equation}
where we defined $\alpha \equiv \cos^2(\varphi) + \frac{f_1}{f_5} \sin^2(\varphi), ~~B\equiv \sqrt{\frac{f_1}{f_5}}\frac{1}{\alpha} \sin(\varphi)\cos(\varphi)= {\sqrt{f_1 f_5} \sin(\varphi) \cos(\varphi) \over f_5 \cos^2(\varphi) + f_1 \sin^2(\varphi)}$.
All products of Gamma matrices above can be simultaneously diagonalized. We will denote the eigenvalues of $\Gamma^{\hat{0} \hat{5}}, \Gamma^{\hat{6} \hat{8}}, \Gamma^{\hat{7} \hat{9}}, \Gamma^{\hat{r} \hat{\zeta} \hat{\phi_1} \hat{\phi_2}}$ by $\pm n_1, \pm i n_2, \pm i n_3, \pm n_4$ respectively. The condition $\Gamma^{11} \epsilon = -\epsilon$ subjects these to the constraint $\prod n_1 n_2 n_3 n_4 = -1$.

Diagonalizing the matrix above is then equivalent to diagonalizing the two matrices:
\begin{equation}
\begin{split}
M_1 &= n_1 \sigma_1 - i B (n_2 +  n_3) \sigma_3 , \\
M_2 &= n_4 \sigma_1 + i B (n_2 + n_3) \sigma_3 .
\end{split}
\end{equation}
Both these matrices have eigenvalues $\pm \sqrt{1 - B^2(n_2 + n_3)^2}$. In particular, when $n_2 n_3 = 1 = -n_1 n_4$, there are $8$ spinors that simultaneously satisfy the two equations
\begin{equation}
\label{bnsprojections}
\begin{split}
&\left( \Gamma^{\hat{0}\hat{5}} \otimes \sigma_1 + B \left( \Gamma^{\hat{6}\hat{8}}+\Gamma^{\hat{7}\hat{9}}\right)\otimes \sigma_3 \right) \epsilon = {f_5 \cos^2{\varphi} - f_1 \sin^2{\varphi} \over f_5 \cos^2{\varphi} + f_1 \sin^2{\varphi}} ~\epsilon  ,\\
&\left( \Gamma^{\hat{r}\hat{\zeta}\hat{\phi_1}\hat{\phi_2}} \otimes \sigma_1 - B \left( \Gamma^{\hat{6}\hat{8}}+\Gamma^{\hat{7}\hat{9}}\right)\otimes \sigma_3 \right) \epsilon = -{f_5 \cos^2{\varphi} - f_1 \sin^2{\varphi} \over f_5 \cos^2{\varphi} + f_1 \sin^2{\varphi}}~ \epsilon .
\end{split}
\end{equation}
These two equations are consistent with $\Gamma^{11} \epsilon = -\epsilon$ and satisfy the equation \eqref{dilatinoBNS}.  They also imply  $\Gamma^{6 7 8 9} \epsilon = \epsilon$.

Hence, we have shown that the background defined by \eqref{bnsturnedon} 
preserves 8 supersymmetries that are parameterized by the projection conditions above. 
Notice that {\em none} of these spinors can be preserved by a probe D1 brane or a probe D5 brane. For arbitrary unit tangent vectors of the worldvolume ${\bf \hat{v}_1}, {\bf \hat{v}_2}$, a probe D1 brane preserves the spinors that have $\Gamma_{\bf {\hat{v}_1}} \Gamma_{\bf{\hat{v}_2}} \otimes \sigma_1 \psi = \psi$. In the two dimensional space specified by \eqref{spinor2compmtext} these spinors are eigenspinors of $\sigma_1$.  Hence none of them coincide with the spinors that are preserved in the background above that are eigenspinors of $\sigma_1 \pm 2 i B \sigma_3$. The same argument
works to show that no probe D5 branes or bound states of D1 and D5 branes can be 
supersymmetric in this background.

Now, consider the near-horizon limit of the geometry \eqref{bnsturnedon}. In this limit, the equation above simplifies dramatically and it is easy to convince oneself that the only projection that survives above is $\Gamma^{6 7 8 9} \epsilon = \epsilon$. There are 16 spinors that satisfy this equation. 
Hence, this is consistent with the `doubling' of supersymmetries that is associated with the appearance of a conformal symmetry in the near-horizon limit. One may now naively suspect, that in the near-horizon a probe D-string could maintain some supersymmetries. 

In the superconformal algebra, there are two types of supercharges. Conventionally, these are denoted by $Q$ -- with a charge under dilatation of $+{1 \over 2}$ -- and $S$ with a dilatation charge $-{1 \over 2}$. Now, to be BPS, we want a brane to preserve some $Q$ charges (in the superconformal algebra all primary states, whether of short representations or not are annihilated by the $S$'s). 
To determine which supercharges are $Q$ and which are $S$ in the near-horizon, we consider the $\hat{r}$ component of the Gravitino equation in the near-horizon limit. 

The Gravitino equation reads 
\begin{equation}\label{gravitinoequationtext}
  \left[
    \partial_M + \frac{1}{4} w_M^{BC} \Gamma_{BC}
    +\frac{1}{8} H_{MAB}\Gamma^{AB} \otimes \sigma_3
    + \frac{1}{16} e^\phi
      \sum_{n=1}^{5} \frac{(-1)^{n-1}}{(2n-1)!}
      G_{A_1 \ldots A_{2n-1}} \Gamma^{A_1 \ldots A_{2n-1}}
      \Gamma_M \otimes \lambda_n
  \right] \epsilon = 0.
\end{equation}
where $w_M^{B C}$ is the spin connection. In the near-horizon the $r$ component of this equation is, for the background above:
\begin{equation}
{\partial \epsilon \over \partial r} - {1 \over 2 r}\left[ \Gamma^{\hat{0} \hat{5}} {(\mu_5 \cos^2{\varphi} - \mu_1 \sin^2{\varphi}) \sigma_1  -  \sqrt{\mu_5 \mu_1} \cos \varphi \sin \varphi (\Gamma^{\hat{6} \hat{8}} + \Gamma^{\hat{7} \hat{9}}) \otimes (i \sigma_2) \over \mu_5 \cos^2{\varphi} + \mu_1 \sin^2{\varphi}}\right] \epsilon = 0 .
\end{equation}
If we impose $n_2 n_3 = 1$(as the dilatino equation tells us to), the square bracket on the right has eigenvalues $\pm 1$.
Somewhat more remarkably, the eigenvalue $+1$ occurs when the projection condition \eqref{bnsprojections} is satisfied. This means that the $Q$'s in the near-horizon are the same as the $Q$'s in the bulk. The new supercharges are the $S$'s. From the argument above, we now see a D-string or a D5 brane cannot be BPS even 
in the near-horizon.
The argument for global AdS is very similar to the near-horizon argument above and instead of repeating it here, 
we will proceed to verify our results using a charge analysis.

\subsection{Charge Analysis}
In this section, we will use the DBI action to verify the results that we obtained above. For global AdS, we find the interesting result that there are still solutions to the equations of motion that preserve the Killing vector ${\bf n}$ but these solutions are no longer BPS. 

We start by considering the extremal D1-D5 geometry. From the formulae in \eqref{bnsturnedon}, we see that 
\begin{equation}
\label{cfivetns}
\begin{split}
C^{(2)}_{t 5} &= {f_5 \cos^2 \varphi - f_1 \sin^2 \varphi \over f_1 f_5} , \\
e^{-\phi} G_{5 5}  &= {Z \over f_1 f_5} = {f_5 \cos^2 \varphi + f_1 \sin^2 \varphi \over f_1 f_5}.
\end{split}
\end{equation}
We see that the ratio between the components of the ${C^{(2)}}$ field and the metric has been 
spoilt. This effect is quite general and is the same as what we should expect if turn on a 
theta angle.
Now, the equation of motion \eqref{eqmotion} for $r$ receives contributions from
 the following terms. (1) $X^{M} = x_5$,~ $X^{N} = x_5$ and (2)$X^{M} = x_5,~X^{N} = \tau$.
Since, now $e^{-\phi} G_{55}+C^{(2)}_{5 t} \neq 0$, the only way to force our solutions to obey these equations is to set $(x_5)' = 0$. This confirms the expectation that, in the D1-D5 geometry, the supersymmetric brane probe solutions vanish if we move on the moduli space. It is easy to repeat the argument above 
to show that the same result also holds true in the D1-D5-P geometry.

The situation in global AdS is more interesting. When we take the near-horizon limit
of \eqref{bnsturnedon} and translate to global coordinates, we find the metric
\begin{equation}
\label{metricbnsnearhorizon}
\begin{split}
e^{-\phi} G_{M N} d x^M d x^N  &=  Q_5' \left( -\cosh^2 \rho d t^2 + \sinh^2 \rho d \theta^2 + d \rho ^2 + d \zeta^2 + \cos^2 \zeta d \phi_1^2 + \sin^2 \zeta d \phi_2^2\right) \\ &+d z^i d z^i, \\
\end{split}
\end{equation}
and RR 2-form components
\begin{equation}
\begin{split}
&C^{(2)}_{\phi_1 \phi_2} = - Q_5'(1 - \epsilon^2) {\cos (2 \zeta) \over 2}, \\
&C^{(2)}_{t \theta} =  Q_5'(1 - \epsilon^2) {\cosh(2 \rho) - 1 \over 2}, 
\end{split}
\end{equation}
where
\begin{equation}
\begin{split}
Q_5' &= \mu_5 \cos^2 \varphi + \mu_1 \sin^2 \varphi, \\
\epsilon^2 &= {2 \mu_1 \sin^2 \varphi \over \mu_5 \cos^2 \varphi + \mu_1 \sin^2 \varphi}.
\end{split}
\end{equation}
The equation of motion for $\rho$ now receives contributions from: $(1) X^{M} = \theta, X^{N} = \theta~(2) X^{M} = \theta, X^{N} = \tau$, while the equation of motion for $\zeta$ receives contributions from $(1) X^{M} = \phi_1, X^{N} = \phi_1 ~(2) X^{M} = \phi_2, X^{N} = \phi_2 ~(3) X^{M} = \phi_1, X^{N} = \phi_2 ~(4) X^{M} = \phi_2, X^{N} = \phi_1.$ The identities we need are
\begin{equation}
\begin{split}
e^{-\phi} G_{\theta \theta} + C^{(2)}_{\theta t} &=  \epsilon^2 G_{\theta \theta} = Q_5'\epsilon^2 \sinh^2 {\rho} ,\\
e^{-\phi} G_{\phi_1 \phi_1} + C^{(2)}_{\phi_1 \phi_2} &=  {Q_5' \over 2} (1 + \epsilon^2  \cos(2 \zeta)), \\
e^{-\phi} G_{\phi_2 \phi_2} + C^{(2)}_{\phi_2 \phi_1} &= {Q_5' \over 2} (1 - \epsilon^2 \cos(2 \zeta)) . \\
\end{split}
\end{equation}
The equations of motion are then satisfied if
\begin{equation}
\label{constraint}
\begin{split}
\sinh{2 \rho}  \theta' &= 0 ,\\
\sin (2 \zeta)  (\phi_1' - \phi_2') &= 0 . \\
\end{split}
\end{equation}
The first equation requires us to stay at a constant point in $\theta$. The second equation requires $\phi_1' = \phi_2'$. With these constraints, one can find solutions of the form \eqref{solutions} to the equations of motion. 

Unfortunately, these solutions do not maintain the BPS bound.  Generalizing 
the formulae of table \ref{global}, we find that
\begin{equation}
\label{newmomenta}
\begin{split}
P_{t} &= {-Q_5' \over 2 \pi}  \gamma \cosh^2 \rho ,\\
P_{\theta} &= {Q_5' \over 2 \pi} \gamma \sinh^2 \rho ,\\
\tilde{P}_{\phi_1} &= {Q_5' \over 2 \pi} \left(\gamma \cos^2 \zeta - \phi_1' \cos^2{\zeta} - {1 - \epsilon^2 \over 2} \cos (2 \zeta) \phi_2' + {1 - \epsilon^2 \over 2} \phi_2' \right) ,\\
\tilde{P}_{\phi_2} &= {Q_5' \over 2 \pi} \left(\gamma \sin^2 \zeta - \phi_2' \sin^2 \zeta + {1 - \epsilon^2 \over 2} \cos (2 \zeta) \phi_1' + {1 - \epsilon^2 \over 2}  \phi_1'\right).
\end{split}
\end{equation}
Substituting, $\phi_1' = w = \phi_2'$, we find that
\begin{equation}
\label{energyincrease}
E - L - J_1 - J_2 = - \int \left(P_{t} + P_{\theta} + \tilde{P}_{\phi_1} + \tilde{P}_{\phi_2}\right)~ d \sigma  = Q_5'\epsilon^2 w.
\end{equation}

So, the energy of these solutions increases as we move off the special submanifold in moduli space where the anti self-dual NS-NS fluxes and theta angles are set to zero. Equation \eqref{energyincrease} tells us how this happens as a function of the distance in moduli space from the special submanifold.

\section{Semi-Classical Quantization}
\label{quantizationsection}
The phase-space of a theory is isomorphic to the space of all its classical solutions. Using the Lagrangian, we can equip this space with a symplectic form that we can invert to calculate Dirac brackets. Then, by promoting Dirac brackets to commutators, we can use the set of classical solutions to canonically quantize the theory. The advantage of this approach is that it is covariant and that it allows us to restrict attention to special sectors of phase space by identifying the corresponding sector of classical solutions.\begin{footnote}{This is valid only if the symplectic form does not mix a solution that belongs to this subset with a solution that doesn't.}\end{footnote}  
This technique has a long history and the first published reference to it, known to us, is by Dedecker \cite{dedecker1953cvf}. Later, this was studied in  \cite{goldschmidt1973hcf,Kijowski:1973gi,gawedzki1974cfl,szczyrba1976sss,garcia570rss}and then brought back into use in the eighties by \cite{zuckerman1987apa,crnkovic1987cdc}. We refer the reader to \cite{Lee:1990nz} for a nice exposition of this method.  

In this section, we will show how this procedure can be implemented for supersymmetric brane probes propagating in the {\underline{near-horizon}}
region of the D1-D5 system. 
As we explained earlier, this study has limited physical relevance because it has been argued that the extremal D1-D5 geometry is not the dual to any particular Ramond vacuum of the boundary CFT but should be thought of as an average
over all Ramond vacua. 
In fact, even classically, we see that our probes in global AdS have the striking feature that they are generically bound the center of AdS. On quantization we would expect these to give rise to `discrete' states. This is in sharp contrast to what we find 
by quantizing probes in the extremal D1-D5 background where all the states that we 
obtain are at the bottom of a continuum. Since, the Ramond and NS sectors of the boundary theory are related by `spectral flow' on the 
boundary, this bolsters the argument above that the extremal D1-D5 geometry is only an `average' geometry and that we should really consider probes about the geometries described in \cite{Lunin:2001fv,Lunin:2002bj,Lunin:2002iz}

Nevertheless, we include this study as an example of how these supersymmetric solutions may be quantized. A detailed study of the quantization of probes in global AdS is left to \cite{Raju}.

Consider the near-horizon limit of the D1-D5 system. Let us define $y= {\alpha' l^2  \over r}$ where $l^2$ is a constant
defined in the next equation. In the near horizon our background is
\begin{equation}\label{dodfb} \begin{split} ds^2&= l^2 \alpha'  \left( {-dt^2 +d x^2  \over
y^2} + {dy^2 \over y^2}  + d \omega_3^2 \right) + \sqrt{Q_1 \over
Q_5 v} ds_{int}^2,    \\ e^{-2 \phi}&= {Q_5 v \over g^2 Q_1},  \\ G^{(3)}& =
 Q_5 \alpha' \sin (2 \zeta) d \zeta \wedge d \phi_1 \wedge d \phi_2 - {2 Q_5 \alpha' \over y^3 } dy \wedge dt
\wedge dx_5, \\ {C^{(2)}}&={-Q_5 \alpha' \over 2} \cos 2 \zeta d \phi_1
\wedge d \phi_2 + {Q_5 \alpha' \over y^2}  dt \wedge d x_5 ,\\
l^2&={g \over \sqrt{v}} \sqrt{Q_1 Q_5}. \\ 
\end{split}
\end{equation}
The momentum conjugate to $y$ is 
\begin{equation}
\label{py}
P_{y} = -{Q_5 \over 2 \pi} {y' \over y^2} .
\end{equation}

The near horizon geometry of the background described above would have
been $AdS_3$ in Poincare coordinates, had the D1-branes and D5-branes
not been on a circle. Adding in the circle identification, we simply
get the orbifold of $AdS_3$ by a (Poincare) shift, i.e. the zero mass
BTZ black hole.

Recall, from section \ref{boundstates}, that we can treat all probes,
D-strings or bound states of p D1 branes and q D5 branes on the same
footing by performing the replacements \eq{tension-renorm-general}
\begin{equation}
Q_5 \rightarrow k = p (Q_5 - q) + q (Q_1 - p), ~~ M_{\rm int} \rightarrow {\cal M}_{p,q} .
\end{equation}
where ${\cal M}_{p,q}$ is the instanton moduli space of p instantons in a $SU(q)$ theory. 

The symplectic form, $\Omega$ on the space of solutions is given by 
\begin{equation}
\label{symplectic}
\Omega = \int \delta P_{M} \wedge \delta X^{M} \, d \sigma ,
\end{equation}
where $\delta$ may be thought of as an exterior derivative in the space of solutions. 
Recall, the discussion in subsection \ref{probe}. Apart from fixing $t = \tau$ we can use diffeomorphism invariance to set 
\begin{equation}
x_5 = w \sigma .
\end{equation}

The formula for the spacetime energy becomes
\begin{equation}
\begin{split}
E &= {k \over w} \int {d \sigma \over 2 \pi} ~\left( {y'^2 \over y^2} + \cos^2 \zeta \phi_1'^2 + \sin^2 \zeta \phi_2'^2 + \zeta'^2 + {g_{a b}^{\rm int} (z^a)' (z^b)' \over k g \alpha'}\right) \cr
&= {E_y + E_{S^3} + E_{\rm int} \over w} .
\end{split}
\end{equation}

Since we have fixed both $t$ and $x_5$, the $\delta P_{5} \wedge \delta x_5 + \delta P_{t} \wedge \delta t$ terms drop out of the symplectic form, which then becomes:
\begin{equation}
\label{sympoincare}
\begin{split}
\Omega &= \int \left( \delta P_y \wedge \delta y + \delta P_{\phi_1} \wedge \delta \phi_1 + \delta P_{\phi_2} \wedge \delta \phi_2 + \delta P_{\zeta} \wedge \delta \zeta + \delta P_{i}^{\rm int} \wedge \delta x^i \right) \, d \sigma \\
&= \Omega_y + \Omega_{S^3} + \Omega_{\rm int} .
\end{split}
\end{equation}

Now, if we define $y=e^{\rho}$, we find that
\begin{equation}
\label{uonetheory}
\begin{split}
\delta P_{y} \wedge \delta y &=  {-k \over 2 \pi} \delta \rho' \wedge \delta \rho ,\\
E_y &= {k \over 2 \pi} \int (\rho')^2 d \sigma .
\end{split}
\end{equation}
We can now expand $\rho$ in modes
\begin{equation}
\rho = {1 \over \sqrt{2 k |n|}} \rho_{n} \exp{i n \sigma} .
\end{equation}
This leads to the Dirac brackets and Hamiltonian
\begin{equation}
\label{poisson}
\begin{split}
\{\rho_{n}, \rho_{-n}\}_{\rm D.B} &= i , ~~n > 0 \\
E_y &= \sum_{n \in \mathbb{Z}} {1 \over 2} n |\rho_n|^2 .
\end{split}
\end{equation}
We can promote these Dirac brackets to commutators to get an infinite sequence of harmonic oscillators. We can think of these oscillators as coming from the left-moving part of a free boson. Roughly, the anti-holomorphic oscillators have been set to zero by supersymmetry. Moreover, the zero modes that tie the left and right movers together are also absent from the expression \eqref{poisson}.  

Now we turn to $\Omega_{S^3}$. We can map the $S^3$ into an $SU(2)$ group element using
\begin{equation}
g = e^{i {\phi_1 - \phi_2 \over 2} \sigma_3 } e^{i \zeta \sigma_2} e^{i {\phi_1 + \phi_2 \over 2} \sigma_3} .
\end{equation}
Now, introduce light-cone coordinates on the worldsheet $x^{\pm} = \tau \pm \sigma$.  Consider the $WZW$ action
\begin{equation}
S = {-k \over 4 \pi} \int d^2 x Tr\{ (g^{-1} \partial_{M} g)^2 \} + k \Gamma^{\rm SU(2)}_{WZ}.
 \end{equation}
where $\Gamma^{\rm SU(2)}_{WZ}$ is the standard Wess Zumino term for the $SU(2)$ model \cite{Witten:1983ar}.  The symplectic form and energy obtained from the action above by restricting to solutions that satisfy $\partial_{+} g = 0$ coincides with $\Omega_{S^3}$ and $E_{S^3}$. Roughly speaking, we have the `left-moving'' part of the $SU(2)$ WZW model. 

The quantum WZW model has a current algebra and states in its Hilbert space break up into representations of this algebra. Each representation is identified by its affine primary $[j]$ \cite{DiFrancesco:1997nk}. The number of affine primaries is finite and $j \in \{0, {1 \over 2}, \ldots {k \over 2}\}$. What primaries occur in the spectrum above? If we consider the limit of large $k$, the WZW model describes three free bosons. If we were to quantize three bosons, $X^i(\sigma, \tau)$, using the symplectic form $\int d (X^i)' \wedge d X^i$, we would project out all right moving oscillators and all zero mode-motion. This suggests that the only affine primary in the spectrum is $[0]$.

We can obtain this result another way by using the fact that the spectrum of the $SU(2)$ model comprises the affine primaries $\sum_{j = 0}^{k/2} [j]_{\rm left} \times [j]_{\rm right}$. Since, here we have restricted the right moving-sector to be trivial, the only left-moving primary that can occur is $[0]$. 

Finally, we turn to the internal degrees of freedom that correspond to fluctuations on the internal manifold. Just as above the symplectic form $\Omega_{\rm int}$ and $E_{\rm int}$ give rise to the left-moving sector of the non-linear sigma model on ${\cal M}_{p,q}$. We will denote this Hilbert space,  which corresponds to the holomorphic part of the trivial zero mode sector of the sigma model on ${\cal M}_{p,q}$ by $H^0({\cal M}_{p,q})$. 
 
To conclude, we have found that the quantization of D-strings in the near-horizon of the D1-D5 system yields the left-moving part of the $R \times SU(2) \times {\cal M}_{p,q}$ sigma model defined on a circle of length $2 \pi w$. We need to sum over all $w$ to obtain the physical spectrum. 

The theory above is the Ramond sector of the theory of `long-strings' studied in \cite{Seiberg:1999xz,Maldacena:2000hw,Maldacena:2000kv}(A closely related theory was studied in \cite{Callan:1991ky,Callan:1991dj,Callan:1991at}).  There, it is shown how the $R \times SU(2)$ theory on the worldsheet may be embedded into a {\em spacetime} $N=4$ superconformal algebra with central charge $6 (k - 1)$. The $N=4$ superconformal algebra on ${\cal M}_{p,q}$ carries over to spacetime. 

It is important to note that we {do not} sum over spin structures in
the worldsheet theory. The fermions are always in the Ramond sector.
The second important feature of the spectrum above is that it is at
the bottom of a continuum of non-supersymmetric states. 
We can always move
infitesimally away from supersymmetry by turning on the continuous
momentum modes of $\rho$. This means that the Hilbert space we
obtained above is of measure `zero' in the full quantum theory.

\section{Results and Discussion}
\label{discussionsection}
In this paper we studied brane probes in  (a)the extremal D1-D5 background, (b) the extremal D1-D5-P background, (c) the smooth geometries of Lunin and Mathur with the same charges as the D1-D5 background and (d) global $AdS_3 \times S^3 \times T^4/K3$. In the first three backgrounds, states that satisfy $E - L = 0$ preserve the right moving supercharges. The charge $-(E - L)$ is generated by the vector ${\partial \over \partial t} + {\partial \over \partial x_5}$ and we found that D-strings that maintained this vector tangent to their worldvolume at all points preserved all right moving supersymmetries. The three backgrounds above preserve 8 supersymmetries and the supersymmetric probes preserve ${1 \over 2}$ of these. In global $AdS_3 \times S^3 \times T^4/K3$, the right moving BPS relation is $-(E - L - J_1 - J_2) = 0$. This combination of charges is generated by the vector ${\partial \over \partial t} + {\partial \over \partial \theta} + {\partial \over \partial \phi_1} + {\partial \over \partial \phi_2}$ and we found that D strings that keep this vector tangent to their worldvolume at all points preserve 4 right moving supersymmetries (this makes them ${1 \over 4}$ BPS in this background). This fact allowed us to parameterize all supersymmetric D string probes in these backgrounds by their initial profiles. This result is summarized in equation \eqref{solutions}. 

D5 branes with self-dual gauge fields on their worldvolumes, that preserve the Killing vector above, are also supersymmetric. These gauge fields correspond to a dissolved D1 charge on the D5 worldvolume, so we interpreted supersymmetric probes of this kind as supersymmetric bound states of D1 and D5 branes. We found that these bound state probes could be described in a unified $1+1$ dimensional framework described by equations \eqref{effectivePdfivebrane} and \eqref{effectiveactionPdfivebrane}. This allowed us to treat them on the same footing as D1 branes.

In global $AdS$, and the corresponding Lunin-Mathur solution, the probes we found could not escape to infinity for a generic assignment of charges. This indicates that upon quantization they give rise to discrete bound states that contribute to the BPS partition function of string theory on this background. A detailed investigation of this is left to \cite{Raju}. The fact that this structure of classical bound states is not seen in the extremal D1-D5 geometry provides further evidence for the argument that this background is not the correct dual to any Ramond vacuum in the boundary CFT.

In Section \ref{movingoffspecial}, we showed that these supersymmetric probes vanished if we turned on an anti-self-dual NS-NS field or theta angle. This means that the BPS partition function jumps as we move off the special point in moduli space where these background moduli are set to zero. This issue
is discussed further in \cite{Raju}. We note that this result is 
similar to the result that the ${1 \over 8}$ and ${1 \over 16}$ BPS partition functions of ${\cal N}=4$ SYM theory on $S^3 \times R$ jump as soon as we turn on a 't Hooft coupling but are not further renormalized \cite{Kinney:2005ej}. Finally, in section \ref{quantizationsection}, we quantized the supersymmetric probes above in the near-horizon of the extremal D1-D5 geometry to obtain `long-string' states at the bottom of a continuum of non-supersymmetric states. 
%The investigation of quantum bound states in global AdS is left to \cite{Raju}. 

It would be interesting to find smooth supergravity solutions that
correspond to the probes above. It is possible that these solutions
could be generated by using the profiles we find in the programme of
\cite{Lunin:2001fv,Lunin:2002bj}. An ensemble of energetic spinning
probes may be a useful representation of the $BTZ$ black hole. An
indication of this was seen in Section \ref{susydstringsdbi}.  Now, in
the probe approximation, we can have many probes moving in $AdS_3$
that are simultaneously supersymmetric. In global $AdS$ our analysis
indicates that these probes would all be bound to $AdS$ and hence
exist at a finite distance determined by their charges.  If these
probes have large values of $p,q$, they have many internal degrees of
freedom that could give rise to a macroscopically measurable
degeneracy. This suggests the interesting possibility that there may
be multi-black hole solutions in global $AdS_3 \times S^3 \times
T4/K3$.  Similar ideas have been proposed by de Boer
\cite{DeBoer:2007} and Sundborg \cite{Mansson:2000sj}.

\section*{Acknowledgements}
We are very grateful to S. Minwalla for collaboration and for extensive 
discussions throughout this project.
We would also like to thank  M. Berkooz, A. Dabholkar, S. Das, A. Dhar, J. de Boer, F. Denef, R. Gopakumar, L. Grant,  M. Guica, S. Kim, J. Maldacena, S. Mukhi, S. Nampuri, K. Narayan, N. Suryanarayana, D. Tong, S. Trivedi, S. Wadia and especially S. Lahiri and K. Papadodimas for helpful discussions. G.M. would
like to thank Department of Physics and Astronomy, University of Lexington,
Kentucky, US and Perimeter Institute of Theoretical Physics, Waterloo, Ontario, Canada for hospitality where part of this work was done.

\section*{Appendices}
\appendix
\section{Miscellaneous Technical Details}
\label{miscdetails}
\subsection{Inverse of the Born-Infeld Matrix}
\label{appendixinverse}
The matrix D in \eqref{Ddefined} is simple to invert. We will only be interested in the first row and column, so we list those below:
\begin{equation}\label{Dinv}
\begin{split} \sqrt{-|D|} D^{\tau  \alpha} &= \{ -{ \beta
F_{\sigma i} F_{\sigma}^{i}+ h_{\sigma \sigma} (\beta^2 + {1 \over 2}
|F|^2) \over h_{\tau \sigma}} , \beta^2+{1 \over 2} |F|^2 , \\ &\left. -\beta {F_{\sigma 1}}  -{F_{12}} {F_{\sigma 2}}
-{F_{13}} {F_{\sigma 3}}-{F_{14}} {F_{\sigma 4}} , {F_{12}}
{F_{\sigma 1}}  -\beta {F_{\sigma 2}} -{F_{14}} {F_{\sigma
3}}+{F_{13}} {F_{\sigma 4}}, \right. \\ &\left. F_{13} {F_{\sigma
1}}+F_{14} F_{\sigma 2}-\beta F_{\sigma 3}-F_{12} F_{\sigma 4},
F_{14} {F_{\sigma 1}} -{F_{13}} {F_{\sigma 2}} +{F_{12}} {F_{\sigma
3}}-\beta F_{\sigma 4} \right\} \\ 
\sqrt{-|D|} D^{\alpha \tau} &=  \{ -{
\beta F_{\sigma i} F_{\sigma}^{i}+ h_{\sigma \sigma} (\beta^2 + {1
\over 2} |F|^2) \over h_{\tau \sigma}},
\beta^2+{1 \over 2} |F|^2, \\ &\left. \beta
{F_{\sigma 1}}-F_{12} {F_{\sigma 2}}-{F_{13}} {F_{\sigma 3}}-F_{14}
{F_{\sigma 4}}, F_{12} {F_{\sigma 1}}+\beta {F_{\sigma 2}}-F_{14}
{F_{\sigma 3}}+F_{13} {F_{\sigma 4}}, \right.  \\ &\left.F_{13}
{F_{\sigma 1}}+{F_{14}} {F_{\sigma 2}}+\beta {F_{\sigma 3}}-F_{12}
{F_{\sigma 4}}, F_{14} {F_{\sigma 1}}-F_{13} {F_{\sigma 2}}+F_{12}
{F_{\sigma 3}}+\beta {F_{\sigma 4}} \right\}
\end{split}
\end{equation}

\subsection{Vielbeins}
\label{allvielbeins}
In this subsection, we list our vielbein conventions for the backgrounds 
considered above. 

\subsubsection{D1-D5:}

The metric is given in Table \ref{donefive}. The Vielbein is defined by:
\begin{equation}
\label{allvielbeinsd1d5}
\begin{split}
e^{\hat{t}} &= (f_1 f_5)^{-{1 \over 4}} d t, ~~~ 
e^{\hat{5}} = (f_1 f_5)^{-{1 \over 4}} d x_5,~~~ e^{\hat{r}} = (f_1 f_5)^{1 \over 4} d r, \\
e^{\hat{\zeta}} &= (f_1 f_5)^{1 \over 4} r d \zeta,~~~ e^{\hat{\phi}_1} = (f_1 f_5)^{1 \over 4} r \cos \zeta,~~~e^{\hat{\phi}_2} = (f_1 f_5)^{1 \over 4} r \sin \zeta, ~~~e^{a} = {e^{\phi \over 2} \over \sqrt{g}} d z^{a} .
\end{split}
\end{equation}

\subsubsection{D1-D5-P:}
The metric is given in Equation \ref{donefivep}. The Vielbein is defined by:
\begin{equation}
\label{allvielbeinsd1d5p}
\begin{split}
e^{\hat{t}} &= (f_1 f_5)^{-1/4} \left( (1 - {r_p^2 \over r^2})^{1 \over 2} d t - {{r_p^2 \over r^2} \over \sqrt{1 - {r_p^2 \over r^2}}} d x_5 \right), ~~~ 
e^{\hat{5}} = (f_1 f_5)^{-1/4} (1 - {r_p^2 \over r^2})^{-1/2} d x_5,\\
e^{\hat{r}} &= (f_1 f_5)^{1 \over 4} d r, ~~~
e^{\hat{\zeta}} = (f_1 f_5)^{1 \over 4} r d \zeta,~~~ e^{\hat{\phi}_1} = (f_1 f_5)^{1 \over 4} r \cos \zeta,~~~e^{\hat{\phi}_2} = (f_1 f_5)^{1 \over 4} r \sin \zeta,\\
e^{a} &= {e^{\phi \over 2} \over \sqrt{g}} d z^{a} .
\end{split}
\end{equation}

\subsubsection{Lunin-Mathur:}

The metric is given by \eqref{luninmetric}. The Vielbein is defined by:
\begin{equation}
\label{allvielbeinslunin}
\begin{split}
e^{\hat{t}} &= \left({H \over 1 + K}\right)^{1 \over 4} \left( d t - A_{\hat{i}} d x^{\hat{i}} \right),~~~ e^{\hat{5}} = \left({H \over 1 + K}\right)^{1 \over 4} \left( d x_5 + B_{\hat{i}} d x^{\hat{i}} \right),\\
e^{{\hat{m}}} &= \left({H \over 1 + K}\right)^{-1/4} d x^{\hat{m}},~~~e^{{\hat{a}}} = \{H (1 + K)\}^{1 \over 4} dx^{{\hat{a}}} .
\end{split}
\end{equation}

\subsubsection{Global AdS:}

The metric is defined in Table \ref{global}. The Vielbein is defined by:
\begin{equation}
\begin{split}
e^{\hat{t}} &= l \cosh{\rho} d t, ~~~ e^{\hat{\theta}} = l \sinh{\rho} d \theta,~~~ \\
e^{\hat{\zeta}} &= l d \zeta,~~~ e^{\hat{\phi}_1} =  l \cos \zeta,~~~e^{\hat{\phi}_2} = l \sin \zeta, ~~~e^{\hat{a}} = \sqrt{Q_1 \over Q_5 v} d z^{a} .
\end{split}
\end{equation}
\section{Proof of the classical \label{sec:bound-proof} energy bound}

We will use the notation \begin{equation}\label{gennotation}
 \begin{split} \theta'& = w,
\phi_1'= w_1, \phi_2'= w_2, \\ x &= \sinh^2\rho, s = \sin^2\zeta, \\
{\cal A}^2 &= \rho'^2 + \zeta'^2 + X'^2 \end{split}
\end{equation} In general these
quantities depend on $\sigma$.

Note that \begin{equation}\label{genpzero}
 E = {Q_5\over 2 \pi} \int d\sigma f ,
\end{equation}
where \begin{equation}\label{fdef}
 \begin{split} f & \equiv { a\, sx + a_1 x + a_2 s + b
\over c_1 x + c_2 s + d} , \\ a &= w_2^2 - w_1^2 - w(w_2 - w_1) , \\
a_1 &= \asq + w^2 + w_1^2 - w w_1, a_2 = w_2^2 - w_1^2 ,\\ b &= \asq
+ w_1^2 , \\ c_1 &= w, c_2 = w_2 - w_1, d = w_1 .\end{split}
\end{equation}

The variables $(s,x), 0\le s \le 1, 0 \le x < \infty$ span the
rectangle ABCD, where \begin{equation}\label{defabcd}
 A= (s,x)= (0,0), B=(1,0),
C=(1,\infty), D=(0,\infty) .
\end{equation}

It is possible to prove that a function $f$ of the form \eqref{fdef}
attains its minimum (with respect to the variables $s,x$) at one of
the four vertices A,B,C or D. 

Hence the minimum value of $f$ is \begin{equation}\label{fmin}
 f_{min} = min\{f_A, f_B,
f_C, f_D\} .
\end{equation}

We will assume that the $w$'s ($w,w_1,w_2$) are non-negative
(consistent with supersymmetry as discussed in the previous
subsections). We will also assume that not all $w$'s are
simultaneously zero (so that the induced metric in
\eqref{induced-metric} is nonsingular); $\asq$ can be zero or
non-zero.

In the generic case when the $w$'s ($w,w_1,w_2$) as well as $\asq$
are non-vanishing, the values of $f$ at the four vertices are 
\begin{equation}\label{fabcd}
 \begin{split} f_A &= {b\over d}= w_1 + {\asq\over w_1} ,\\
f_B &= {a_2 + b\over c_2+ d}= w_2 + {\asq\over w_2} , \\ f_C &= {a+a_1
\over c_1} = {3 w\over 4} + {1\over w} [ \asq + (w_2 - {w\over
2})^2] , \\ f_D &= {a_1 \over c_1} = {3 w\over 4} + {1\over w} [ \asq
+ (w_1 - {w\over 2})^2] . \end{split}
\end{equation} Note that for $w,w_1,w_2,\asq$
all non-vanishing \begin{equation}\label{fabcdmin}
 \begin{split} f_A & \ge  w_1 , \\ f_B &
\ge  w_2 , \\ f_C & \ge  {3\over 2} w_2 ,\\ f_D & \ge  {3\over 2} w_1 .
\end{split}
\end{equation} The minimum value of $f_C$ is obtained for $w= 2 w_2$, and that of
$f_D$ is obtained for $w = w_1$.

In the above discussion we worked at a fixed $\sigma$. If
$w,w_1,w_2,\asq, s,x$ are independent of $\sigma$, the above bounds
\eqref{fabcdmin}, \eqref{fmin} for the function imply similar bounds
for $-P_t$ (see \eqref{genpzero}). Thus, suppose that the minimum
value of $f$ is $f_A$. In that case we get \begin{equation}\label{genbounda}
E \ge
w_1 Q_5
\end{equation} Now since $w_1 \equiv \phi_1'>0$ is independent of
$\sigma$, it has to be a positive integer, since 
\begin{equation}\label{periodic}
\int_0^{2\pi} d\sigma w_1 = \phi_1(\sigma = 2\pi) - \phi_1(\sigma=0)
= n_1 2\pi, n_1 \in Z_+
\end{equation} Note that we are considering all $w$'s to
be positive at the moment.

Thus \eqref{genbounda} is consistent with the bound \eqref{bound} we
found in the special cases.

The special cases in which some of the quantities $w,w_1,w_2,\asq$
vanishes can be understood as limits of \eqref{fabcd} or can be
dealt with separately. The conclusion about the bound remains the
same.

\noindent\underbar{Dependence on $\sigma$}

In the most general case, $w,w_1,w_2,\asq, s,x$ depend on $\sigma$.
It can be shown that even in this case the bound \eqref{bound} for
$-P_t$ is satisfied. As an example, suppose that the minimum value
of $f$ occurs at the point $A$ for some subset $I_1$ of $0 \le
\sigma < 2\pi$ and the minimum switches to $B$ in the remaining part
$I_2$ of $0 \le \sigma < 2\pi$. Thus 
\begin{equation}\label{sigmadependence}
 \begin{split}
{E \over Q_5/(2\pi)} &\ge \int_{I_1} d\sigma w_1 +   \int_{I_2}
d\sigma w_2 \\ &\ge \int_{I_1 + I_2}  d\sigma w_1 = 2\pi n_1
\end{split}
\end{equation}
since by hypothesis $w_2 > w_1$ in $I_2$. Here $n_1$ is the integer
winding number of the string around $\phi_1$.

\mysec{Summary}

We have proved in this section that the classical energy of an arbitrary
supersymmetric configuration satisfies the lower bound \eq{bound}.
The essential reason why the bound exists, as clear from the proof
above, is that supersymmetry allows only non-negative winding of the
string along $\phi_1, \phi_2, \theta$. Furthermore, we do not allow
all the winding numbers to be zero simultaneously (so that det $h$
remains non-zero).

\section{\label{sec:gauge}Gauge-invariant Noether charges}

In this section, we address the issue of the apparent
dependence of the Noether charges in Table \ref{global}
on the gauge choice of the two-form potential $B$.

Note that, like in case of the Dirac monopole potential
$A_i$ on $S^2$, the magnetic part of the two-form potential $B$
\[
\frac{B_{\rm mag}}{\alpha'} = -\frac12 Q_5(\cos 2\zeta + b)
d\phi_1 \wedge d\phi_2,~~ b = {\rm constant}, 
\]
cannot be globally defined with a fixed value of $b$ on $S^3$.
For $B$ to be non-singular, we must
have $b=-1$ in a neighbourhood of $\zeta=\pi/2$, and
$b=1$ in a neighbourhood of  $\zeta=0$.

In an overlap of such neighbourhoods, we have an
ambiguity in the choice of $b$, and we must ensure that
Noether charges and BPS relations are gauge-invariant.

We find below that the BPS relations are indeed
written in terms of gauge-invariant Noether charges
(obtained from the ``gauge-invariant momenta'' 
$\tilde P$ below) which are defined as
follows.
\bea
&&
E - L - J_1 - J_2 = - \int d\sigma [P_t +
P_\theta + \tilde P_{\phi_1} +  \tilde P_{\phi_2}]=0 ,
\nn
&&
\tilde P_{\phi_1}:= P_{\phi_1} + \frac{(b+1)Q_5}{4\pi}\phi_2'
=\{ P_{\phi_1} 
- \frac{1}{2\pi\alpha'}C^{(2)}_{\phi_1\phi_2}\phi_2'
\}
-  \frac{Q_5}{4\pi}[\cos 2\zeta -1]\phi_2' ,
\nn
&&
\tilde P_{\phi_2}:= P_{\phi_2} -\frac{(b-1)Q_5}{4\pi}\phi_1'
=\{P_{\phi_2} 
+ \frac{1}{2\pi\alpha'}C^{(2)}_{\phi_1\phi_2}\phi_1'\}
+ \frac{Q_5}{4\pi}[\cos 2\zeta +1]\phi_1' .
\label{gauge-inv-mom-p1-p2}
\eea
Here the expressions in 
$\{ \}$ are the so-called ``mechanical momenta''
(cf. $p_i - A_i$) which are also gauge-invariant, but
are different from the ones ($\tilde P$) entering the BPS relation.

\subsection{Derivation of the gauge-invariant ``momenta''
from a Bogomolnyi relation}

We will consider the Bogomolnyi bound for D1-branes in
global coordinates. Consider  
the following motion of the D1-brane (this is
sufficiently general for our purposes here)
\bea
&&
t=\tau, \theta=\tau, \zeta=const, \rho=const, \phi_{1,2}=w_{1,2} \sigma
+ \phi_{1,2}(\tau); 
\eea
We will show that the Bogomolnyi bound involves the ``gauge-invariant
momenta'' $\tilde P_{\phi_{1,1}}$, thus justifying their definition
which we introduced above.

We list below, for such motion, the Lagrangian, the canonical momenta
and the canonical Hamiltonian
\bea
&& 
c=\cos\zeta, s=\sin\zeta, v= w_2 \dot \phi_1 - w_1 \dot \phi_2,
 \lambda= Q_5/(2\pi),
\beta = c^2 w_1^2 + s^2 w_2^2, \Delta = \beta - c^2 s^2 v^2,
\nn
&&
L= -\lambda[ \sqrt{\Delta} + \frac12 v (\cos 2\zeta + b) ] ,
\nn
&&
P_{\phi_1}= \lambda w_2[ \frac{v c^2 s^2}{\sqrt{\Delta}} 
-\frac12(\cos 2\zeta + b)] ,
\nn
&&
P_{\phi_2}= \lambda w_1 [- \frac{v c^2 s^2}{\sqrt{\Delta}} 
+ \frac12(\cos 2\zeta + b)] ,
\nn
&&
H_{\rm can} = \frac{\lambda \beta}{\sqrt{\Delta}} 
\label{bps-long-eqn} .
\eea
It is easy to see that the momenta satisfy a constraint:
\be
w_1 P_{\phi_1} + w_2 P_{\phi_2}  =0 .
\label{constraint-p1-p2}
\ee
In the expression for the canonical Hamiltonian \eq{bps-long-eqn}, 
$\Delta$ depends on the velocity combination
$v$ which is to be expressed in terms of the (constrained)
momenta $P_{\phi_1}, P_{\phi_2}$. 

The gauge-invariant momenta, \eq{gauge-inv-mom-p1-p2}, are
given by
\bea
&&
\tilde P_{\phi_1}= P_{\phi_1} + \frac\lambda 2 w_2 (b+1)= 
\lambda w_2 s^2[ \frac{v c^2}{\sqrt{\Delta}} +1] ,
\nn
&&
\tilde P_{\phi_2}=  P_{\phi_2} - \frac\lambda 2 w_1 (b-1)=
\lambda w_1 c^2[- \frac{v s^2}{\sqrt{\Delta}} +1] .
\label{gauge-inv-p1-p2-special}
\eea
We now proceed with our analysis of the Bogomolnyi relation.

$\bullet$~~{Case: One of $w_1, w_2$ vanishes.}

We have written
the constraint equation \eq{constraint-p1-p2} 
for $w_1, w_2$ non-zero. The analysis
becomes significantly simpler when either of them vanishes. We
will Consider the case $w_1=0$. Eq. \eq{constraint-p1-p2}
becomes
\bea
&&
\tilde P_{\phi_2}=0 (= P_{\phi_2}) .
\eea
The expressions for the other momentum and the canonical Hamiltonian are
\bea
&&
\tilde P_{\phi_1}= \lambda s w_2 [ \frac{c^2 \dot\phi_1}{\sqrt{1 - c^2 
{\dot\phi_1}^2}} + s] ,
\nn
&&
H_{\rm can} = \lambda s w_2 \frac1{\sqrt{1 - c^2 {\dot\phi_1}^2}} .
\eea
Eliminating $\dot\phi_1$ between $\tilde P_{\phi_1},H_{\rm can}$ we get
\bea
&&
(\frac{H_{\rm can}}{\lambda w_2})^2 = s^2 + 
\frac1{c^2}(\frac{\tilde P_{\phi_1}}{\lambda w_2} - s^2)^2
\nn
&&
= 2\frac{\tilde P_{\phi_1}}{\lambda w_2} - 1 + 
\frac1{c^2} (\frac{\tilde P_{\phi_1}}{\lambda w_2} -1)^2
\label{w1-0-bogo}
\eea
In a sector with a given  gauge-invariant ``charge''
$\tilde P_{\phi_1}$ the minimum value of $H_{\rm can}$ is obtained
for
\bea
\frac{\tilde P_{\phi_1}}{\lambda w_2} = 1 ,
\eea
where 
\bea
H_{\rm can, BPS}= \lambda w_2 = \tilde P_{\phi_1} .
\eea
Note that it is the gauge-invariant $\tilde P_{\phi_1}$ 
that appears in the BPS relation, as promised.

The case $w_2=0$ can be similarly computed. Again it is the
gauge-invariant $\tilde P_{\phi_2}$  that appears in the BPS
relation.

$\bullet$ {Case: both $w_1, w_2$ non-zero.}

We define a canonical transformation
\bea
&& \phi = w_2 \phi_1 - w_1 \phi_2, \Phi = \frac{\phi_2 - \phi_1}{w_2 - w_1} ,
\nn
&& p= \frac{P_{\phi_1} + P_{\phi_2}}{w_2 - w_1}, 
P = w_1 P_{\phi_1} + w_2 P_{\phi_2} .
\label{can-p-P}
\eea
The constraint, encountered in \eq{constraint-p1-p2}, becomes
$P=0$. The gauge transformation generated by the constraint
can be fixed by putting $\Phi=0$. Note that although
$\Phi$ is not a periodic coordinate, the constraint  $\Phi=0$
is well-defined. We have assumed here $w_1 \ne w_2$; the case
$w_1 = w_2$ can be dealt with similarly by an 
appropriate canonical transformation.

We denote $v \equiv \dot \phi$. From  we get\eq{bps-long-eqn}
\bea
&& p = \lambda[ \frac{c^2 s^2 v}{\sqrt{\beta - c^2 s^2 v^2}} - 
\frac12 (b + \cos 2 \zeta)] ,
\nn
&& H =\lambda \frac\beta{\sqrt{\beta - c^2 s^2 v^2}} .
\eea
The Bogomolnyi bound must be saturated when $v= w_2 - w_1$ 
(which follows from $\dot\phi_1 = \dot\phi_2 =1$). When
we substitute this in the above equation, we get
\bea
&& H = (w_2 - w_1)p',
\nn
&& p':= p + \frac{\lambda}2[ b + \frac{w_2 + w_1}{w_2 - w_1}] .
\label{gen-gauge-inv-p}
\eea  
The top line is the BPS relation and the second line
defines the gauge-invariant momentum. The definition of $p'$ agrees
with the the gauge-invariant momenta \eq{gauge-inv-p1-p2-special}, in 
the sense that if we replace $P_{\phi_i}$ by
$\tilde P_{\phi_i}$ in the definition of $p$ in \eq{can-p-P},
we recover the expression for $p'$ as given above.

This proves the expression for the gauge-invariant
momenta \eq{gauge-inv-mom-p1-p2} for non-zero $w_1, w_2$.

\section{Killing Spinor Equations for D$1$-D$5$ Systems}
\label{appD1D5}
In this appendix, we write down and solve Killing spinor equations
for the naive D$1$-D$5$ system. The general Killing spinor
equations are shown in section \ref{appD1D5KSE}.
We then proceed to write down and solve the dilatino equation, both in the bulk
and in the near-horizon limit, in section \ref{appD1D5dilatino}.
The gravitino equation is similarly written down and solved in the bulk
and in the near-horizon limit in section \ref{appD1D5gravitino}.

We then go on to investigate what happens when we add
momentum to the D1-D5 system in section
\ref{appD1D5p}.
Finally, we write down and solve the dilatino Killing spinor equation
for the Lunin-Mathur geometries in section \ref{appLM}.

Although the results we derive below are quite well known we 
found it surprisingly difficult to find their explicit
derivations in the supergravity literature. So, we hope that these
explicit calculations will be useful for the reader. The reader may also
find the references \cite{Bergshoeff:1992cw,Bergshoeff:1994dg,Bergshoeff:1997kr,Bergshoeff:1999bx,Papadopoulos:2003jk,Lu:1998nu} useful.

\subsection{Killing Spinor Equations}\label{appD1D5KSE}
In this section, we write down the Killing spinor equations
for type IIB string theory. Consider a type IIB two-component spinor
\begin{equation}\label{spinor2comp}
  \epsilon=\left(
    \begin{array}{cc}
      \epsilon_1 \\
      \epsilon_2
    \end{array}
  \right),
\end{equation}
which satisfies $\Gamma^{11} \epsilon = - \epsilon$.
The dilatino Killing spinor equation is 
\begin{equation}\label{dilatinoequation}
  \left[
    \partial_M \phi \Gamma^M + \frac{1}{12} H_{MAB}\Gamma^{MAB} \otimes \sigma_3
    +\frac{1}{4} e^\phi
    \sum_{n=1}^{5} \frac{(-1)^{n-1}(n-3)}{(2n-1)!}
    G_{A_1 \ldots A_{2n-1}} \Gamma^{A_1 \ldots A_{2n-1}} \otimes \lambda_n
  \right] \epsilon = 0,
\end{equation}
where $\lambda_n = \sigma_1$ for $n$ even, and
$\lambda_n = i\sigma_2$ for $n$ odd. The $\{\sigma_i\}$, $i=1,2,3$ are
the Pauli matrices.
$H$ and $G$ are the NS-NS and R-R field strengths, and $\phi$
denotes the dilaton. 
Similarly, the gravitino Killing spinor equation is
\begin{equation}\label{gravitinoequation}
  \left[
    \partial_M + \frac{1}{4} w_M^{BC} \Gamma_{BC}
    +\frac{1}{8} H_{MAB}\Gamma^{AB} \otimes \sigma_3
    + \frac{1}{16} e^\phi
      \sum_{n=1}^{5} \frac{(-1)^{n-1}}{(2n-1)!}
      G_{A_1 \ldots A_{2n-1}} \Gamma^{A_1 \ldots A_{2n-1}}
      \Gamma_M \otimes \lambda_n
  \right] \epsilon = 0.
\end{equation}
Throughout this appendix, we will find it useful to divide
the coordinates $M=0,\ldots,9$ into $\mu=0,5$ ($x^0$ and $x_5$), $m=1,2,3,4$ ($r,\zeta,\phi_1,\phi_2$)
and $a=6,7,8,9$ (the torus directions). We also need to define
$r^2 \equiv x_m x^m$.

\subsection{The D1-D5 system}
\subsubsection{The Dilatino Equation}\label{appD1D5dilatino}
In this section, we write down the dilatino equation 
for the D1-D5 system, and solve it to find the corresponding
projection conditions. We do this both in the bulk and in the
near-horizon limit.
This system is defined by
the following metric, dilaton and R-R background\footnote{
Note that, to avoid cluttering the notation, we are here using a form of the
metric and the dilaton
in which constant factors have been scaled away.
This convention
is used throughout appendices \ref{appD1D5} and \ref{appglobal}.
}:
\begin{equation}\label{D1D5background}
\begin{split}
  ds^2 & = \frac{1}{\sqrt{f_1 f_5}} dx_\mu dx^\mu + \sqrt{f_1 f_5} dx_m dx^m + \sqrt{\frac{f_1}{f_5}} dx_a dx^a, \\
  e^{2\phi} & = \frac{f_1}{f_5}, \\
  G^{(3)} & = - \frac{f_1'}{f_1^2} dr \wedge dx^0 \wedge d x_5 + Q_5 \sin(2 \zeta) d\zeta \wedge d \phi_1 \wedge d \phi_2, \\
  G^{(7)} & = Q_1 \sin (2 \zeta) d \zeta \wedge d \phi_1 \wedge d \phi_2 \wedge \prod d \Sigma_a
              - f_5' f_5^{-2} dr \wedge d x^0 \wedge d x_5 \wedge \prod d \Sigma_a,
\end{split}
\end{equation}
where we used the definitions
\begin{equation}\label{f1f5}
\begin{split}
  f_1 & = 1 + \frac{Q_1}{r^2}, \\
  f_5 & = 1 + \frac{Q_5}{r^2}.
\end{split}
\end{equation}
Hence, the dilatino equation (\ref{dilatinoequation}) in the bulk is
\begin{equation}\label{D1D5dilatino}
\begin{split}
  \left[
  f_1^{-5/4} f_5^{-1/4} f_1' \Gamma^{\hat{r}} \left(
    \one - \Gamma^{\hat{0}\hat{5}} \otimes \sigma_1
  \right)
  + f_5^{-5/4}f_1^{-1/4}f_5' \Gamma^{\hat{r}} \left(
    - \one - \Gamma^{\hat{r}\hat{\zeta}\hat{\phi_1}\hat{\phi_2}} \otimes \sigma_1
  \right)
  \right] \epsilon = 0 \\
  \Rightarrow
  \Gamma^{\hat{6}\hat{7}\hat{8}\hat{9}} \epsilon = + \epsilon,
  \Gamma^{\hat{0}\hat{5}} \otimes \sigma_1 \epsilon = \epsilon.
\end{split}
\end{equation}
Note that we find the expected projection conditions
\eq{dosusy} and \eq{dfsusy} for this background.

We now want to investigate what happens to the dilatino equation
(\ref{D1D5dilatino}) in the near-horizon limit $r \rightarrow 0$.
In this limit, the equations (\ref{f1f5}) become
\begin{equation}
\begin{split}
  f_1 & \rightarrow \frac{Q_1}{r^2}, \\
  f_5 & \rightarrow \frac{Q_5}{r^2}.
\end{split}
\end{equation}
Consequently, some terms in (\ref{D1D5dilatino}) cancel,
and we are left with
\begin{equation}\label{D1D5dilatinoNHG}
  \left[
    \Gamma^{\hat{r}\hat{0}\hat{5}} + \Gamma^{\hat{\zeta}\hat{\phi_1}\hat{\phi_2}}
  \right] \otimes \sigma_1 \epsilon = 0
  \Rightarrow
  \Gamma^{\hat{6}\hat{7}\hat{8}\hat{9}} \epsilon = + \epsilon,
\end{equation}
which is the dilatino equation in the near-horizon limit.
Note that one of the projection conditions have dropped out,
i.e. we get the expected doubling of supersymmetries
in the near-horizon geometry.
\subsubsection{The Gravitino Equation}\label{appD1D5gravitino}
In this section, we find and solve the gravitino equation
for the background (\ref{D1D5background}). We will again
do this both for the bulk and in the near-horizon limit,
beginning with the bulk.
First, note that we can define the vielbeins
\begin{equation}
\begin{split}\label{vielbeins}
  e^{\hat{\mu} } & = (f_1 f_5)^{-1/4} dx^{\mu}, \\
  e^{\hat{m}} & = (f_1 f_5)^{+1/4} dx^{m}, \\
  e^{\hat{a}} & = \left(\frac{f_1}{f_5}\right)^{+1/4} dx^{a}.
\end{split}
\end{equation}
Thus, the corresponding spin connections are
\begin{equation}\label{D1D5spinconnections}
\begin{split}
  w^{\hat{\mu} \hat{n}} & = - \frac{1}{4r} (f_1 f_5)^{-3/2} (f_1 f_5)' \left[ x^n dx^{\mu} \right] , \\
  w^{\hat{m} \hat{n}} & = + \frac{1}{4r} (f_1 f_5)^{-1} (f_1 f_5)' \left[ x^{n} dx^{m} - x^{m} dx^{n} \right] , \\
  w^{\hat{a} \hat{n}} & = + \frac{1}{4r} f_1^{-1} f_5^{+1/2} \left( \frac{f_1}{f_5} \right)' \left[ x^{n} dx^{a} \right].
\end{split}
\end{equation}
To simplify the notation, we have defined
$r^2 = x_m x^m$.
Using these spin connections and the background (\ref{D1D5background})
in the gravitino equation (\ref{gravitinoequation}), we get
\begin{equation}\label{D1D5gravitino}
\begin{split}
  \left[ \frac{D}{Dx^{M}} + \frac{1}{8} f_1^{1/2}f_5^{-1/2} \left[ f_1' f_1^{-7/4} f_5^{+1/4} \Gamma^{\hat{r}\hat{0}\hat{5}}
  + (f_1 f_5)^{-3/4} f_5' \Gamma^{\hat{\zeta}\hat{\phi_1}\hat{\phi_2}}
  \right]
  \Gamma_{M} \otimes \sigma_1 \right] \epsilon & = 0 \\
  \Rightarrow
  \partial_M \epsilon & = 0,
\end{split}
\end{equation}
which is the gravitino equation in the bulk.
Note that $\Gamma^{\hat{r}}=\Gamma^{\hat{m}}\frac{x^m}{r}$.
Due to cancellation of terms,
we conclude that the Killing spinor $\epsilon$ is just a constant
expressed in terms of vielbeins corresponding to Cartesian coordinates.

We now want to investigate what happens to the bulk gravitino equation
(\ref{D1D5gravitino}) in the near-horizon limit $r \rightarrow 0$.
In this limit, the spin connections (\ref{D1D5spinconnections}) become:
\begin{equation}
\begin{split}
  w^{\hat{\mu} \hat{n}} & \rightarrow (Q_1 Q_5)^{-1/2} \left[ x^{n} dx^{\mu} \right], \\
  w^{\hat{m} \hat{n}} & \rightarrow \frac{1}{r^2} \left[ x^{m} dx^{n} - x^{n} dx^{m} \right], \\
  w^{\hat{a} \hat{n}} & \rightarrow 0.
\end{split}
\end{equation}
Inserting these spin connections and the background (\ref{D1D5background})
into equation (\ref{gravitinoequation}),
we now find the gravitino equation in the near-horizon limit,
\begin{equation}\label{D1D5gravitinoNHG}
\begin{split}
  \left[
    \frac{D}{Dx^M} - \frac{1}{4} (Q_1 Q_5)^{-1/4} \left[
      \Gamma^{\hat{r}\hat{0}\hat{5}} + \Gamma^{\hat{\zeta}\hat{\phi_1}\hat{\phi_2}}
    \right] \Gamma_M \otimes \sigma_1
  \right] \epsilon & = 0 \\
  \Rightarrow
  \partial_a \epsilon & = 0.
\end{split}
\end{equation}
As a consistency check, the gravitino equation (\ref{D1D5gravitinoNHG})
is indeed equivalent to Mikhailov's equation (\ref{Mikhailov}).
For the torus coordinates, there is again a cancellation
making the spinor $\epsilon$ constant in those directions.

In the near-horizon limit, where we find an $AdS_3 \times S^3$ structure
it is convenient to move to `polar' coordinates: $(r, x^0, x_5)$ for the $AdS_3$
and $(\zeta, \phi_1, \phi_2)$ for the $S^3$. In this basis, the killing spinors
do have a non-trivial dependence on these 6 coordinates.

We therefore want to find this dependence by solving
the gravitino equation
(\ref{D1D5gravitinoNHG}). We proceed as follows.
The $S^3$ and $(r,x^0,x_5)$ parts can be analyzed separately.
In fact, the $S^3$ part is identical to (\ref{polargravitinoequationss3})
in section \ref{appglobalsolving}, and can be solved as detailed
in that section. The final solution to that part is
given by equation (\ref{solutionpolargravitinoequationss3}).

The $(r,x^0,x_5)$ part is
\begin{equation}\label{cartesiangravitinoequationsrtt}
\begin{split}
  \left[ \frac{\partial}{\partial r} \mp \frac{1}{2r} \Gamma^{\hat{0}\hat{5}} \right] \epsilon & = 0, \\
  \left[ \frac{\partial}{\partial x^0} + r D \right] \epsilon & = 0, \\
  \left[ \frac{\partial}{\partial x_5} \pm rD \right] \epsilon & = 0,  
\end{split}
\end{equation}
where we defined
\begin{equation}
  D = \frac{1}{2} (Q_1 Q_5)^{-1/2} \left( \Gamma^{\hat{0}\hat{r}} \pm \Gamma^{\hat{5}\hat{r}} \right) .
\end{equation}
The split signs correspond to
eigenvalues of $\sigma_1$, i.e.
$\sigma_1 \epsilon = \pm \epsilon$.
As a consistency check, it can be verified that these three operators commute.
They can be solved analogously to the $S^3$ part, using the relation
\begin{equation}
  \exp \left[ \mp \frac{\delta}{2} \Gamma^{\hat{0} \hat{5}} \right]
  r \left( \Gamma^{\hat{0}} \pm \Gamma^{\hat{5}}\right)
  \exp \left[ \pm \frac{\delta}{2} \Gamma^{\hat{0} \hat{5}} \right]
  = r e^{-\delta} \left( \Gamma^{\hat{0}} \pm \Gamma^{\hat{5}}\right),
\end{equation}
to move factors of the type
$\exp \left[ \mp \frac{\delta}{2} \Gamma^{\hat{0} \hat{5}} \right]$
through the $D$.
The solution to (\ref{cartesiangravitinoequationsrtt}) is
\begin{equation}\label{rx0x5solution}
\begin{split}
  \epsilon(r,x^0,x_5) &= M_1(r,x^0,x_5) \epsilon_0, \;
\\M_1(r,x^0,x_5) 
&\equiv 
  \exp \left[ \pm \frac{1}{2} \ln (r) \Gamma^{\hat{0}\hat{5}} \right]
  \exp \left[ -\frac{1}{2} (Q_1 Q_5)^{-1/2} (x^0 \pm x_5) (\Gamma^{\hat{0}\hat{r}} \pm \Gamma^{\hat{5}\hat{r}}) \right],
\end{split}
\end{equation}
where $\epsilon_0$ is a constant spinor on $(r,x^0,x_5)$.
Hence, the full solution to (\ref{D1D5gravitinoNHG}) is obtained
by combining the $S^3$ solution (\ref{solutionpolargravitinoequationss3})
with the $(r,x^0,x_5)$ solution (\ref{rx0x5solution}), i.e.
\begin{equation}\label{D1D5gravitinoNHGsolution}
\begin{split}
  \epsilon(r,x^0,x_5,\zeta,\phi_1,\phi_2) & = M_1(r,x^0,x_5) 
M_2(\zeta,\phi_1,\phi_2) \epsilon_0 \\
  M_2(\zeta,\phi_1,\phi_2) & = \exp \left[ \pm \frac{1}{2} \zeta \Gamma^{\hat{\phi_1}\hat{\phi_2}} \right]
  \exp \left[ +\frac{1}{2} (\phi_2 \mp \phi_1) \Gamma^{\hat{\zeta}\hat{\phi_2}} \right],
\end{split}
\end{equation}
where $\epsilon_0$ is a constant spinor on $r,x^0,x_5,S^3$
and $M(r,x^0,x_5) $ is already defined in \eq{rx0x5solution}.
Note that half of the spinors which satisfy (\ref{D1D5dilatino}) are still
constants in vielbeins corresponding to Cartesian coordinates,
as in (\ref{D1D5gravitino}). The reason is that the projection
$\Gamma^{\hat{0}\hat{5}} \otimes \sigma_1 \epsilon = + \epsilon$
makes the $x^0, x_5$ dependence drop out.

\subsection{The D1-D5-P system}\label{appD1D5p}
In this section, we investigate the D1-D5-P system,
i.e. the D1-D5 system with momentum $p$ added.
The background is in fact almost identical to (\ref{D1D5background}),
only the metric is changed to \eq{donefivep} which
can be rewritten as 
\begin{equation}\label{D1D5pmetric}
  ds^2 = -(A dt+ B dx_5)^2 + C^2 dx_5^2 + \sqrt{f_1 f_5} dx_m dx^m + \sqrt{\frac{f_1}{f_5}} dx_a dx^a, \\
\end{equation}
where we defined
\begin{equation}
\begin{split}
  A&= (f_1 f_5)^{-1/4}\sqrt{1 - K}, \\
  B&= (f_1 f_5)^{-1/4}K/\sqrt{1 - K}, \\
  C&= (f_1 f_5)^{-1/4}/\sqrt{1 - K},
\end{split}
\end{equation}
where $K=r_p^2/r^2$.
As can be verified using (\ref{D1D5background}) and (\ref{D1D5pmetric})
in (\ref{dilatinoequation}), the dilatino equation remains on
the form (\ref{D1D5dilatino}).

To obtain the gravitino equation, we again use vielbeins
(\ref{vielbeins}) and corresponding spin connections
(\ref{D1D5spinconnections}). However, in the $t$ and $x_5$ 
directions, we instead use
\begin{equation}\label{vielbeinsD1D5p}
\begin{split}
  e^{\c t}&= A dt + B dx_5, \\
  e^{\c x} &= C dx_5, \\
\end{split}
\end{equation}
The relevant new spin connections are
\begin{equation}\label{D1D5psc}
\begin{split}
  \omega^{\c x \c m} & = \frac{x^m}{r} 
(f_1 f_5)^{-1/4}
  [- {\cal A}  e^{\c t} + C'/C\  e^{\c x}]=\frac{x^m}{r} 
(f_1 f_5)^{-1/4}
  [- {\cal A} (A dt + B dx_5) + C' dx_5], \\
  \omega^{\c x \c t}&= (f_1 f_5)^{-1/4}{\cal A} \frac{x^m}{r}  e^{\c m}
  = {\cal A} dr, \\
  \omega^{\c t \c m}&=\frac{x^m}{r}  
(f_1 f_5)^{-1/4}[A'/A\  e^{\c t}+ {\cal A} e^{\c x}]
  = \frac{x^m}{r} (f_1 f_5)^{-1/4}[A'/A\ (A dt + B dx_5)+ 
  {\cal A} C dx_5],
\end{split}
\end{equation}
where we defined
\begin{equation}
  {\cal A} = \half \frac{A B'- A'B}{AC}= \frac{K'}{1- K}.
\end{equation}
Using (\ref{D1D5background}), (\ref{D1D5spinconnections}), (\ref{D1D5pmetric})
and (\ref{D1D5psc}),
we can now obtain the $M=x_5$ component of the gravitino equation
(\ref{gravitinoequation}), which is
$\hat{D_5} \epsilon=0$,
where 
\bea
&\hat{D_5} = \del_5 - \frac18 \frac{(f_1 f_5)^{-1/2}}{\sqrt{1 - K}}
\bigg[[ \frac{(f_1 f_5)'}{f_1 f_5} - 2 K']\Gamma^{\c x \c r}+ 
[2K' - K \frac{(f_1 f_5)'}{f_1 f_5}]\Gamma^{\c t \c r}
- \frac{(f_1 f_5)'}{f_1 f_5}(\Gamma^{\c r \c t}- K 
\Gamma^{\c r \c x}) \otimes \sigma_1 \bigg]
\nn
& = \del_5 - \frac18 \frac{(f_1 f_5)^{-1/2}}{\sqrt{1 - K}}
\bigg[\frac{(f_1 f_5)'}{f_1 f_5}\Gamma^{\c x \c r}
\big(\one - \Gamma^{\c0\c5}\otimes \sigma_1 \big)
- K\frac{(f_1 f_5)'}{f_1 f_5} \Gamma^{\c t \c r}
\big(\one - \Gamma^{\c0\c5}\otimes \sigma_1 \big) 
- 2K' \Gamma^{\c x \c r}\big(\one + \Gamma^{\c0\c5}\big)\bigg]
\nn
\eea 
Hence, using the additional constraint
\begin{equation}\label{D1D5pconstraint}
\Gamma^{\c0\c5}\epsilon= - \epsilon
\end{equation}
in addition to the ones obtained in equation (\ref{D1D5dilatino}),
we find that
\[
\del_5 \epsilon=0.
\]
Similarly, using (\ref{D1D5dilatino}) and (\ref{D1D5pconstraint}), the $M=t$ component of the gravitino equation becomes
\[
\del_0 \epsilon=0.
\]
However, the $M=m$ ($m=1,2,3,4$) component leads to
\bea
& [\del_r -\frac{1}{4} \frac{K'}{(1 - K)}\Gamma^{\c 0 \c 5}]\epsilon=
[\del_r + \frac{1}{4} \frac{K'}{(1 - K)}]\epsilon=0,
\nn
\eea
using (\ref{D1D5pconstraint}), so the 
spinor must have a non-trivial dependence on $r$, 
\bea\label{D1D5pspinor}
\epsilon(r)= (1- K)^{-1/4}\epsilon_0.
\eea
For the remaining components, the gravitino equation
just reduces to $\partial_M \epsilon = 0$.
Hence, the full Killing spinor is just
given by (\ref{D1D5pspinor}) with $\epsilon_0$ a constant
with respect to the vielbeins (\ref{vielbeins})
(except the $x^0$ and $x_5$ directions)
and (\ref{vielbeinsD1D5p}) (the $t$ and $x$ directions).

\subsection{Lunin-Mathur Geometries}\label{appLM}
In this section, we write down and solve the dilatino Killing spinor
equation for the Lunin-Mathur geometries. This serves to verify that these geometries do preserve the same supersymmetries as the naive D1-D5 geometry. 

Using the metric in \eqref{luninmetric}, we may first define a vielbein in terms of the following orthonormal 1-forms:
\begin{equation}
\label{vielbeinlunin}
\begin{split}
 e^{\hat{t}} &= \left({H \over 1 + K}\right)^{1 \over 4} \left( d t - A_{\hat{n}} d x^{\hat{n}} \right) ,\\
e^{\hat{5}} &= \left({H \over 1 + K}\right)^{1 \over 4} \left( d x_5 + B_{\hat{n}} d x^{\hat{n}} \right) ,\\
e^{{\hat{m}}} &= \left({H \over 1 + K}\right)^{-1/4} d x^{\hat{m}} ,\\
e^{{\hat{a}}} &= \{H (1 + K)\}^{1 \over 4} dx^{{\hat{a}}} .
\end{split}
\end{equation}
The field strength may now be computed in terms of the RR 2-forms given in \eqref{luninmetric} and we find:
\begin{equation}
\label{threeformlm}
\begin{split}
G^{(3)} &= d C^{(2)} = \left({H \over 1 + K}\right)^{3\over 4} \epsilon^{\hat{n}}_{\hat{m} \hat{l} \hat{p}} \partial_n H^{-1} e^{\hat{m}} \wedge e^{\hat{l}} \wedge e^{\hat{p}} \\
&+ \left({1 + K \over H}\right)^{1 \over 4} \partial_n {1 \over (1 + K)} e^{\hat{t}} \wedge e^{\hat{5}} \wedge e^{\hat{n}} + {1 \over 1 + K} \left(-\partial_m B_n e^{\hat{5}} - \partial_m A_n e^{\hat{t}}\right) \wedge e^{\hat{n}} \wedge e^{\hat{m}} .
\end{split}
\end{equation}
Notice, that under Poincare duality:
\begin{equation}
\begin{split}
&G^{(7)} = *_{10} G^{(3)} \\
&= \left[ \left({H \over 1 + K}\right)^{3\over 4} \epsilon^{\hat{n}}_{\hat{m} \hat{l} \hat{p}} \partial_n K e^{\hat{m}} \wedge e^{\hat{l}} \wedge e^{\hat{p}}  +  \left({1 + K \over H}\right)^{1 \over 4} \partial_n H e^{\hat{t}} \wedge e^{\hat{5}} \wedge e^{\hat{n}} \right. \\
&\left. + H \left(-\partial_m B_n e^{\hat{5}} - \partial_m A_n e^{\hat{t}}\right) \wedge e^{\hat{n}} \wedge e^{\hat{m}}\right] \wedge d x^6 \wedge d x^7 \wedge d x^8 \wedge d x^9 .
\end{split}
\end{equation}
This justifies the result \eqref{cprimeluninmathur} for $C'^{(2)}$ (recall by definition $G^{(7)} = d C'^{(2)} \wedge d x^6 \wedge \ldots \wedge d x^9$) which we claimed was obtained by interchanging $H \leftrightarrow {1 \over 1 + K}$ in $C^{(2)}$.

Now, substituting the result \eqref{threeformlm} into the dilatino equation  \eqref{dilatinoequation} we find that it becomes
\begin{equation}
\begin{split}
  & H^{1/4}(1+K)^{-5/4} (\partial_m K) \left[ \one - \Gamma^{\hat{0}\hat{5}} \otimes \sigma_1 \right] \epsilon \\
  & + H^{5/4}(1+K)^{-1/4} \left[ -(\partial_m H^{-1})\Gamma^{\hat{m}} - (*_4 dH^{-1})_{mnp} \Gamma^{\hat{m}\hat{n}\hat{p}} \otimes \sigma_1 \right] \epsilon \\
  & + H^{3/4}(1+K)^{-3/4} \Gamma^{\hat{0}} \left[ \Gamma^{\hat{m}\hat{n}} (\partial_m B_n) - \Gamma^{\hat{m}\hat{n}\hat{0}\hat{5}} (\partial_m A_n) \otimes \sigma_1 \right] \epsilon = 0 ,
\end{split}
\end{equation}
which is satisfied when in addition, to the properties of the metric \eqref{luninmetric}, we use the projections:
\begin{equation}
\begin{split}
  \Gamma^{\hat{6}\hat{7}\hat{8}\hat{9}} \epsilon & = \epsilon, \\
  \Gamma^{\hat{0}\hat{5}} \otimes \sigma_1 \epsilon & = \epsilon.
\end{split}
\end{equation}
This confirms the result we quoted in Section \ref{Killingspinor}

\section{Killing Spinor Equations in Global $AdS$}\label{appglobal}
In this appendix, we analyze the Killing spinor equations
in global $AdS$. In sections \ref{appglobaldilatino}
and \ref{appglobalgravitino}, we write down the
dilatino and gravitino equations, respectively.
We proceed to solving the gravitino equation in section
\ref{appglobalsolving}.
Finally, we compare the results to those of Mikhailov
in section \ref{appglobalMikhailov}.

  \subsection{The Dilatino Equation}\label{appglobaldilatino}
In this section, we will find the dilatino equation
and the corresponding projection conditions in global $AdS$.
The background we are working with is defined by
the following metric, dilaton and the 3-form R-R field strength:
\begin{equation}\label{backgroundAdS}
\begin{split}
  ds^2 & = (Q_1 Q_5)^{1/2} \left[
    -\cosh^2(\rho) dt^2 + \sinh^2(\rho) d\theta^2 + d\rho^2 \right] \\
    & + (Q_1 Q_5)^{1/2} \left[ d \zeta^2 + \cos^2(\zeta) d\phi_1^2 + \sin^2(\zeta) d\phi_2^2
    \right] + ds_{T^4}^2 \\
  e^{2 \phi} & = \frac{Q_1}{Q_5}, \\
  G^{(3)} & = +Q_5 \sinh{\rho} dt \wedge d \theta \wedge d \rho
              +Q_5 \sin(2 \zeta) d \zeta \wedge d\phi_1 \wedge d\phi_2.
\end{split}
\end{equation}
Using this in equation (\ref{dilatinoequation}),
we find the dilatino equation
\begin{equation}\label{AdSdilatino}
  \left[
    \Gamma^{\hat{\rho}\hat{t}\hat{\theta}} + \Gamma^{\hat{\zeta}\hat{\phi_1}\hat{\phi_2}}
  \right] \otimes \sigma_1 \epsilon = 0
  \Rightarrow
  \Gamma^{\hat{6}\hat{7}\hat{8}\hat{9}} \epsilon = \epsilon.
\end{equation}
Note that it implies the usual torus projection.

  \subsection{The Gravitino Equation}\label{appglobalgravitino}
In this section, we want to find the gravitino equation in
the global $AdS$ background (\ref{backgroundAdS}).
We begin by defining the vielbeins
\begin{equation}
\begin{split}
  e^{\hat{t}} & = (Q_1 Q_5)^{+1/4} \cosh(\rho) dt, \\
  e^{\hat{\theta}} & = (Q_1 Q_5)^{+1/4} \sinh(\rho) d\theta, \\
  e^{\hat{\rho}} & = (Q_1 Q_5)^{+1/4} d\rho, \\
  e^{\hat{\zeta}} & = (Q_1 Q_5)^{+1/4} d\zeta, \\
  e^{\hat{\phi_1}} & = (Q_1 Q_5)^{+1/4} \cos(\zeta) d\phi_1, \\
  e^{\hat{\phi_2}} & = (Q_1 Q_5)^{+1/4} \sin(\zeta) d\phi_2.
\end{split}
\end{equation}
Thus, the corresponding non-vanishing spin connections are
\begin{equation}\label{sc}
\begin{split}
  w^{\hat{t} \hat{\rho}} & = \sinh(\rho) dt, \\
  w^{\hat{\theta} \hat{\rho}} & = \cosh(\rho) d\theta, \\
  w^{\hat{\phi_1} \hat{\zeta}} & = -\sin(\zeta) d\phi_1, \\
  w^{\hat{\phi_2} \hat{\zeta}} & = \cos(\zeta) d\phi_2.
\end{split}
\end{equation}
Using the background (\ref{backgroundAdS}) and the spin connections
(\ref{sc}) in equation (\ref{gravitinoequation}),
we find the gravitino equation
\begin{equation}\label{AdSgravitino}
  \left[
    \frac{D}{Dx^M} - \frac{1}{4} (Q_1 Q_5)^{-1/4}
      \left[ \Gamma^{\hat{\rho}\hat{t}\hat{\theta}} + \Gamma^{\hat{\zeta}\hat{\phi_1}\hat{\phi_2}}
      \right] \Gamma_M \otimes \sigma_1
  \right] \epsilon = 0.
\end{equation}
As a consistency check, the gravitino equation (\ref{AdSgravitino})
is indeed equivalent to Mikhailov's equation (\ref{Mikhailov}).
We show how to solve the gravitino equation
(\ref{AdSgravitino})
in section \ref{appglobalsolving}.

\subsection{Solving the Gravitino Equation}\label{appglobalsolving}
In this section, we show how to solve
the gravitino equation (\ref{AdSgravitino}).
We proceed as follows.
In fact, the $S^3$ and $AdS_3$ parts split,
and can be analyzed separately.
We begin with the $S^3$ part, which is
\begin{equation}
\begin{split}\label{polargravitinoequationss3}
  \left[ \frac{\partial}{\partial \zeta} \mp \frac{1}{2} \Gamma^{\hat{\phi_1}\hat{\phi_2}} \right] \epsilon & = 0, \\
  \left[ \frac{\partial}{\partial \phi_1} + A \right] \epsilon & = 0, \\
  \left[ \frac{\partial}{\partial \phi_2} \mp A \right] \epsilon & = 0,
\end{split}
\end{equation}
where we defined
\begin{equation}
  A \equiv \frac{1}{2} \sin(\zeta) \Gamma^{\hat{\zeta} \hat{\phi_1}} \pm \frac{1}{2} \cos(\zeta) \Gamma^{\hat{\zeta}\hat{\phi_2}}.
\end{equation}
As a consistency check, it can be verified that
these three operators commute. The split signs correspond to
eigenvalues of $\sigma_1$, i.e.
$\sigma_1 \epsilon = \pm \epsilon$.
The first equation of (\ref{polargravitinoequationss3}) implies that
\begin{equation}
  \epsilon(\zeta,\phi_1, \phi_2) = \exp\left[\pm \frac{1}{2} \zeta \Gamma^{\hat{\phi_1}\hat{\phi_2}} \right] \Psi(\phi_1,\phi_2).
\end{equation}
The second equation of (\ref{polargravitinoequationss3}) then implies that
\begin{equation}
  \Psi(\phi_1, \phi_2) = \exp \left[ \mp \frac{1}{2} \phi_1 \Gamma^{\hat{\zeta}\hat{\phi_2}}  \right] \chi(\phi_2),
\end{equation}
where we used the relation
\begin{equation}
  \exp \left[ \pm \frac{1}{2} \zeta \Gamma^{\hat{\phi_2}\hat{\phi_1}} \right]
  \left(
    \cos(\zeta) \Gamma^{\hat{\zeta}\hat{\phi_2}} \pm \sin(\zeta) \Gamma^{\hat{\zeta}\hat{\phi_1}}
  \right)
  \exp \left[ \mp \frac{1}{2} \zeta \Gamma^{\hat{\phi_2}\hat{\phi_1}} \right]
  = \Gamma^{\hat{\zeta}\hat{\phi_2}}
\end{equation}
to move the factor $\exp \left[ \pm \frac{1}{2} \zeta \Gamma^{\hat{\phi_1}\hat{\phi_2}} \right]$
through the $A$.
The third equation of (\ref{polargravitinoequationss3}) similarly implies that
\begin{equation}
  \chi(\phi_2) = \exp \left[ +\frac{1}{2} \phi_2 \Gamma^{\hat{\zeta}\hat{\phi_2}}\epsilon_0 \right],
\end{equation}
where $\epsilon_0$ is a constant spinor on $S^3$.
So the solution to (\ref{polargravitinoequationss3}) is
\begin{equation}\label{solutionpolargravitinoequationss3}
  \epsilon(\zeta,\phi_1,\phi_2) =
  \exp \left[ \pm \frac{1}{2} \zeta \Gamma^{\hat{\phi_1}\hat{\phi_2}} \right]
  \exp \left[ +\frac{1}{2} (\phi_2 \mp \phi_1) \Gamma^{\hat{\zeta}\hat{\phi_2}} \right]
  \epsilon_0.
\end{equation}
We now proceed to the $AdS_3$ part, which is
\begin{equation}
\begin{split}\label{polargravitinoequationsads3}
  \left[ \frac{\partial}{\partial \rho} \mp \frac{1}{2} \Gamma^{\hat{t}\hat{\theta}} \right] \tilde{\epsilon} & = 0, \\
  \left[ \frac{\partial}{\partial t} + B \right] \tilde{\epsilon} & = 0, \\
  \left[ \frac{\partial}{\partial \theta} \mp B \right] \tilde{\epsilon} & = 0,
\end{split}
\end{equation}
where we defined
\begin{equation}
  B \equiv \frac{1}{2} \sinh(\rho) \Gamma^{\hat{\rho} \hat{t}} \pm \frac{1}{2} \cosh(\rho) \Gamma^{\hat{\rho}\hat{\theta}}.
\end{equation}
Again, we can verify that these three operators commute.
We now make the change of variables defined by
\begin{equation}
\begin{split}
  \rho & = i \zeta, \\
  t & = i \phi_1, \\
  \theta & = \phi_2,
\end{split}
\end{equation}
which turn (\ref{polargravitinoequationsads3}) into
\begin{equation}
\begin{split}
  \left[ \frac{\partial}{\partial \zeta} \pm \frac{1}{2} \Gamma^{\hat{\phi_1}\hat{\phi_2}} \right] \tilde{\epsilon} & = 0, \\
  \left[ \frac{\partial}{\partial \phi_1} + C \right] \tilde{\epsilon} & = 0, \\
  \left[ i \frac{\partial}{\partial \phi_2} \mp C \right] \tilde{\epsilon} & = 0,
\end{split}
\end{equation}
where we defined
\begin{equation}
  C \equiv \frac{1}{2} \sin(\zeta) \Gamma^{\hat{\zeta} \hat{\phi_1}} \mp \frac{1}{2} \cos(\zeta) \Gamma^{\hat{\zeta}\hat{\phi_2}}.
\end{equation}
This system of equation can be analyzed analogously to the $S^3$ case,
and we find the solution
\begin{equation}
  \tilde{\epsilon}(\rho,t,\theta) =
  \exp \left[ \pm \frac{1}{2} \rho \Gamma^{\hat{t}\hat{\theta}} \right]
  \exp \left[ +\frac{1}{2} (\theta \mp t) \Gamma^{\hat{\rho}\hat{\theta}} \right]
  \tilde{\epsilon_0},
\end{equation}
where $\tilde{\epsilon_0}$ is a constant spinor on $AdS_3$.
Thus, the full solution to (\ref{AdSgravitino}) is
\begin{equation}\label{AdSgravitinosolution}
\begin{split}
  &\epsilon(\zeta,\phi_1,\phi_2,\rho,t,\theta)  = \epsilon(\zeta,\phi_1,\phi_2) \tilde{\epsilon}(\rho,t,\theta) = \\
  & = \exp \left[ \pm \frac{1}{2} \zeta \Gamma^{\hat{\phi_1}\hat{\phi_2}} \right]
  \exp \left[ +\frac{1}{2} (\phi_2 \mp \phi_1) \Gamma^{\hat{\zeta}\hat{\phi_2}} \right]
  \exp \left[ \pm \frac{1}{2} \rho \Gamma^{\hat{t}\hat{\theta}} \right]
  \exp \left[ +\frac{1}{2} (\theta \mp t) \Gamma^{\hat{\rho}\hat{\theta}} \right]
  \epsilon_0,
\end{split}
\end{equation}
where $\epsilon_0$ is a constant spinor on $AdS_3 \times S^3$.

\subsection{Comparison to Mikhailov}\label{appglobalMikhailov}
In this section, we compare our gravitino equations
to those of Mikhailov \cite{Mikhailov:2000ya}.
In particular, Mikhailov writes down the
general form of the equation a spinor in
global $AdS_3 \times S^3$ must satisfy as
\begin{equation}\label{Mikh}
  \left( \frac{D}{Dx^{p}} - \frac{1}{2} \frac{\partial f}{\partial R} \Gamma_{p} \Gamma^{\hat{0}}\Gamma^{\hat{1}}\Gamma^{\hat{2}} \right) \epsilon_{++} = 0,
\end{equation}
where $p=0,1,2$ and $0,1,2$ are the three coordinates on either $AdS_3$ or $S^3$.
The spinor $\epsilon_{++}$ is defined by requiring
that $\sigma_1 \epsilon_{++} = + \epsilon_{++}$.
The equation (\ref{Mikh}) presupposes an
embedding in a higher-dimensional space.
In more detail, we can embed $M=S^3$ or $AdS_3$ in $N=\mathbb{R}^4$ or $\mathbb{R}^{2,2}$ as
\begin{equation}
  ds_N^2 = dR^2 + e^{2f(R)}  ds_{M}^2 = dR^2 + R^2 d\Omega^2_M,
    \Rightarrow  2 f(R) = \log(R^2) \Rightarrow \frac{\partial f}{\partial R} = \frac{1}{R},
\end{equation}
where $R$ is a radial coordinate in $N$.
For us, $R^2=(Q_1 Q_5)^{1/2}$, which means that equation (\ref{Mikh}) becomes
\begin{equation}\label{Mikhailov}
  \left(
    \frac{D}{Dx^p} - \frac{1}{2} (Q_1 Q_5)^{-1/4} \Gamma^{\hat{0}} \Gamma^{\hat{1}} \Gamma^{\hat{2}} \Gamma_p
  \right) \epsilon_{++} = 0.
\end{equation}
Mikhailov's equation (\ref{Mikhailov}) is
indeed equivalent to
(\ref{D1D5gravitinoNHG}) and (\ref{AdSgravitino}).

\bibliographystyle{JHEP}
\bibliography{references}

\end{document}